\documentclass[twocolumn]{aastex6}

\usepackage{multirow}
  
\newcommand{\kepler}{\emph{Kepler}}
\newcommand{\ktwo}{\emph{K2}}
\newcommand{\prae}{Praesepe}
\newcommand{\gpe}{GP--EBOP}
\newcommand{\ksc}{{\sc k2sc}}
\newcommand{\spitzer}{\emph{Spitzer}}

\newcommand{\etop}{\emph{top}}

\newcommand{\ebot}{\emph{bottom}}
\newcommand{\eleft}{\emph{left}}
\newcommand{\eright}{\emph{right}}

\newcommand{\eTop}{\emph{Top}}

\newcommand{\eBot}{\emph{Bottom}}

\newcommand{\teff}{$T_{\rm eff}$}
\newcommand{\logg}{$\log g$}

\newcommand{\linejump}{0.9}

\shorttitle{Low-mass EBs in Praesepe}
\shortauthors{E. Gillen et al.}
\begin{document}

\title{New low-mass eclipsing binary systems in Praesepe discovered by \ktwo}

\author{
Edward Gillen\altaffilmark{1}, 
Lynne A. Hillenbrand\altaffilmark{2}, 
Trevor J. David\altaffilmark{2}, 
Suzanne Aigrain\altaffilmark{3}, 
Luisa Rebull\altaffilmark{4}, 
\\John Stauffer\altaffilmark{4}, 
Ann Marie Cody\altaffilmark{5} 
\& Didier Queloz\altaffilmark{1}}

\indent \altaffiltext{1}{Astrophysics Group, Cavendish Laboratory, J.J. Thomson Avenue, Cambridge CB3 0HE, UK.}
\altaffiltext{2}{Department of Astronomy, California Institute of Technology, Pasadena, CA 91125, USA}
\altaffiltext{3}{Sub-department of Astrophysics, Department of Physics, University of Oxford, Keble Road, Oxford, OX1 3RH, UK}
\altaffiltext{4}{Spitzer Science Center, California Institute of Technology, 1200 E California Blvd., Pasadena, CA 91125, USA}
\altaffiltext{5}{NASA Ames Research Center, Moffet Field, CA 94035, USA}

\email{ecg41@cam.ac.uk}



\begin{abstract}

We present the discovery of four low-mass ($M<0.6$\,$M_\odot$) eclipsing binary (EB) systems in the sub-Gyr old \prae\ open cluster using \kepler/\ktwo\ time-series photometry and Keck/HIRES spectroscopy. We present a new Gaussian process eclipsing binary model, \gpe, as well as a method of simultaneously determining effective temperatures and distances for EBs.
Three of the reported systems (AD 3814, AD 2615 and AD 1508) are detached and double-lined, and precise solutions are presented for the first two. We determine masses and radii to 1--3\% precision for AD 3814 and to 5--6\% for AD 2615. 
Together with effective temperatures determined to $\sim$50\,K precision, we test the PARSEC v1.2 and BHAC15 stellar evolution models. Our EB parameters are more consistent with the PARSEC models, primarily because the BHAC15 temperature scale is hotter than our data over the mid M-dwarf mass range probed.
Both ADs 3814 and 2615, which have orbital periods of 6.0 and 11.6 days, are circularized but not synchronized. This suggests that either synchronization proceeds more slowly in fully convective stars than the theory of equilibrium tides predicts or magnetic braking is currently playing a more important role than tidal forces in the spin evolution of these binaries.
The fourth system (AD 3116) comprises a brown dwarf transiting a mid M-dwarf, which is the first such system discovered in a sub-Gyr open cluster.
Finally, these new discoveries increase the number of characterized EBs in sub-Gyr open clusters by 20\% (40\%) below $M<1.5$\,$M_{\odot}$ ($M<0.6$\,$M_{\odot}$).

\end{abstract}




\section{Introduction}

% Background
Stellar evolution theory underpins much of observational astrophysics, yet significant uncertainties remain at low masses ($M\lesssim 0.8$ $M_{\odot}$) and young ages ($t \lesssim 1$ Gyr).
Unfortunately, this mass and age range is also where observational constraints are scarce. 
The fundamental goal of stellar evolution theory is to accurately predict the observables (radius, temperature and luminosity) for a star of given mass, age and metallicity.
The evolutionary pathway of a star is governed primarily by its mass, which is accessible only through study of gravitational interactions such as in binary or higher order multiple star systems.  For eclipsing binaries (EBs) the ratio of the radii of the stars is attainable.  
Eclipsing binaries are particularly important objects if they are also detected as double-lined systems in spectra, as the individual masses and radii of both stars can be extracted from the combined light curves and radial velocity curves of the system.  Radii can also be measured directly using interferometric techniques, but only for the brightest of nearby stars. When the inferred mass and radius values reach a precision of a few percent or less, they provide one of the strongest observational tests of stellar evolution theory available \citep[e.g.][]{Torres10,Stassun14}. 

% Open clusters
Open clusters are fruitful astrophysical laboratories given that their members share broad coevality, composition and distance. 
The detection of multiple EBs in a given cluster, with each member of each pair sharing the same age and metallicity but spanning a range of masses, offers a particularly strong test of stellar evolution theory. 
The pursuit of EB parameters, among other science goals, has motivated numerous programmes to target open clusters via time-series photometry, e.g. the ground-based Monitor, PTF Orion and YETI projects \citep{Aigrain07,vanEyken11,Neuhauser11}, and space-based observations with CoRoT and \spitzer\ \citep{Gillen14,Morales-Calderon12}.
Furthermore, since March 2014, the re-purposed Kepler mission, \ktwo\ \citep{Howell14}, has targeted a number of star forming regions and young (sub-Gyr) open clusters across the ecliptic for $\sim$80 days each.
To date, the nearby $\rho$ Ophiuchi star forming region and Upper Scorpius young OB association ($\sim$1 and 5-10 Myr, respectively) were observed in Campaign 2, as were the Pleiades and Hyades open clusters ($\sim$125, 600--800 Myr, respectively) in Campaign 4, and Praesepe (600--800 Myr) in Campaign 5.

% Praesepe literature review
The \prae\ open cluster, also known as the Beehive cluster or M44, was targeted by \ktwo\ in Campaign 5 (April--July 2015). \prae\ is a relatively nearby, metal-rich, several hundred Myr cluster hosting $>$1000 high probability members ($>$80\%) and more than 100 candidate members ($>$50\% probability) \citep{Kraus07,Rebull17}. Given its richness and proximity, Praesepe is a well-studied benchmark cluster. 
The parallaxes of bright \prae\ members in the GAIA DR1 suggest a distance of $182.8\pm1.7\pm14$ pc, where the first error represents the uncertainty on the cluster center determination and the second reflects the observed radial spread of high probability members on the sky \citep{vanLeeuwen17}. 
This is in agreement with the commonly quoted Hipparcos distance to the cluster, $181.5\pm6.0$ pc \citep{vanLeeuwen09}. The cluster has a low reddening along the line of sight of $E(B-V) = 0.027\pm0.004$ \citep{Taylor06}.
Metallicity estimates typically fall within the range [Fe/H] $\sim$ 0.12--0.16 \citep[e.g.][]{Boesgaard13,Yang15,Netopil16} but can be as high as [Fe/H] = $0.27\pm0.10$ \citep[e.g.][]{Pace08}.
The age of \prae\ is estimated in the range $\sim$600--900 Myr \citep[e.g.][]{Adams02} with traditional estimates typically falling at the lower end, often through association with the Hyades \citep[e.g.][]{Salaris04}. More recently, however, \citet{Brandt15} included stellar rotation 
to conclude that the upper main sequences of both Praesepe and the Hyades were consistently well-fit at an age of $\sim$750--800 Myr. The age of Praesepe is further discussed in section \ref{age_discussion}.

% Binarity in Praesepe
The binary fraction within the cluster has been extensively studied. \citet{Pinfield03} noted that binaries in Praesepe appear to favor similar-mass systems.
\citet{Boudreault12} focused on the low-mass population, finding binary frequencies of: $25.6\pm3.0$\% between 0.2\,$<$\,M\,$<$\,0.45 $M_{\odot}$, $19.6\pm3.0$\% between 0.1\,$<$\,M\,$<$\,0.2 $M_{\odot}$ and $23.2\pm5.6$\% between 0.07\,$<$\,M\,$<$\,0.1 $M_{\odot}$.
\citet{Wang14} analyzed the full Praesepe membership to find a binary occurrence rate of 20--40\%. Furthermore, a significant population of binaries and higher order systems were identified by \citet{Khalaj13}, who propose a binary fraction of $35\pm5$\% in the mass range 0.6--2.2 $M_{\odot}$, assuming mass-dependent pairing of primary stars following the results of recent star formation simulations \citep[e.g.][]{Bate09}.

This paper presents the characterization of four high-probability, low-mass eclipsing binary members of Praesepe.
\S \ref{objects} describes the sources and previous literature characterization.
In \S \ref{observations} we detail the photometric and spectroscopic observations. In \S \ref{analysis} we present a modified eclipsing binary model for detached systems, \gpe, and describe the light curve and radial velocity analyzes. We then present the results for each system in \S \ref{results}. In \S \ref{discussion} we present an updated method to simultaneously determine the effective temperatures of both stars as well as the distance to an EB system, before discussing these new EBs in the context of calibrating stellar evolution models, and informing tidal evolution theory in close binaries. Finally, we conclude in \S \ref{conclusions}.

%%%%%%%%%%%%%%%%%%%%%%%%%%%%%%%%%%%%%%%%%%%%%%%%%%
\section{New Eclipsing Binaries Among Praesepe Members}
\label{objects}

Half a dozen deep proper motion surveys of Praesepe have been
published since 2000 \citep{Adams02,Kraus07,Baker10,Boudreault12,
Khalaj13,Wang14}.
Three of our four EBs are considered Praesepe members in at least four of
those six studies (AD 3814, 2615 and 3116). Our fourth EB (AD 1508) is 
identified as a Praesepe member in only two of those studies.   

\begin{figure}
  \centering
  \includegraphics[width=0.9\linewidth]{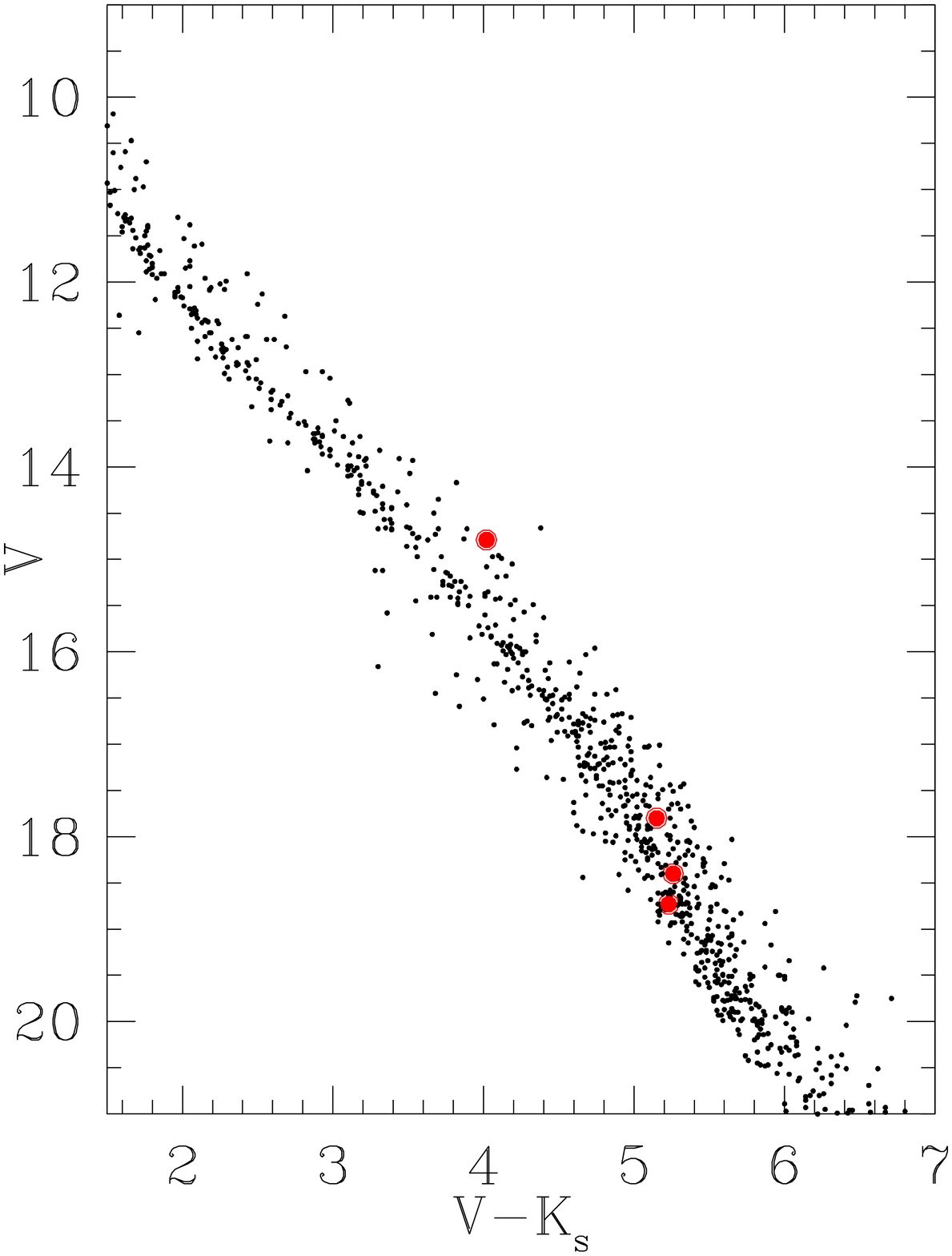}
  \includegraphics[width=0.9\linewidth]{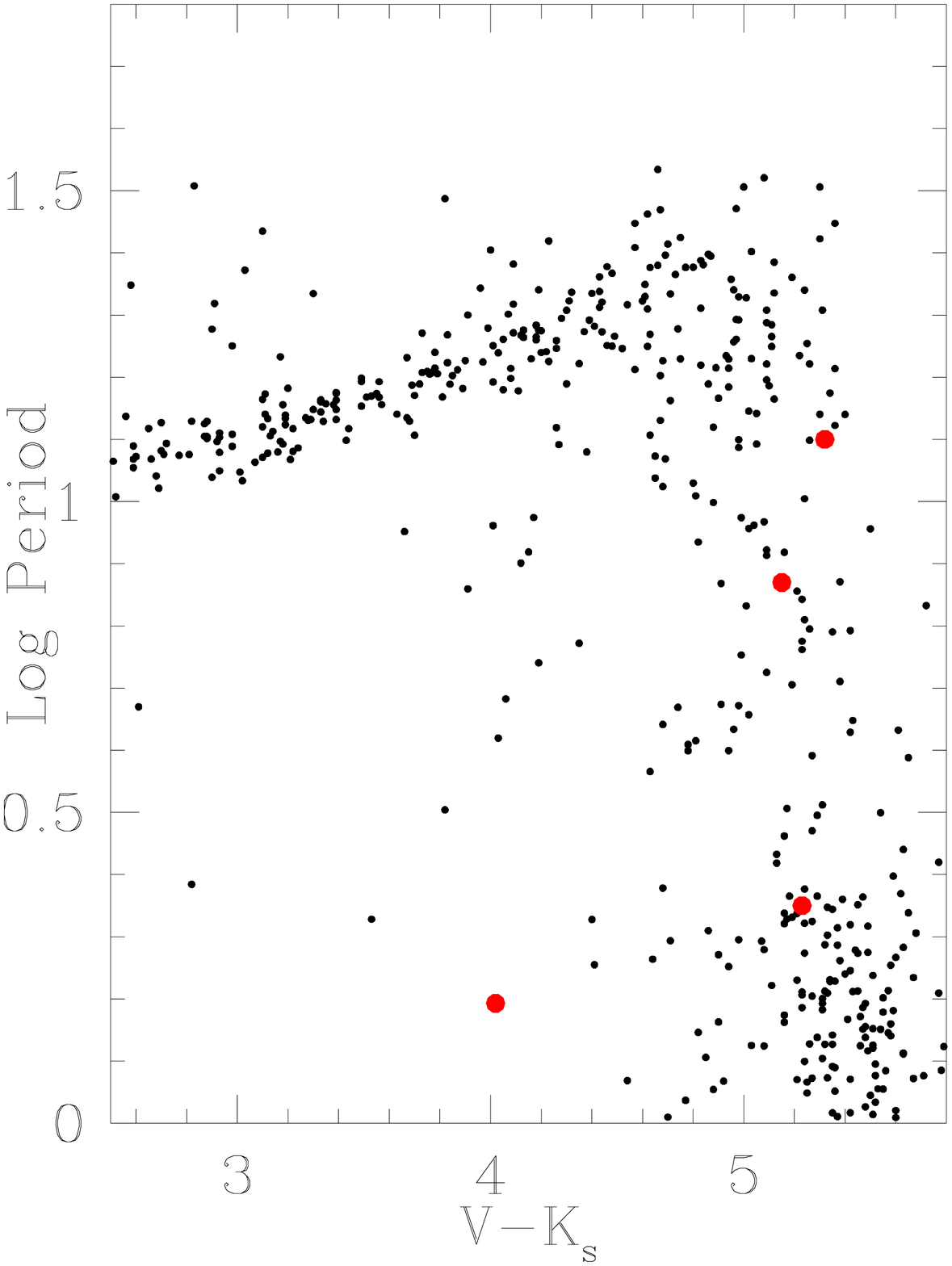}
  \caption{Color -- magnitude diagram (\etop) and color -- rotation period diagram (\ebot) illustrating the location of the four new eclipsing systems relative to the sequence of Praesepe members. \eTop: From brightest to faintest are: AD 1508, AD 3814, AD 2615, and AD 3116 with elevation above the color-magnitude sequence a rough indicator of the mass ratio of a binary system (equal mass ratio produces a 0.75 mag magnitude excess). \eBot: from slowest to fastest rotators are: AD 2615, AD 3814, AD 3116 and AD 1508. }
  \label{cmd}
\end{figure}

In the top panel of Figure~\ref{cmd},
we show where these four objects fall in a V vs. V-K$_s$\ color-magnitude diagram, where we have
derived V-K$_s$\ estimates based on a conversion \citep{Rebull17} 
from G-K$_s$, where G is
the star's magnitude in the Gaia DR1 catalog.   All four stars have photometry
consistent with Praesepe membership.  
AD 1508 is the earliest type (brightest)
of the four; it is located well above the single star main-sequence locus,
suggesting that it is a nearly equal mass binary.  AD 3116 and 3814 are located
nearly on the single star main-sequence locus, and so their binary companions
are presumably very low mass.  AD 2615 is displaced about 0.4 mag above the
single star locus, and so is likely to have an intermediate mass binary
companion.   

Three of the four stars have published spectral types:
AD 3814 - M5 \citep{West11}; AD 2615 - M4.0 \citep{Adams02}, M5 \citep{West11}; and AD 3116 - M4.5 \citep{Adams02}, M3.9 (\citealt{Kafka06}. These spectral types are broadly consistent with their V-K$_s$\ colors. 
All four systems have spectral types estimated from photometry \citep{Kraus07}: AD 3814 - M$3.4\pm0.1$; AD 2615 - M$4.0\pm0.1$; AD 3116 - M$3.9\pm0.1$; and AD 1508 - M$0.1\pm0.1$. As these form a homogeneous set for our EBs, we adopt these spectral types here. 
For each system, properties extracted from the literature are reported in Table \ref{info_tab}.

In the bottom panel of Figure \ref{cmd} we show the Praesepe V vs. V-K$_s$ color vs. rotation period diagram and indicate our four systems in red. Given the wide spread in rotation periods for mid--M dwarfs, ADs 3814, 2615 and 3116 all lie along the single star trend, but the early--M dwarf AD 1508 lies far below the single star trend with a short rotation period.

%%%%%%%%%%%%%%%%%%%%%%%%%%%%%%%%%%%%%%%%%%%%%%%%%%
\section{Observations}
\label{observations}

% ==========================================
\subsection{Photometry}

We proposed targets for the \ktwo\ Campaign 5 observations, which included \prae, as part of the \ktwo\ Young Suns Survey (PI Stauffer). Targets were collated through merging various proper motion surveys \citep{Klein-Wassink27,Jones83,Jones91,Kraus07,Wang14} with published BVRI photometry \citep[][and references therein]{Mermilliod90,Stauffer82}. The K2FOV tool was used to select targets falling `on silicon' and we further limited our proposal to stars with spectral type later than F0 (i.e. possessing outer convective envelopes) and brighter than $R < 17$. This gave 477 high-probability \prae\ targets in total. In addition to our proposed systems we also investigated light curves of Praesepe candidates from other \ktwo\ programs.  

The \ktwo\ observations of \prae\ spanned 27 April -- 10 July 2015 and the FoV centerd on 08:40:38	+16:49:47. Given the typical 30-minute cadence of \kepler\ observations, this resulted in $\sim$3300 data points for each target. Short cadence (1 min) observations are also possible for a small number of targets but all systems presented here were observed in standard long cadence mode. We discuss our method to reduce the \ktwo\ photometry in \S\ref{k2sc}. For objects showing the signatures of eclipses in the \ktwo\ time series photometry,  we cross-referenced the EPIC identifiers with literature information in order to determine basic system properties (see \S\ref{par_est}) and to identify which systems to pursue with high dispersion spectroscopy (see \S\ref{spectroscopy}).

% ==========================================
\subsubsection{\ktwo\ data detrending and eclipse detection}
\label{k2sc}

We started from the Simple Aperture Photometry (SAP) light curves, which were made available at the Mikulski Archive for Space Telescopes (MAST) as part of \ktwo\ Data Release 7\footnote{See {\tt https://keplerscience.arc.nasa.gov/\\k2-data-release-notes.html\#k2-campaign-5} for details.}. We used the \ksc\ pipeline \citep{Aigrain16} to correct the light curves for systematics caused by the quasi-periodic rolling motion of the spacecraft, while preserving the intrinsic variability of the target stars. \ksc\ works by modeling the SAP flux as the sum of two smooth, random functions: one depending on the star's position on the detector, and one depending on time, plus white noise. The position component represents instrumental systematics associated with the satellite's pointing variations (mainly intra- and inter-pixel sensitivity variations), while the time component represents the star's intrinsic variability, plus any long-term instrumental effects not accounted for by the position component. Both components are modeled using Gaussian Process (GP) regression (see \S\ref{gpe_model_sec} for further details and references on GPs). While both components are initially treated as aperiodic, a quasi-periodic GP is automatically used for the time component if the light curve shows any evidence of periodic behavior after a first pass treatment with default parameters.

A careful treatment of outliers ensures that \ksc\ mostly preserves short-duration events such as planetary transits or stellar eclipses. However, once the eclipses were identified (by visual examination) in the four systems discussed in the present paper, their light curves were re-processed using \ksc's periodic mask option. This option enables the user to supply the period, epoch and duration of the eclipses, and any in-eclipse points are then ignored when training the GP model. 
In effect, we are using the \ksc\ GP model to interpolate in both flux and position space to the times affected by the eclipses, thereby providing a model prediction for the total system flux across each eclipse. In our analysis we use the \ksc\ light curve that has been detrended for instrument systematics but which retains the stellar variability component. This allows us to simultaneously model both the stellar variability and eclipses (see \S\ref{analysis}).

% ==========================================
\subsubsection{Estimation of primary star properties from broadband colors}
\label{par_est}

We estimated primary effective temperatures and masses using broadband color relations and absolute magnitudes presented in Table \ref{info_tab}, respectively.

Effective temperatures (\teff) were estimated using the empirical color-\teff\ relations presented in \citet{Mann16} (their eq. 6) and \citet{David16a} (their eq. 1, which is derived from fitting polynomials to the color and temperature data presented in \citet{Pecaut13} for dwarf stars, and is valid for $0.3 < V-K_{s} < 7.0$). These predict primary effective temperatures of $\sim$3250, 3190, 3240 and 3750 K for ADs 3814, 2615, 3116 and 1508, respectively. In \S\ref{SED_sec} we directly determine the effective temperatures of both stars in each EB through modeling their spectral energy distributions (SEDs) and compare our \teff\ values to these empirical predictions in Table \ref{Td_comp_tab}.

We estimated primary masses from absolute K band magnitudes using the semi-empirical relation of \citet{Mann16} (their eq. 10) and the empirical relation of \citet{Benedict16} (their eq. 11). For this, we converted apparent to absolute magnitudes assuming a cluster distance of $182.8\pm14$ pc \citep{vanLeeuwen17} and assumed a reddening along the line of sight of $E(B-V) = 0.027\pm0.004$ \citep{Taylor06}. These two relations predict primary masses of: $\sim$0.43, 0.34, 0.28 and 0.72 $M_{\odot}$ for ADs 3814, 2615, 3116 and 1508, respectively. For AD 1508 we used only the \citet{Mann16} mass prediction as this system lies outside the validity range ($0.1 \lesssim M \lesssim 0.6$ $M_{\odot}$) of the Benedict relation.

We note that these predictions are for single stars and hence are not appropriate for binary systems unless the system magnitudes are dominated by the primary component. Furthermore, these empirical relations are approximations only and are estimated from systems that typically do not contain as high a metallicity as Praesepe ([Fe/H] $\sim$ 0.1--0.27).   
Nonetheless, they serve to highlight the expected temperature and mass regimes of the systems to be analyzed.

\begin{table*}
  \centering
  \caption{Names, coordinates, properties and membership information for the four newly identified EBs.} 
  \label{info_tab}
  \resizebox{\textwidth}{!}{%
  \begin{tabular}{llccccr}
    \hline
    \hline
    \noalign{\smallskip}
    Property  &  Units  &  AD 3814  &  AD 2615 &  AD 3116  &  AD 1508  &  Refs.  \\
    \noalign{\smallskip}
    \hline
    \noalign{\smallskip}
    
    EPIC  &    &    211972086  &  212002525  &  211946007  & 212009427 &   \\ [-0.5ex]
    2MASS  &    &    J08504984+1948364  &  J08394203+2017450  &  J08423943+1924520  &  J08312987+2024374  & \\ [-0.5ex]
    Other names  &    &    ...  &  ... &  HSHJ 430  &  ...  &  1  \\ [-0.5ex]
% non-coordinate
    RA  &  J2000.0  &    08:50:49.84  & 08:39:42.03  & 08:42:39.43  &  08:31:29.87  &   \\  [-0.5ex]
    Dec  &  J2000.0  &    +19:48:36.4  &  +20:17:45.0  &  +19:24:51.9  &  +20:24:37.5  &  \\  [-0.5ex]

 \noalign{\smallskip} \noalign{\smallskip} \noalign{\smallskip}
 
     $u$  &  AB  &   $21.009\pm0.093$   &  $21.747\pm0.185$  &  $22.290\pm0.190$ & $18.102\pm0.014$  &  2   \\ [-0.5ex]
     $g$  &  AB  &    $18.769\pm0.008$  &  $19.416\pm0.012$  &  $19.646\pm0.014$ & $15.540\pm0.004$  &  2    \\ [-0.5ex]
     $r$  &  AB  &    $17.299\pm0.006$  &  $17.905\pm0.007$   &  $18.206\pm0.007$  & $14.151\pm0.004$  &  2   \\ [-0.5ex]
     $i$  &  AB  &    $15.803\pm0.005$   &  $16.324\pm0.004$  &  $16.675\pm0.005$  & $13.700\pm0.001$  &  2    \\ [-0.5ex]
     $z$  &  AB  &    $14.999\pm0.005$  &  $15.456\pm0.006$  &  $15.845\pm0.006$  &  $12.905\pm0.004$ &  2    \\ [-0.5ex]
     
      V  &  Vega  &   17.80  &  18.46  &  18.73  & 14.79  &  3  \\ [-0.5ex] 

     $J$  &  Vega  &    $13.529\pm0.026$  &  $14.027\pm0.021$  &  $14.348\pm0.032$  & $11.674\pm0.022$ &  4  \\ [-0.5ex]     
     $H$  &  Vega  &    $12.911\pm0.024$  & $13.456\pm0.026$  &  $13.769\pm0.037$  & $10.949\pm0.023$ &  4  \\  [-0.5ex]    
    $K_{s}$  &  Vega  &   $12.651\pm0.022$   & $13.136\pm0.034$  &  $13.499\pm0.043$  & $10.767\pm0.020$ &  4  \\  [-0.5ex]    

   WISE 1  &  Vega  &  $12.478\pm0.024$  &  $12.938\pm0.024$  &  $13.299\pm0.029$  &  $10.677\pm0.023$  &  4  \\ [-0.5ex]
   WISE 2  &  Vega  &  $12.291\pm0.026$  &  $12.773\pm0.031$  &  $13.096\pm0.039$  &  $10.638\pm0.021$  &  4  \\ [-0.5ex]

 \noalign{\smallskip}  \noalign{\smallskip} \noalign{\smallskip}
        
    Spectral type  &  M sub-type  &  $3.4\pm0.1$  &  $4.0\pm0.1$  &  $3.9\pm0.3$  & $0.1\pm0.1$  &  5 \\ [-0.5ex]
    H$\alpha$ emission  &  \AA\  & 2.4--3.5, ...  &  3.0--4.3, 10.7  &   3.1--5.2, 4.6  & 2.0--2.1, ... &  6,7 \\  [-0.5ex]
    
    RA proper motion,  $\mu_{\alpha}$   &  mas yr$^{-1}$   &  -37.5  &  -39.3  &  -37.5  & -37.3  &  5  \\ [-0.5ex]
    Dec proper motion, $\mu_{\delta}$   &  mas yr$^{-1}$   &  -14.1  &  -11.6  &  -8.2  & -16.7  &  5   \\ [-0.5ex]
    
    Membership probability   &  \%   &  97.9  &  99.7  &  99.1  &  98.3 &  5  \\
    
    \noalign{\smallskip}
    \hline
  \end{tabular} }
  \begin{list}{}{}  
   \item[\textbf{Notes.}]{The quoted photometric uncertainties are formal measurement errors and hence do not capture the intrinsic variability of these systems. }
     \item[\textbf{References.}] 1. \citet{Hambly95}; 2. Sloan Digital Sky Survey Data Release 13; 3. \citet{Rebull17}; 4. NASA/IPAC Infrared Science Archive; 5. \citet{Kraus07}; 6. This work, with quoted range as measured over the epochs listed in Table~\ref{RVs_tab}; 7. \citet{Adams02}.
  \end{list}
\end{table*}

% =======================================
\subsection{Spectroscopy}
\label{spectroscopy}

We obtained high resolution spectra for each of the identified eclipsing binary systems using the Keck HIRES spectrograph \citep{Vogt94}.
The observations were taken between 2015 December and 2017 January, with the exact epochs along with estimated signal-to-noise ratios and measured radial velocities given in Table \ref{RVs_tab}.
The spectra cover the wavelength range $\approx$4800--9200  \AA\ at a spectral resolution of  $R > 36,000$, and were reduced using the $makee$ software written by Tom Barlow.
We measured radial velocities using the cross correlation techniques within the $fxcor$ task in $IRAF$, with absolute reference to between 3 and 5 (depending on the night) late type radial velocity standards.  The standards and their approximate spectral types include: GJ 514 (M0.5), HD 95650 (M1), LHS 3433B (M2), Gl821 (M2), GJ 408 (M2.5), GJ 176 (M2.5), GJ 109 (M3.5), GJ 402 (M4), Gl 876 (M4), GJ 105B (M4.5), GJ 388 (M4.5), GJ411 (M4.5), GJ 406 (M6.5), with the reference velocities generally taken from \cite{Nidever02}. Telluric-free spectral regions were selected over between 6 and 19 orders (depending on the signal-to-noise of the target spectrum) for cross correlation function fitting.  Depending on the velocity separation of the peaks, they were fit either singly or simultaneously, and depending on the signal-to-noise of the spectrum, the fitting function was either Gaussian or parabolic.
Errors in the quoted radial velocities were determined from the empirical scatter among the measured orders and reference stars for each observation, with some hand editing to remove extreme outliers deriving from particularly poor measurements.  In general, the scatter among the measurements that is quoted as the radial velocity error, is smaller than or comparable to the mean among the errors in the individual measurements over the orders and reference stars included in the quoted radial velocity value.  This gives us some confidence that we are accurately representing the random errors in our methods.

AD 3814, AD 2615, and AD 1508 are detected as double-lined systems, 
with measurable radial velocities for each component at nearly all epochs.
AD 3116, however, presented only a single line set, which we attribute to the primary.
In the double-lined systems, the CCF peak height ratios were used to approximate the light ratio between the two components, which was then applied as a prior in the light curve modeling (see \S\ref{analysis}). 

In addition to the radial velocities, H$\alpha$ equivalent width measurements were made for each EB using the $splot$ task in $IRAF$.  The values quoted in Table \ref{info_tab} represent the combined system and the range records the variability over the various epochs of observation in Table \ref{RVs_tab}.

\begin{table*}  
 \centering  
 \caption{Radial velocities derived from Keck/HIRES spectra for ADs 3814, 2615, 3116 and 1508 (\etop\ to \ebot). } 
 \label{RVs_tab}  
 \begin{tabular}{c c c c r r }  
 \noalign{\smallskip} \noalign{\smallskip} \hline  \hline \noalign{\smallskip} 
   \multicolumn{3}{c}{Epoch}  &  S/N & \multicolumn{2}{c}{RV ~ (km\,s$^{-1}$)} \\
   UT date  &  BJD  &  Phase\,*  & 7500 \AA\ &  Primary~~~~  &  Secondary~~  \\

\noalign{\smallskip} \hline \noalign{\smallskip} \noalign{\smallskip}

\multicolumn{6}{c}{
....................................................... ~ AD 3814 ~ .......................................................
} \\  
 
2015\,12\,24  &  2457381.15090  &  0.607  &16&  $54.08\pm0.77$  &  $-6.83\pm0.93$  \\ [-0.5ex]
2015\,12\,29  &  2457386.14539  &  0.437  &16&  $21.42\pm0.76$  &  $58.70\pm1.10$  \\ [-0.5ex]
2016\,02\,02  &  2457420.89940  &  0.214  &16&  $0.10\pm0.75$   &  $95.82\pm1.13$   \\ [-0.5ex]
2016\,02\,03  &  2457421.92652  &  0.385  &15&  $12.91\pm0.76$  &  $77.18\pm0.93$   \\  [-0.5ex]
2016\,05\,17  &  2457525.80479  &  0.652  &14&  $60.96\pm0.37$  &  $-19.29\pm1.15$   \\  [-0.5ex]
2016\,12\,22  &  2457744.96970  &  0.084  &15&  $15.72\pm0.35$  &  $65.38\pm0.56$  \\ [-0.5ex]
2016\,12\,26  &  2457748.97551  &  0.750  &13&  $68.56\pm1.04$  &  $-28.91\pm1.18$  \\  [-0.5ex]
2017\,01\,13  &  2457766.85771  &  0.723  &12&  $67.17\pm0.40$  &  $-29.35\pm0.56$  \\ [-0.5ex]
 
 \noalign{\smallskip} \noalign{\smallskip} \noalign{\smallskip}

\multicolumn{6}{c}{
....................................................... ~ AD 2615 ~ .......................................................
} \\  
 
2015\,12\,29  &  2457386.16741  &  0.039  &13&  $26.03\pm0.86$  &  $43.50\pm0.77$  \\  [-0.5ex]
2016\,05\,17  &  2457525.78402  &  0.059  &13&  $20.45\pm0.76$  &  $45.69\pm0.80$  \\ [-0.5ex]
2016\,05\,20  &  2457528.78074  &  0.317  &14&  $-1.28\pm0.60$  &  $65.74\pm0.60$   \\ [-0.5ex]
2016\,10\,14  &  2457676.07100  &  0.997  &10&  \multicolumn{2}{c}{$35.22\pm0.29$} \\ [-0.5ex]
2016\,12\,22  &  2457745.03382  &  0.935  &13&  $49.63\pm0.48$  &  $20.66\pm0.41$ \\ [-0.5ex]
2017\,01\,13  &  2457766.90914  &  0.818  &5&  $72.09\pm0.53$  &  $5.63\pm0.60$ \\ [-0.5ex]

  \noalign{\smallskip} \noalign{\smallskip} \noalign{\smallskip}

\multicolumn{6}{c}{
....................................................... ~ AD 3116 ~ .......................................................
} \\  
 
2016\,02\,02  &  2457420.92116  &  0.102  &12&  $26.28\pm0.82$  &  --- ~~~~    \\ [-0.5ex]
2016\,02\,03  &  2457421.90606  &  0.599  &12&  $40.47\pm0.83$  &  --- ~~~~    \\ [-0.5ex]
2016\,05\,17  &  2457525.76250  &  0.978  &12&  $39.75\pm0.59$  &  --- ~~~~    \\ [-0.5ex]
2016\,05\,20  &  2457528.75747  &  0.488  &13&  $27.46\pm0.54$  &  --- ~~~~    \\ [-0.5ex]
2016\,10\,14  &  2457676.09435  &  0.796  &13&  $55.96\pm0.30$  &  --- ~~~~    \\ [-0.5ex]
2016\,12\,22  &  2457744.98886  &  0.542  &12&  $31.91\pm0.44$  &  --- ~~~~    \\ [-0.5ex]
2017\,01\,13  &  2457766.87986  &  0.583  &6&  $37.23\pm0.77$  &  --- ~~~~    \\ [-0.5ex]

  \noalign{\smallskip} \noalign{\smallskip} \noalign{\smallskip}

\multicolumn{6}{c}{
....................................................... ~ AD 1508 ~ .......................................................
} \\  

2016\,12\,22  &  2457745.047527495  &  0.971  &40&  $50.62\pm1.40$  &  $16.36\pm1.57$ \\ [-0.5ex]
2016\,12\,26  &  2457748.953315984  &  0.479  &30&  $21.66\pm2.79$  &  $42.37\pm3.23$ \\ [-0.5ex]
2017\,01\,13  &  2457766.842635326  &  0.970  &40&  $52.56\pm2.19$  &  $18.69\pm1.82$ \\ [-0.5ex]

 \noalign{\smallskip} \noalign{\smallskip}\noalign{\smallskip}  
 \hline  
 \end{tabular}  
\begin{list}{}{}  
\item[*] Phase is defined relative to primary eclipse.   
\end{list}  
 \end{table*}

 \begin{figure}
   \centering
   \includegraphics[width=\linewidth]{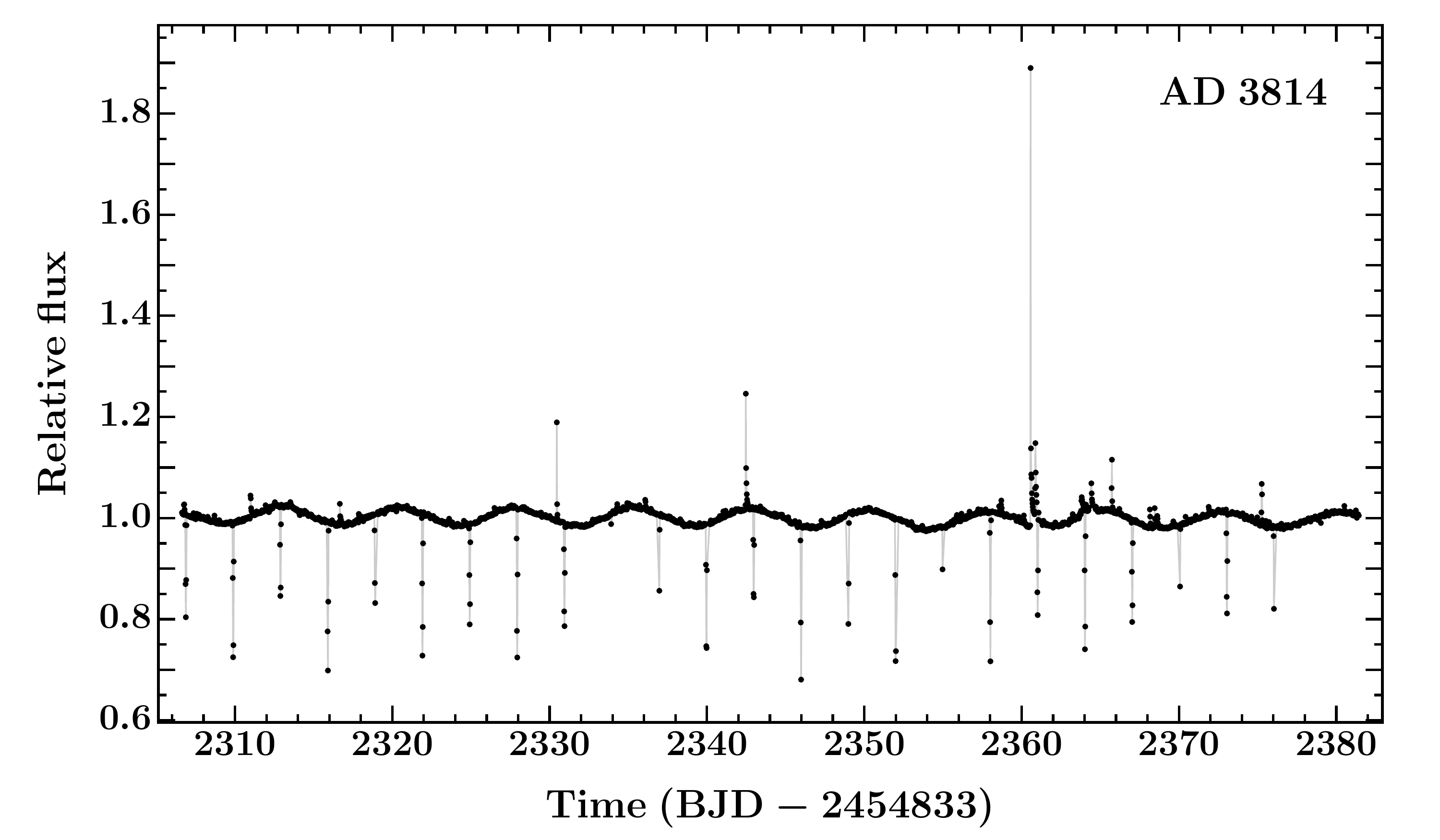}
    \includegraphics[width=\linewidth]{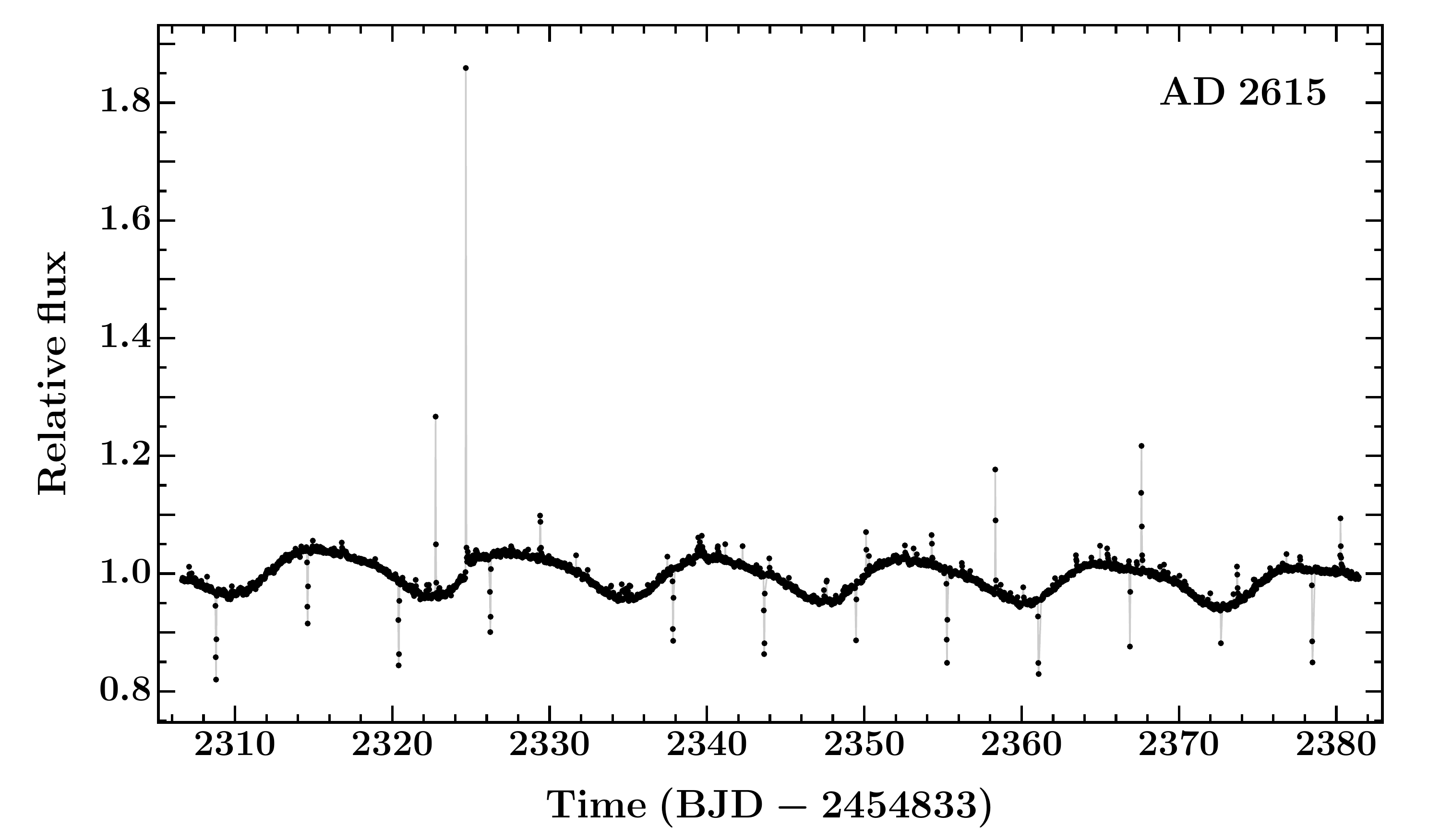}
    \includegraphics[width=\linewidth]{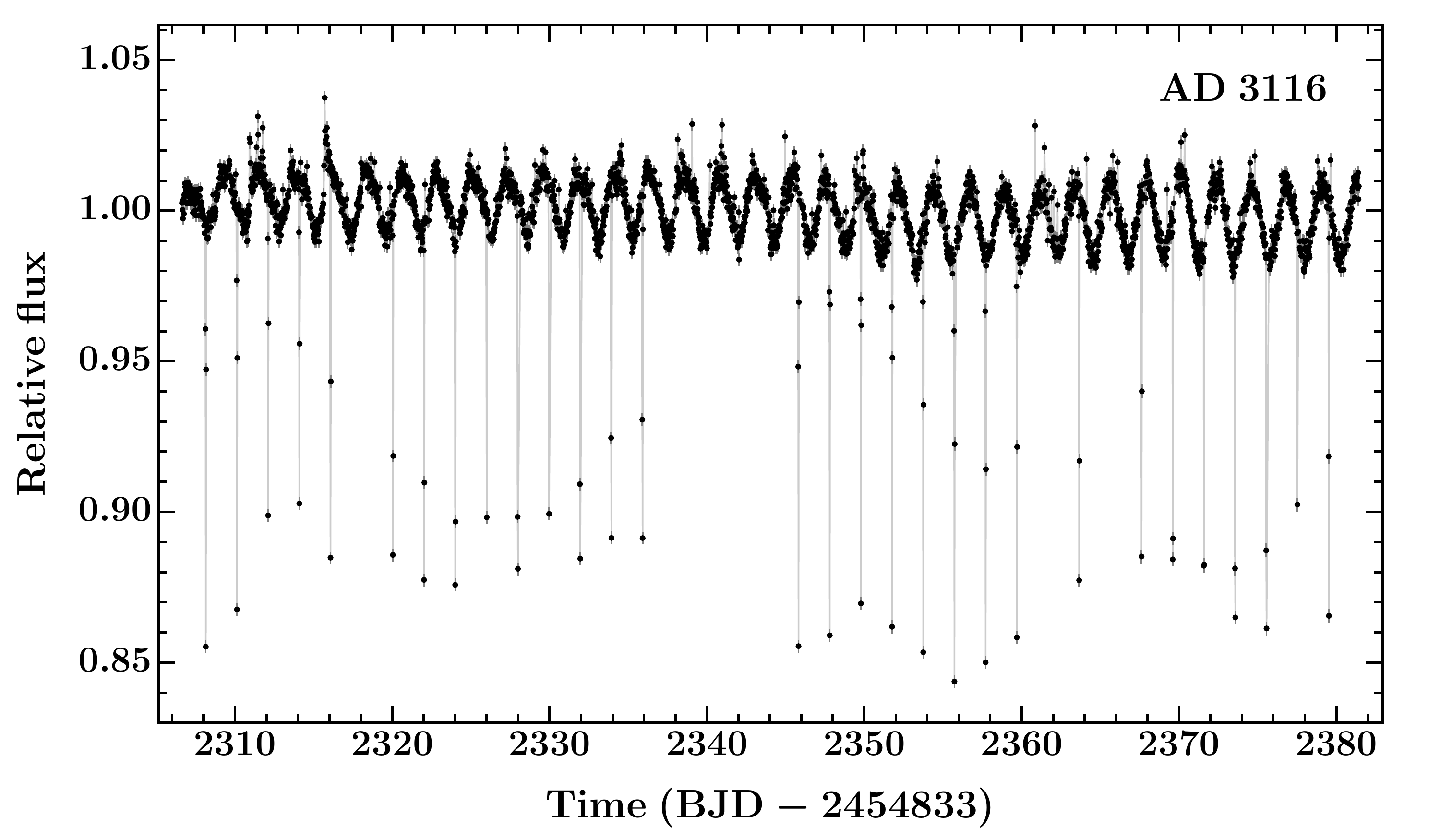} 
    \includegraphics[width=\linewidth]{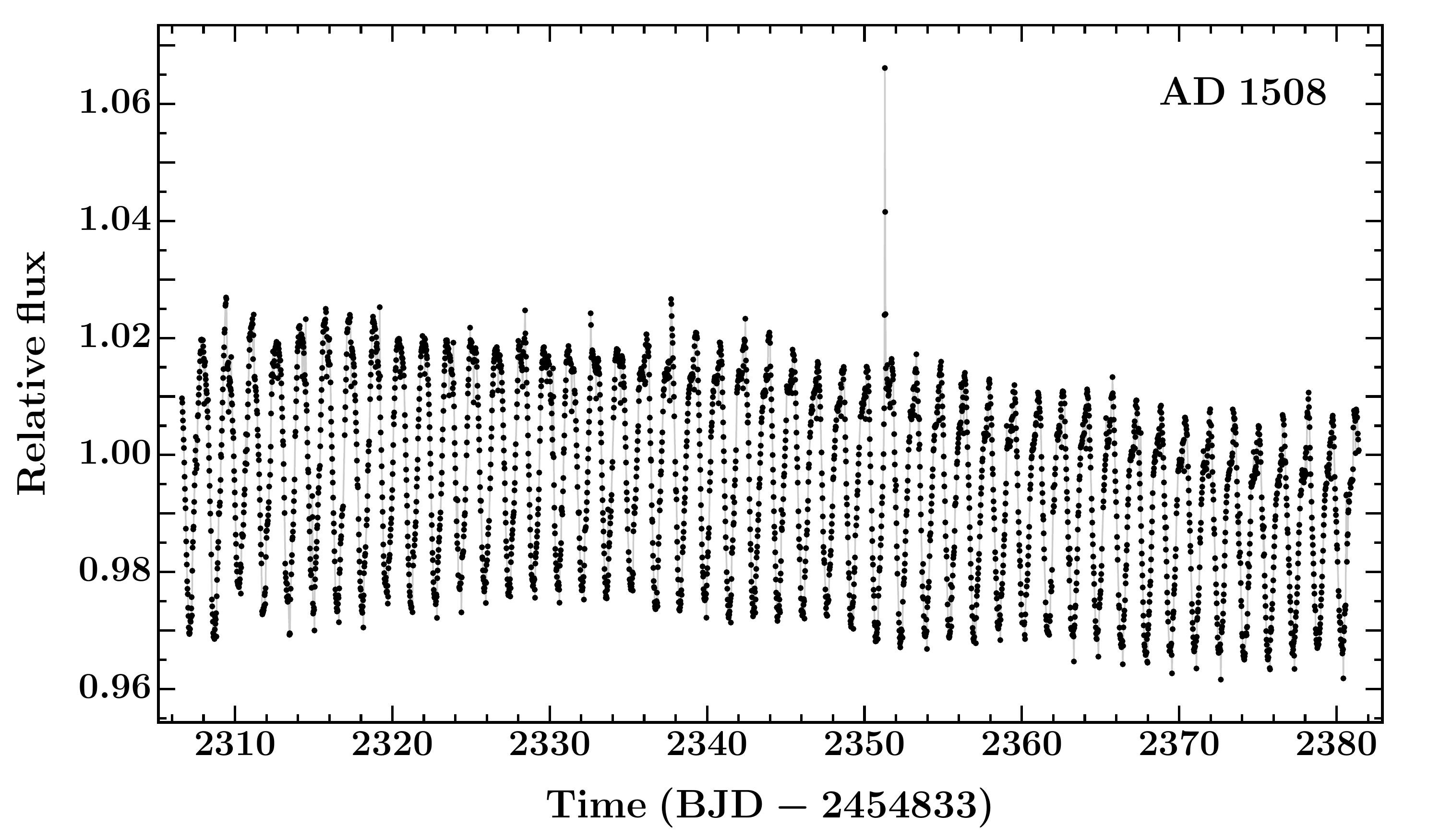}    
    \caption{Systematics-detrended \ksc\ PDC light curves of ADs 3814, 2615, 3116 and 1508 (\etop\ to \ebot). Each system shows out-of-eclipse variations arising from evolving starspot modulation upon which the stellar eclipses are superposed. Numerous stellar flares are visible throughout the observations, most notably on ADs 3814 and 2615, including one in each system reaching a relative flux $\gtrsim$1.8. Missing eclipses, as seen in AD 3116, are an artifact present in the PDC light curves. 
}
   \label{raw_LCs}
\end{figure}

\begin{figure}
   \centering
   \includegraphics[width=\linewidth]{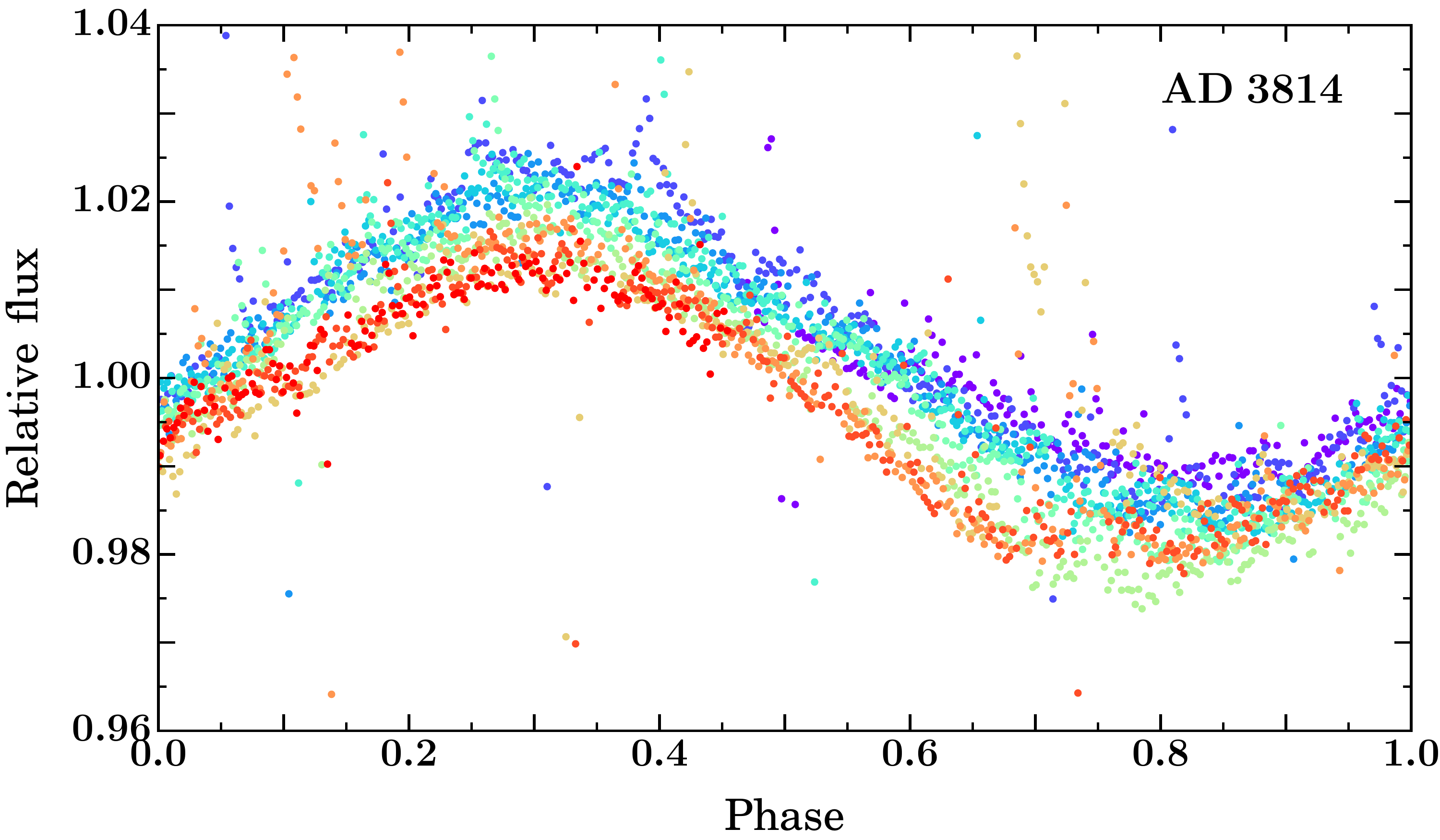} 
    \includegraphics[width=\linewidth]{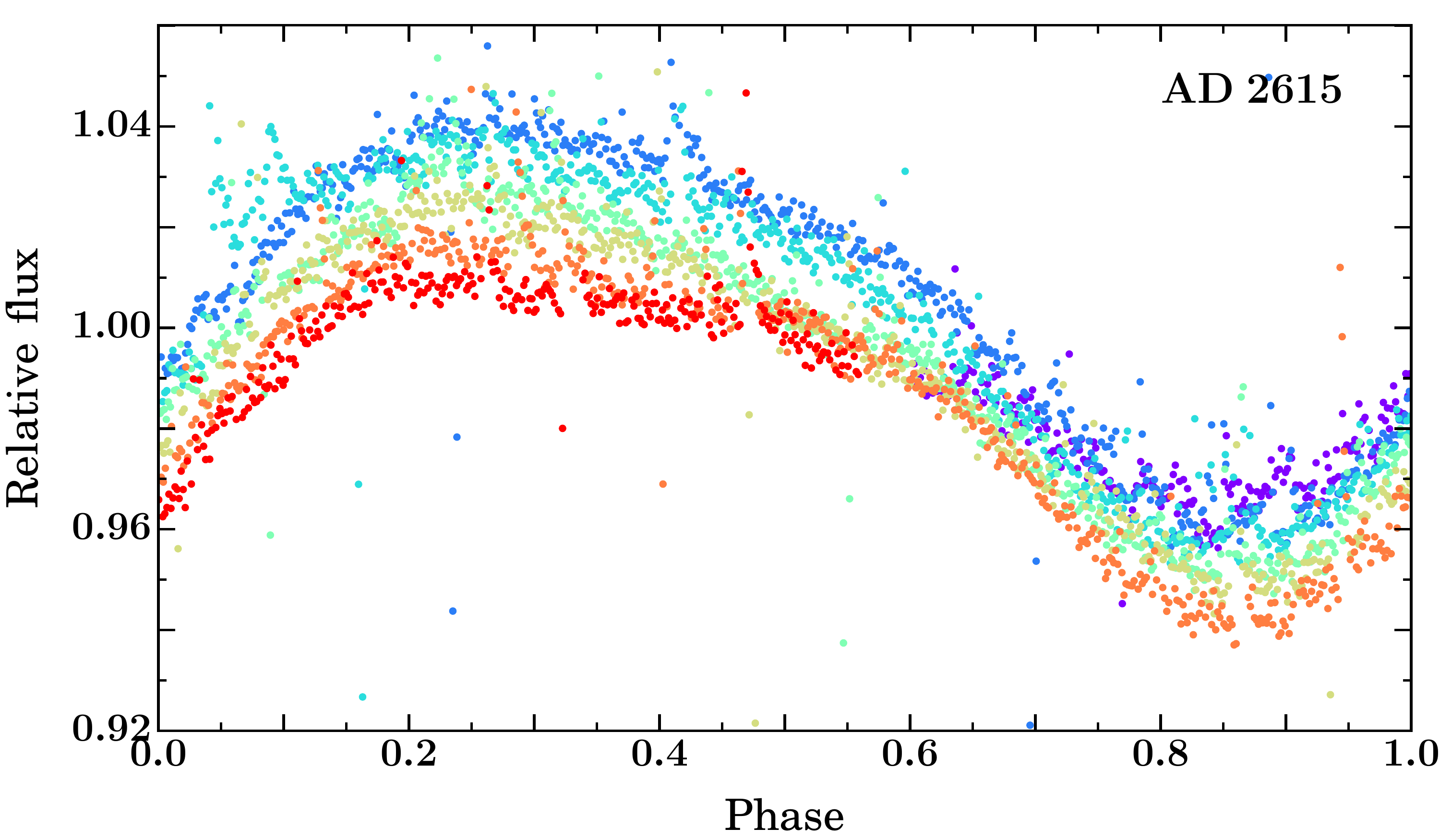}
    \includegraphics[width=\linewidth]{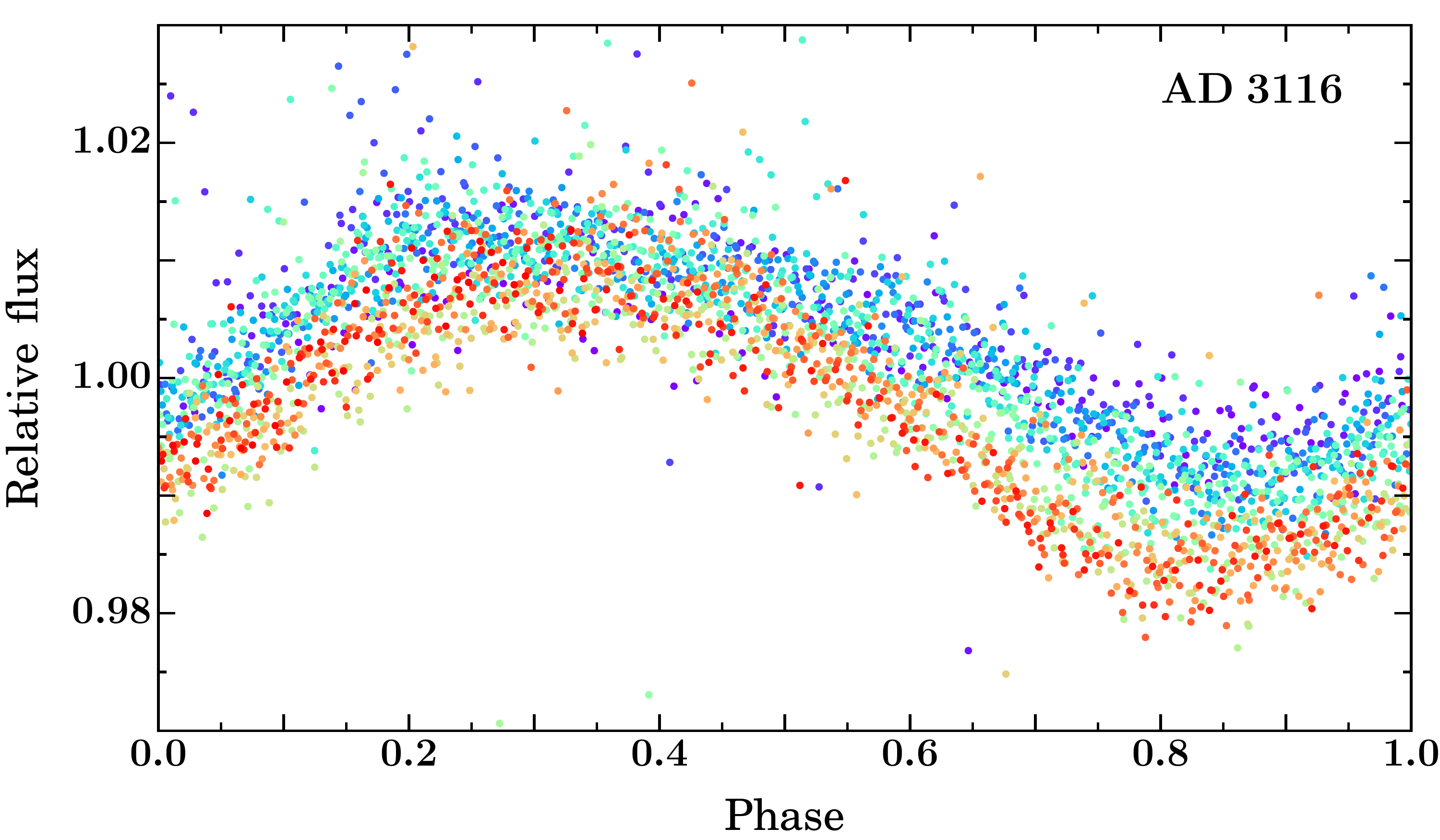} 
    \includegraphics[width=\linewidth]{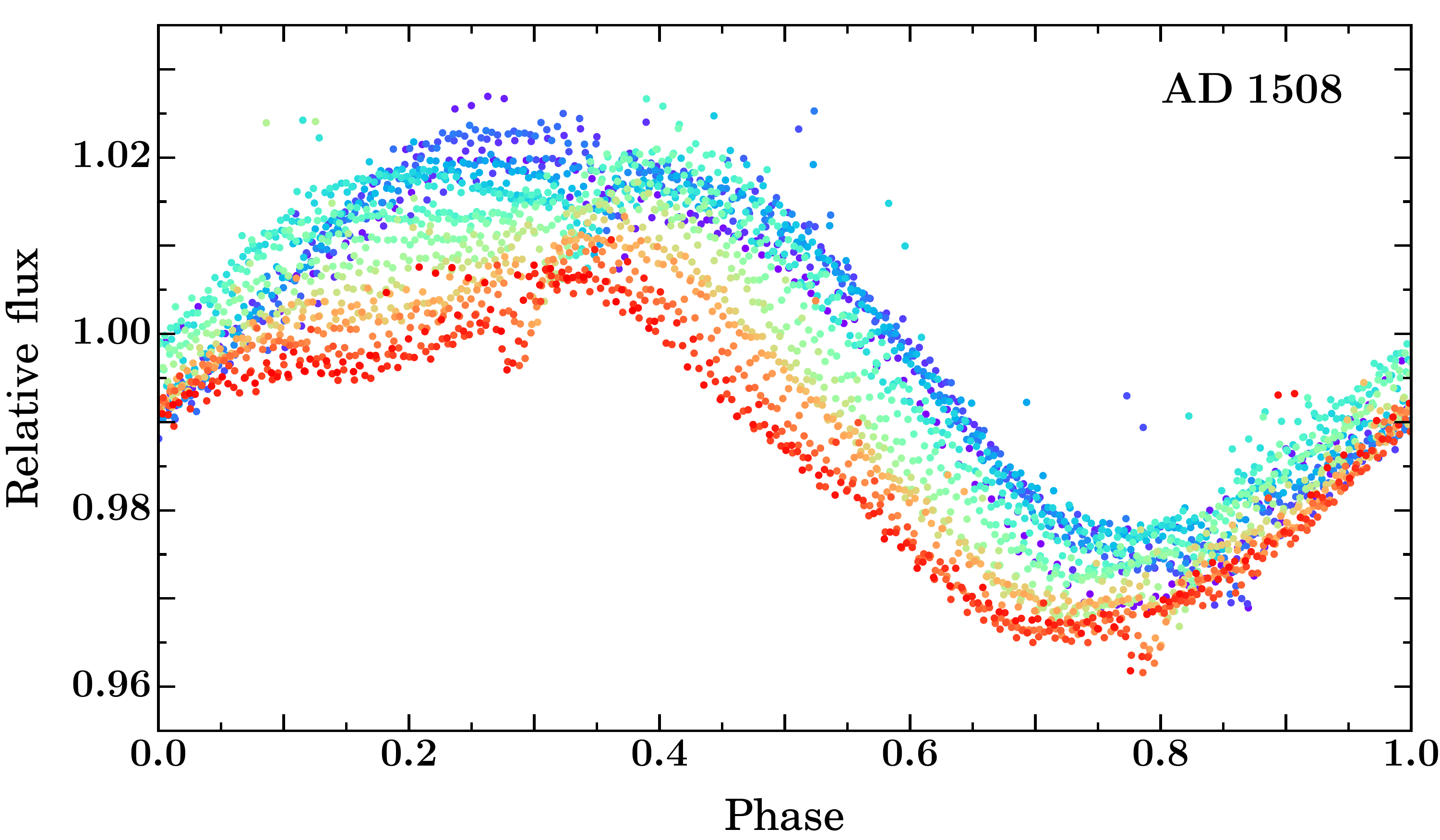}    
    \caption{Evolution of the photometric modulation observed in ADs 3814, 2615, 3116 and 1508 (\etop\ to \ebot). The systematics-detrended \ksc\ light curves shown in Figure \ref{raw_LCs} have been folded on the period of the observed variability. The rainbow color scheme highlights the evolution through the 75 day campaign (\emph{beginning} to \emph{end}, violet to red). For ADs 3814, 2615, 3116 and 1508, the number of periods folded upon are 11, 7, 34 and 49, respectively, which simply reflects the orbital periods of the systems. }
   \label{phot_var}
\end{figure}

%%%%%%%%%%%%%%%%%%%%%%%%%%%%%%%%%%%%%%%%%%%%%%%%%%
\section{Analysis with the \gpe\ model}
\label{analysis}

Both young and low-mass stars typically display photometric and spectroscopic modulation arising from the longitudinal inhomogeneity of active regions on the stellar surface, with activity timescales a strong function of stellar mass. In close binaries ($P \lesssim 15$ days), activity levels are generally observed to be higher than in their single star counterparts. 
This variability is important to properly account for when analyzing the observed stellar eclipses since it can subtly modify the detailed shape of individual eclipses. Ideally, therefore, we would model the stellar variability at the same time as fitting for the eclipses and, in doing so, propagate any uncertainties in the variability modeling through into the posterior distributions for the EB parameters. This approach motivated the development of a new eclipsing binary model, \gpe, which we use here to characterize the new \prae\ EBs by simultaneously modeling the \ktwo\ light curves and Keck/HIRES radial velocity measurements, accounting for activity-induced effects.  The method is distinct from those that account for stellar variability by detrending first and then modeling eclipses second.

\subsection{\gpe}
\label{gpe_model_sec}

\gpe\ comprises a central eclipsing binary (EBOP) model coupled with a Gaussian process (GP) model, which has an MCMC (Markov chain Monte Carlo) wrapper. It can be used to model both eclipsing binary systems and transiting planets: we use it here in its first capacity but note its tested ability to model planet transits \citep[e.g.][]{Pepper17}.
Below we briefly describe the main components of the model:

\begin{itemize}

% EB component
\item \emph{EB component}. The EB model is a modified version of the (JKT)EBOP family of models, which was first presented in \citet{Irwin11}. Each star is modeled as a sphere when computing light curves from the eclipses and as a biaxial spheroid for the calculation of reflection and ellipsoidal effects. This model is able to compute light ratios and radial velocities, and can correct for the ``classical'' light travel time across the system.

Differing from previous EBOP-based models, this implementation uses the analytic method of \citet{Mandel02} for the quadratic limb darkening law. \gpe\ utilizes the LDtk toolkit \citep{Parviainen15}, which allows uncertainties in the stellar parameters (effective temperature, surface gravity and metallicity) to be propagated through the PHOENIX stellar atmosphere models \citep{Husser13} and into priors on the limb darkening coefficients. Limb darkening parameterization within the fitting process follows the triangular sampling method of \citet{Kipping13}.

% GP component
\vspace{1.5mm}
\item \emph{GP component}. The GP model utilizes the {\tt george} package\footnote{\href{http://dan.iel.fm/george}{http://dan.iel.fm/george}} \citep{Ambikasaran14} and is used to model the out-of-eclipse (OOE) photometric data. A detailed description of Gaussian process regression is beyond the scope of this paper but the interested reader is referred to \citet{Roberts12} for a gentle introduction, \citet{Rasmussen06} for a more detailed entry, \citet{Aigrain12} for application to stellar light curves and \citet{Gillen14} for application to eclipsing binary light curves and cross-correlation functions.   

A simple way to view GPs is to think of them as a means of modeling a light curve by parameterizing the covariance between pairs of flux measurements, rather than explicitly specifying a functional form of model to fit the data. In this Gaussian process model, the joint distribution of the observed flux measurements is taken to be a multivariate Gaussian, whose covariance matrix is populated through a covariance function that depends on the observation times.  As such, a GP is distribution over functions. When the parameters of a GP (called hyperparameters) are varied, we step through function space rather than the more familiar parameter space of conventional methods.

Crucially for our application, the power of GP regression is that we obtain an uncertainty on the prediction for the OOE variability across each eclipse, which we can then propagate through into our posterior distributions for the EB parameters.

% MCMC wrapper
\vspace{1.5mm}
\item \emph{MCMC wrapper}. \gpe\ explores the posterior parameter space using the the Affine Invariant MCMC method, as implemented in {\tt emcee} \citep{Foreman-Mackey13}.

\end{itemize}

% ============================
\subsection{Light curves}
\label{sec_light_curves}

The K2 light curves are a timeseries of flux measurements. \gpe\ models the light curves by assuming the joint distribution over the flux measurements ${\bf F}$ is given by a multivariate Gaussian whose mean function $\mu$ is an eclipse model and whose covariance matrix ${\bf K}$ is described by a Gaussian process:
\begin{equation}
{\bf F} \sim \mathcal{N} (\mu, {\bf K}).
\end{equation}
The elements of the covariance matrix ${\bf K}$ are given by:
\begin{equation}
{\bf K}_{ij} =  k(t_{i},t_{j})  +  k_{w}(i,j)
\end{equation}
where the first term represents the specific kernel chosen to describe to the out-of-eclipse variations and the second term describes the white noise component. 

Figure \ref{raw_LCs} shows the raw light curves of the four new EBs and Figure \ref{phot_var} shows these phase-folded on the photometric variability period. The OOE light curves of all four systems presented here display evolving starspot modulation with characteristic amplitudes, periods and evolutionary timescales. To model these smoothly evolving data, therefore, we chose a GP with a quasi-periodic Exponential Sine Squared kernel (hereafter QPESS). This is a periodic kernel that is allowed to evolve over time, i.e. mimicking evolving starspot modulation. The QPESS kernel has the required flexibility to explain the large-scale flux variations in the OOE light curves. It is given by:
\begin{equation}
k{\scriptscriptstyle \rm QPESS} (t_{i}, t_{j})    =   A^{2}   \exp \left[  -\Gamma  \sin^2 \left(  \frac{\pi \left| {t_{i}-t_{j}} \right|}{P}  \right)   - \frac{ \left( {t_{i}-t_{j}} \right)^{2} }{2l^{2}}  \right] .
\end{equation}
The first exponential describes the periodic component and the second the evolution of the periodic signal. $A$ is the characteristic amplitude of the variations, $\Gamma$ is the scale of the correlations, $P$ the period of the oscillations and $l$ the evolutionary timescale. $t_{i}$ and $t_{j}$ represent example times of two flux measurements within the time series. The resulting periods (Table~\ref{lc_model_tab}) differ from those reported by \cite{Rebull17} (based on Lomb-Scargle techniques) at about the $\sim$1\% level. The white noise term is given by:
\begin{equation}
k_{w}(i,j) = \sigma^{2} \delta_{ij}
\end{equation}
where $\sigma$ is the standard deviation and $\delta_{ij}$ is the Kronecker delta function.
Within \gpe\ the white noise term is incorporated via a multiplicative scale factor on the observational uncertainties, as {\tt george} adds these scaled uncertainties in quadrature to the diagonal of the covariance matrix.

We model the \ksc\ light curves that have been detrended for instrument systematics but which still contain stellar activity variations. After visual inspection of the SAP and PDC \ksc\ light curves we opted to work with the PDC versions as these display lower point-to-point scatter.

As can be seen in Figure \ref{raw_LCs}, numerous stellar flares are present throughout the light curves. Flares were treated in two ways depending on whether or not they affected the stellar eclipses. Those that did not were automatically removed using the following method: the light curve was smoothed using a running median filter, which was followed by running sigma cuts to identify flares. The data before and after the flare peak was removed until the light curve returned to the smoothed light curve value. 
Flares affecting the stellar eclipses were treated more carefully: as even a detailed modeling would not correct the photometry to a precision required to include in our eclipse modeling, we opted to conservatively mask out the affected data via visual inspection. 
The resulting light curves, which were modeled in our analysis, can be seen in Figures \ref{3814_LC}, \ref{2615_LC} and \ref{3116_LC} for ADs 3814, 2615 and 3116, respectively. The light curve of AD 1508 was treated slightly differently as only a preliminary solution is presented here (see \S\ref{1508_results} for details).

The full light curves (eclipses and out-of-eclipse variability) and radial velocity variations were simultaneously modeled by \gpe\, stepping through the parameter space 50,000 times with each of 144 `walkers'. The first 25,000 steps were discarded as burn-in and the remainder of each chain was thinned following inspection of the autocorrelation lengths for each parameter. To account for the $\sim$30 minute cadence of \ktwo\ observations, \gpe\ was supersampled at 1 minute cadence and numerically integrated to the \ktwo\ sampling for model evaluation. The uncertainties on the limb darkening coefficients were inflated by a factor of 30, above the uncertainties derived from the PHOENIX models. 
This inflation factor was determined by comparing quadratic limb darkening coefficients of LDtk, \citet{Claret12} and \citet{Sing10} for common \teff\, \logg\ and metallicity values in a representative range for our EBs across the {\it Kepler} bandpass. We used the spread in their predictions, and applied a further increase to account for systematic uncertainties in M-dwarf model atmospheres, to determine our inflation factor. Reflection effects and gravity brightening were not included in the modeling. The former is accounted for by the GP model and the latter makes no significant difference to the model posterior distributions, which we tested by performing model runs with different gravity brightening exponent ($\beta$) values. We note that \citet{Alencar97} found that $\beta$ ranges between 0.2 and 0.4 for stars with temperatures between $3700 \leqslant T \leqslant 7000$ K and that the typical \citet{Lucy67} value of $\beta=0.32$ best describes stars with $T = 6500$ K.

% ============================
\subsection{Radial velocities}

The Keck/HIRES RVs were modeled using Keplerian orbits simultaneously with the \ktwo\ light curves. Spectroscopic light ratios (available for three of the four systems presented here) were estimated from cross-correlation peak heights and applied as priors on the light curve model component. This can help break the well-known degeneracy between the radius and surface brightness ratios, which can often be a limiting factor in the individual radius estimates for near equal-mass EBs.

An RV jitter term, incorporated in \gpe, was used to allow the uncertainties on the Keck/HIRES RV measurements to be scaled, if necessary. This helps account for additional variations arising from e.g. stellar activity and instrument systematics. This jitter term is added in quadrature to the observational uncertainties. When RVs from multiple instruments are obtained, \gpe\ can scale the uncertainties for each instrument individually and account for offsets between different instrument RV zero points.

\begin{table*}  
 \centering  
 \caption{Spectroscopic light ratios and quadratic limb darkening priors applied in the \gpe\ modeling for ADs 3814, 2615, 3116 and 1508. } 
 \label{priors_tab}  
 \begin{tabular}{c @{\hskip 10mm} c c @{\hskip 10mm} c @{\hskip 8mm} c c @{\hskip 8mm} c l }  
 \noalign{\smallskip} \noalign{\smallskip} \hline  \hline \noalign{\smallskip} 
   System  &  \multicolumn{2}{l}{Spectroscopic light ratio ~~~~~~~~ }  & \multicolumn{5}{c}{Limb darkening coefficients and assumed model atmosphere parameters\,*} \\
     & BJD &  $l_{\rm sec}/l_{\rm pri}$  &  Component  &  $\mu$  &  $\mu'$  &  \teff\ (K)  & \logg\ (cgs)  \\

\noalign{\smallskip} \hline \noalign{\smallskip} \noalign{\smallskip}
\multirow{2}{*}{AD 3814}  &  \multirow{2}{*}{245\,7766.9}  &  \multirow{2}{*}{$0.41\,^{+0.25}_{-0.19}$} &  Pri  &  $0.46 \pm 0.13$  & $0.21 \pm 0.46$  &  $3200 \pm 200$  &  $4.9 \pm 0.1$ \\
  &  &  &  Sec  &  $0.49 \pm 0.24$  &  $0.23 \pm 0.76$  &  $3100 \pm 200$  &  $5.0 \pm 0.1$  \\

\noalign{\smallskip} \noalign{\smallskip}
AD 2615  &  245\,7766.9  &  $1.13\,^{+0.24}_{-0.20}$ &  Pri \& Sec  &  $0.46 \pm 0.13$  &  $0.21 \pm 0.46$  &  $3200 \pm 200$  &  $4.9 \pm 0.1$  \\

\noalign{\smallskip} \noalign{\smallskip}
\multirow{2}{*}{AD 3116}  &  \multirow{2}{*}{---}  &  \multirow{2}{*}{---} &  Pri  &  $0.46 \pm 0.13$  &  $0.21 \pm 0.46$  &  $3200 \pm 200$  &  $4.9 \pm 0.1$  \\
  &  &  &  Sec  &  $0.68 \pm 0.17$  &  $0.17 \pm 0.46$  &  $2500 \pm 200$  &  $5.0 \pm 0.1$  \\

\noalign{\smallskip} \noalign{\smallskip} 
AD 1508  &  245\,7745.0  &  $0.63\,^{+0.41}_{-0.26}$ &  Pri \& Sec  &  $0.47 \pm 0.14$  &  $0.20 \pm 0.31$  &  $3700 \pm 200$  &  $4.8 \pm 0.1$  \\

 \noalign{\smallskip} \noalign{\smallskip}\noalign{\smallskip}  
 \hline  
 \end{tabular}  
 \begin{list}{}{}  
 \item[*] $\mu$ and $\mu'$ are the coefficients for the linear and quadratic terms, respectively, of the quadratic limb darkening law. All limb darkening coefficients were computed assuming $Z = 0.14 \pm 0.05$.
 \end{list}  
 \end{table*}

%%%%%%%%%%%%%%%%%%%%%%%%%%%%%%%%%%%%%%%%%%%%%%%%%%
\section{Results}
\label{results}

The K2 light curves and Keck/HIRES radial velocity measurements of the four new EBs (ADs 3814, 2615, 3116 and 1508) were modeled with \gpe; the results for each system are discussed in turn below. Throughout our analysis we define the primary as the star that, when occulted, gives the deepest eclipse, and the secondary as the occulting star. We note that these adjectives do not necessarily imply that the primary star is the more massive or brighter star, as we find to be the case with AD 2615.

% ========================================================================
\subsection{AD 3814}
\label{3814_results}

AD 3814 has been extensively studied in the literature. The M3.4 spectral type, broadband photometric magnitudes and colors, and proper motion give AD 3814 a high probability of cluster membership. 
Figure \ref{3814_LC} shows the \ktwo\ light curve used in the modeling after flares were removed. Three eclipses were masked in the flare removal process (see section \ref{sec_light_curves}): two secondary eclipses at rBJD\footnote{rBJD = BJD $-$ 2454833.}$\sim$2315 and 2361, and one primary eclipse at rBJD$\sim$2364. The red line and pink shaded region indicate the mean and 2$\sigma$ uncertainty of the posterior \gpe\ eclipse model, which is able to reproduce both the eclipses and the slowly evolving starspot modulation. 

Detrending with respect to the GP component and phase-folding on the binary orbital period allows us to inspect the shape of the eclipses in detail. These are shown in Figure \ref{3814_eclipses}, where the top panel displays the full phase-folded light curve and the bottom panels show zooms around primary and secondary eclipses (\eleft\ and \eright, respectively).
There is clear evidence of increased scatter in the residuals across each eclipse, which is presumably due to uncorrected differential starspot effects. Starspots on the background star will have a differential effect on the eclipse shape, with the eclipse being shallower if starspots on the background star are preferentially occulted by the foreground star and deeper if the unspotted photosphere is preferentially occulted. As the timescale for such differential effects are much faster than the typical starspot modulation observed out of eclipse, the QPESS kernel will struggle to account for this effect given its covariance properties, which constrain it to smooth variations. Instead, the GP will opt to inflate its uncertainty due to the increased scatter. One could theoretically include an additional kernel within the GP model to try and account for such differential effects across eclipses, but this is beyond the scope of the current work.

The 8 Keck/HIRES RVs were modeled simultaneously with the \ktwo\ light curve. The resulting phase-folded RV orbit is shown in Figure \ref{3814_RV} (primary in red and secondary in blue). The colored lines and shaded regions indicate the median and 2$\sigma$ uncertainties on the posterior orbits of the two stars, which are well-fit to the observed RVs. The systemic velocity of the system is $V_{\rm sys} = 33.60 \pm 0.24$ km\,s$^{-1}$ (dashed gray line), which is consistent with the recessional velocity of the cluster $V_{\rm rec}$ $\sim$33--35 km\,s$^{-1}$ \citep[e.g.][]{vanLeeuwen09,Quinn12,Yang15} and hence provides further evidence of cluster membership. We note that the residuals in the phase-folded RV plot display an interesting structure. Inspection of the RV residuals in time, however, does not suggest any long term trend indicative of a tertiary companion, which is consistent with the lack of a detectable tertiary peak in the cross-correlation function. Possible explanations for the residuals are issues with the absolute radial velocity calibration, the RV stability of the reference standards, or the precise placement of the target star in the center of the slit. \gpe\ attempts to account for this unknown noise component by including an additional jitter term that acts to scale the observational uncertainties. We note that if the origin of this noise component were known, it may be possible to model directly within the fit, but this is beyond the scope of the present analysis.

Figure \ref{3814_geom} depicts the system, to scale, at both primary and secondary eclipse, indicating the geometry responsible for the observed eclipses and RV variations. The model parameters and 1 sigma uncertainties for AD 3814 are presented in the first results column of Table \ref{lc_model_tab}. The light curve and RV modeling with \gpe\ yields masses and radii for each star: the primary and secondary masses are $0.3813 \pm 0.0074$ and $0.2022 \pm 0.0045$ M$_{\odot}$ with corresponding radii of $0.3610 \pm 0.0033$ and $0.2256\,^{+0.0 063}_{-0.0049}$ R$_{\odot}$. The masses of both components are constrained to 2\% and the primary and secondary radii to 1\% and 3\%, respectively. The fundamental parameters are compatible with the estimated M3.4$\pm$0.1 spectral type and the primary mass estimate from section \ref{par_est}. The masses, radii and effective temperatures (derived in \S \ref{SED_sec}) of AD 3814 are compared to the current suite of stellar evolution models in section \ref{model_comp}.

We applied a prior on the system light ratio and priors on the quadratic limb darkening coefficients (see Table \ref{priors_tab}). The light ratio was determined from the cross-correlation peak height ratio in a HIRES spectrum taken close to quadrature, which is acceptable as the HIRES spectral range is a reasonable match to the \ktwo\ bandpass. 
We note that the degeneracy between the surface brightness and radius ratios is not apparent in our posteriors, although it is not expected to be significant in this system given the mass and brightness ratios.

We conclude by noting that this system would benefit from a more detailed modeling of the individual eclipses, incorporating a full starspot model, to assess whether the large-scale underlying starspot distribution can be reconstructed from the eclipses, which track different longitudes on the stellar surfaces over the K2 run.

% ========================================================================
\subsection{AD 2615}
\label{2615_results}

AD 2615 is an M4.0 high probability member of Praesepe. The analysis presented here is consistent with the photometric, spectroscopic and membership information from previous studies.  
The light curve of AD2615 that was used in the modeling is shown in Figure \ref{2615_LC}. One secondary eclipse, at rBJD$\sim$2367, was masked following the flare removal process (see section \ref{sec_light_curves}). The red line and pink shaded region represent the mean and 2$\sigma$ uncertainty of the posterior \gpe\ eclipse model. As with AD 3814, the model is able to capture both the stellar eclipses and the evolving starspot modulation. The model's predictive power can be seen before and after the light curve, where it is able to predict the form of the evolving modulation pattern, given the covariance properties of the data; this also drives the motivated prediction and uncertainty across each eclipse.

Figure \ref{2615_eclipses} shows the phase-folded light curve, which has been detrended with respect to the GP component. The eclipse model is an acceptable fit to the data. There is no clear evidence for increased scatter in the residuals, which suggests that the geometry of the eclipses does not preferentially track bright or dark regions on the stellar surfaces, perhaps because the underlying starspot distribution in AD 2615 is more homogeneous than in AD 3814.

Figure \ref{2615_RV} shows the phase-folded RV orbit (red for primary and blue for secondary). The 5 HIRES RVs of both stars are well-fit by the Keplerian model. The 2$\sigma$ uncertainties on the orbits (red and blue shaded regions) increase around quadrature, as expected. The systemic velocity of $V_{\rm sys} = 34.91 \pm 0.39$ km\,s$^{-1}$ (dashed gray line) is compatible with the cluster's recessional velocity, providing further kinematic evidence of cluster membership. We note that a sixth RV observation was conducted but lay too close to primary eclipse to disentangle the two stellar components and hence was not used in the fit. In principle, we could determine an upper limit on the separation of the two stars and use this as an additional constraint in the modeling. However, at phase = 0.997, the solution is already tightly constrained and hence this upper limit would not place useful constraints on our existing solution. We further note that spectral disentangling may offer an interesting alternative route of RV determination for this system, which could utilize this sixth observation. While traditional spectral disentangling techniques require many high SNR spectra, powerful new techniques are emerging designed for fewer and lower SNR spectra \citep[e.g.][]{Czekala17}. It would be interesting to compare the standard CCF-based RV determination with these new spectral disentangling techniques, but this is beyond the scope of the present paper.

Figure \ref{2615_geom} depicts the system, to scale, at primary and secondary eclipse, showing the configuration responsible for the observed eclipses. The medians and 1$\sigma$ uncertainties of the \gpe\ model posteriors are reported in Table \ref{lc_model_tab} (second results column). The primary and secondary masses are $0.212 \pm 0.012$ and $0.255 \pm 0.013$ M$_{\odot}$, with corresponding radii of $0.233 \pm 0.013$ and $0.267 \pm 0.014$ R$_{\odot}$. We remind the reader that we define the primary star as the star which, when occulted, gives the deeper eclipse, but that this does not necessarily mean it is the more massive or brighter of the two components, as indeed is the case in this system. The masses and radii are constrained to 6\% for the primary and 5\% for the secondary. This system would benefit from additional RVs around quadrature to increase the precision of the mass determination. The fundamental parameters are compatible with the estimated M4.0 spectral type but the mass of either component is lower than the estimate from the system's absolute K-band magnitude (Section \ref{par_est}), presumably because this system is a near equal mass binary and so both stars contribute significantly to the K-band flux, resulting in an overestimated single-star mass. The masses, radii, and effective temperatures (c.f. \S \ref{SED_sec}) are compared to stellar evolution models in section \ref{model_comp}.

We applied priors on the system light ratio and stellar limb darkening coefficients (see Table \ref{priors_tab}). Even though the system is near-equal mass and brightness, our spectroscopic light ratio was able to break the degeneracy between the surface brightness and radius ratios, which can be a limiting factor in determining radii in such systems.

% ========================================================================
\subsection{AD 3116}
\label{3116_results}

AD 3116 is an M3.9 high probability member of Praesepe. The system sits at the bottom of the cluster sequence (see Figure \ref{cmd}) suggesting the secondary component contributes little optical light to the total system flux and hence is comparatively low-mass. 

Analysis of the \ktwo\ light curve and 7 HIRES spectra reveals the system to be single-lined with eclipses visible only on the primary component, consistent with its position in color-magnitude space. Secondary spectroscopic lines could not be detected, even after dividing the two spectra and looking for similar but weaker patterns in the CCF, which suggests the secondary contributes very little ($<20-35$\%) to the system's optical light. Given the lack of a detectable secondary eclipse and secondary RV orbit, the data alone are not able to constrain the solution precisely. There exist two families of solutions: one consisting of a small secondary that fully transits and the other a larger secondary on a grazing trajectory. The primary RV orbit requires the secondary to be eclipsed, and hence both models find a negligible surface brightness ratio in the \kepler\ band to remain consistent with the lack of a detectable secondary eclipse. For the solution comprising a large ($R_{\rm sec}/R_{\rm pri} \gtrsim 1$), grazing secondary, this would require an  unusual object possessing a very low temperature given its radius. Inspection of the system mass function revealed that the secondary lay in the brown dwarf regime ($M_{\rm sec} \sim$55 $M_{\rm Jup}$), which further supported the solution comprising a small, fully-transiting secondary. 
We tested the reliability of the primary RV solution to see if individual RVs close to the systemic velocity (i.e. which could be biased by low-level secondary light) may be affecting the eccentricity of the RV orbit and hence the system parameters. We removed all bar the three RVs closest to quadrature and, as expected, the model converged again on a solution requiring the secondary to be eclipsed. This, combined with the small primary RV semi-amplitude and lack of secondary spectroscopic lines, rules out a scenario where the secondary is of comparable size and brightness to the primary but there is no secondary eclipse due to the eccentricity of the orbit. All available information and tests pointed towards a very low-mass, small and cool secondary component.

We therefore chose to place loose uniform priors on the radius ratio and surface brightness ratio to encourage the solution towards a physically sensible secondary component. These priors were $0.0 \leqslant R_{\rm sec}/R_{\rm pri} \leqslant 0.46$ and $0.0 \leqslant J_{\rm K2} \leqslant 0.25$ which, given the expected primary star properties (c.f. \S\ref{par_est}) and secondary star mass estimate, act simply to exclude physically implausible solutions and do not act to constrain the remaining physically plausible solutions. We performed further tests allowing $R_{\rm sec}/R_{\rm pri}$ and $J_{\rm K2}$ to extend up to 0.75 and 0.35, respectively, but find consistent posterior values.

The model fit is shown in Figures \ref{3116_LC}--\ref{3116_geom}, whose descriptions are the same as for ADs 3814 and 2615 in sections \ref{3814_results} and \ref{2615_results} above. The model is a good fit to the primary eclipse and large-scale evolving starspot structure in the \ktwo\ light curve (Figure \ref{3116_LC}). We note that two primary eclipses, at rBJD$\sim$2319 and 2350, were masked following the flare removal process (see section \ref{sec_light_curves}). Figure \ref{3116_eclipses} shows the phase-folded and GP-detrended light curve: the primary eclipse is well-fit, although there is a modest increase in the residual scatter, which is larger than in AD 2615 but smaller than in AD 3814. The RV data suggests a moderately eccentric orbit ($e \sim 0.15$; see Figure \ref{3116_RV}) with a systemic velocity of $V_{\rm sys} = 34.93\,^{+0.61}_{-0.53}$ km\,s$^{-1}$ (dashed gray line). This is consistent with the cluster's recessional velocity, providing additional kinematic evidence of cluster membership. 

Using the empirical relations of \citet{Benedict16}, and assuming the \citet{vanLeeuwen09} cluster distance of $181.5\pm6.0$ pc, the $K_{\rm s}$ magnitude of AD 3116 implies a primary mass of $M_{\rm pri} = 0.28 \pm 0.02$ $M_{\odot}$, where the uncertainty arises equally from the empirical relation scatter and our assumed 0.1 mag uncertainty on the quoted $K_{\rm s}$ value. We checked this value using the empirical relations of \citet{Mann16} and find $M_{\rm pri} \sim 0.29$ $M_{\odot}$, consistent with the Benedict et al. value. Taking the Benedict value, the mass function from our final solution then yields $M_{\rm sec} = 54.2\pm4.3$ $M_{\rm Jup}$. This is one of only $\sim$20 known transiting brown dwarfs \citep[e.g.][]{Csizmadia16,Nowak16,Bayliss16} and the primary component is one of only three M-dwarfs known to host a transiting brown dwarf. Furthermore, this is only the second known transiting brown dwarf in an open cluster (i.e. where the age is well-constrained), and the first younger than a Gyr.

Figure \ref{3116_geom} shows the system geometry at primary and secondary eclipses. That the brown dwarf is fully occulted yet shows no detected signature in the \ktwo\ band theoretically allows us to place an upper limit on the optical reflected light and hence albedo of the object. Using the \citet{Mann16} empirical relations to estimate the primary radius, and hence secondary radius and semi-major axis from our light curve modeling, we can estimate the system scale. This then allows us to compute the angle on the sky that the brown dwarf subtends as seen from the primary. With $R_{\rm sec} \sim 0.11$ $R_{\odot}$ and $a \sim 4.7$ $R_{\odot}$, the brown dwarf intercepts $\sim$0.007\% of the visible light from the primary star. Therefore, even if the brown dwarf reflected all incident flux (i.e. an albedo of 1), we would not detect a drop in flux in the \ktwo\ light curve when the brown dwarf is occulted. 

We applied priors on the limb darkening coefficients (see Table \ref{priors_tab}). The secondary temperature was set to be as low as the PHOENIX models allow but is likely still too high (see Table \ref{Td_comp_tab}). However, as the secondary gives no detectable eclipse it makes no significant difference to the presented solution. Given the system is single-lined we did not place a prior on the system light ratio in the \ktwo\ band.

% ========================================================================
\subsection{AD 1508}
\label{1508_results}

AD 1508 is a high probability M0.1 member of Praesepe, which sits high above the cluster sequence (see Figure \ref{cmd}), suggesting a near-equal mass system. The preliminary analysis presented here is consistent with this picture.
The \ktwo\ light curve of AD 1508 (see Figures \ref{raw_LCs} and \ref{phot_var}; bottom panels) is dominated by evolving starspot modulation at the few percent level. Very shallow grazing eclipses are also present with a depth of less than 1\%. We obtained only three RVs for this system, which unfortunately fall close to primary and secondary eclipses (see Table \ref{RVs_tab}). Given this, and the shallow eclipses, a precise solution is not possible. Instead, we provide our initial analysis and offer the system to the community for further pursuit. 

The \ktwo\ light curve and three Keck/HIRES RVs were simultaneously modeled with \gpe. However, given the preliminary nature of the modeling, and unlike the other three systems, we opted to simplify the light curve analysis by performing an initial detrending of the starspot modulation and then modeled the residuals with \gpe\ to analyze the stellar eclipses. To do this, the out-of-eclipse light curve was flattened through two iterations of a cubic basis spline with knots every 2 hours and rejection of 0.5-$\sigma$ outliers. 
Figure \ref{1508_LC} shows the resulting detrended light curve that was modeled with \gpe. Low-level (likely systematic) residual variations are present, which show a relatively rough behavior. Accordingly, we chose a Matern-3/2 kernel for the GP component, which is given by:
\begin{equation}
k_{\rm M32} (t_{i}, t_{j}) = A^{2} \left(1+\frac{\sqrt{3} \left| {t_{i}-t_{j}} \right| }{l}\right) \exp\left(-\frac{\sqrt{3} \left| {t_{i}-t_{j}} \right| }{l}\right)
\end{equation}
where $A$ is the amplitude and $l$ the characteristic timescale of the variations.

Detrending with respect to the GP component and phase-folding on the orbital period, as shown in Figure \ref{1508_eclipses}, we see that the eclipses are well-fit by the model. 
There is no significant evidence of increased scatter across the eclipses. 
We note that the light curve of AD 1508 appears noisy in comparison to the other systems discussed here, even though it is significantly brighter. This is simply because the plot scales in Figures \ref{1508_LC} and \ref{1508_eclipses} are small as the eclipses are shallow and the starspot modulation has already been detrended for. It is not a reflection of the true noise level in this system: the point-to-point scatter of all systems discussed here decreases with system brightness, as expected.

The phase-folded RV orbit is shown in Figure \ref{1508_RV} which, given only three RVs at non-optimal phases, is not well-constrained. This is reflected in the large 2$\sigma$ uncertainties on the posterior orbits (red and blue for the primary and secondary stars, respectively). Nonetheless, the systemic velocity is relatively well-constrained at $V_{\rm sys} = 33.1\pm1.7$ km\,s$^{-1}$, which is consistent with the cluster recessional velocity and hence provides further kinematic evidence of \prae\ membership. Figure \ref{1508_geom} shows the system, to scale, at primary and secondary eclipse. The shallow eclipses simply result from the very grazing trajectory of the stellar orbits, as viewed from Earth.

The median and 1$\sigma$ uncertainties resulting from our preliminary analysis are reported in Table \ref{lc_model_tab} (fourth results column). Given the available data, significant uncertainties exist in the derived masses and radii. The primary and secondary masses are $0.45\,^{+0.19}_{-0.14}$ and $0.53\,^{+0.22}_{-0.16}$ M$_{\odot}$ with corresponding radii of $0.549\,^{+0.099}_{-0.082}$ and $0.454\,^{+0.094}_{-0.101}$ R$_{\odot}$. The solution is currently limited by the lack of RV constraints and future analysis would benefit from additional RV measurements, especially around quadrature. Nonetheless, the fundamental parameters are compatible with the estimated M0.1$\pm$0.1 spectral type and the primary mass estimate from section \ref{par_est}. Given the existing uncertainties we do not compare this system to stellar evolution models in section \ref{model_comp}.

We applied priors on the system light ratio and limb darkening coefficients (see Table \ref{priors_tab}). Although large uncertainties remain, the spectroscopic light ratio was able to break the degeneracy between the surface brightness and radius ratios, which can be a limiting factor in determining individual radii in near-equal mass and brightness systems.

%%%%%%%%%%%%%%%%%%%%%%%%%%%%%%%%%%%%%%%%%%%%%%%%%%
\section{Discussion}
\label{discussion}

The direct determination of fundamental stellar parameters offers an opportunity to test stellar evolution models. The fundamental predictions of these models are the radius and \teff\ for a star of given mass and metallicity as a function of age. Ideally, therefore, we would be able to determine the mass, radius and \teff\ of both stars as, together, these offer a particularly strong test of stellar evolution theory. However, while the masses and radii of stars in EBs naturally fall out of the joint light curve and radial velocity modeling, estimating effective temperatures is more challenging. In \S\ref{SED_sec} we present a method of simultaneously estimating the effective temperature of both stars, and the distance to the system in a manner that makes full and correct use of the light and radial velocity constraints. We then compare our \teff's and distances to empirical \teff\ relations and to previous distance estimates to \prae. In \S\ref{model_comp} we compare our masses, radii and \teff's to the predictions of stellar evolution models for individual systems and also place the newly characterized EBs in the context of other known low mass EBs and briefly discuss the constraints that can be placed on the  age of Praesepe. Through this model comparison, and in \S\ref{sync} where we comment on the synchronization of the new EBs, we discuss several astrophysical implications of our findings.

\subsection{Simultaneous determination of effective temperatures and distance from the spectral energy distribution}
\label{SED_sec}

\begin{figure*}
\centering
  \includegraphics[width=0.43\linewidth]{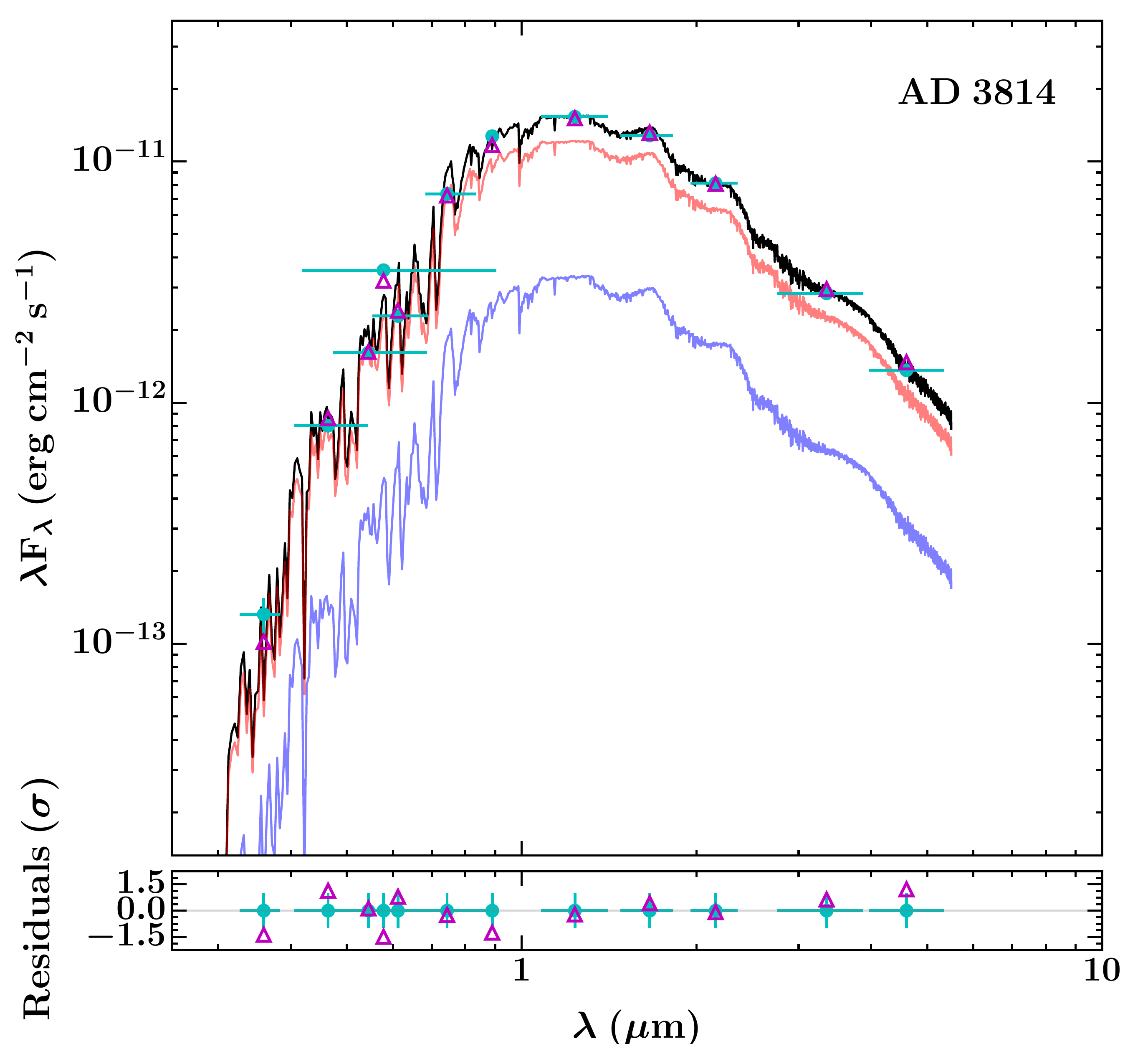}
  \includegraphics[width=0.43\linewidth]{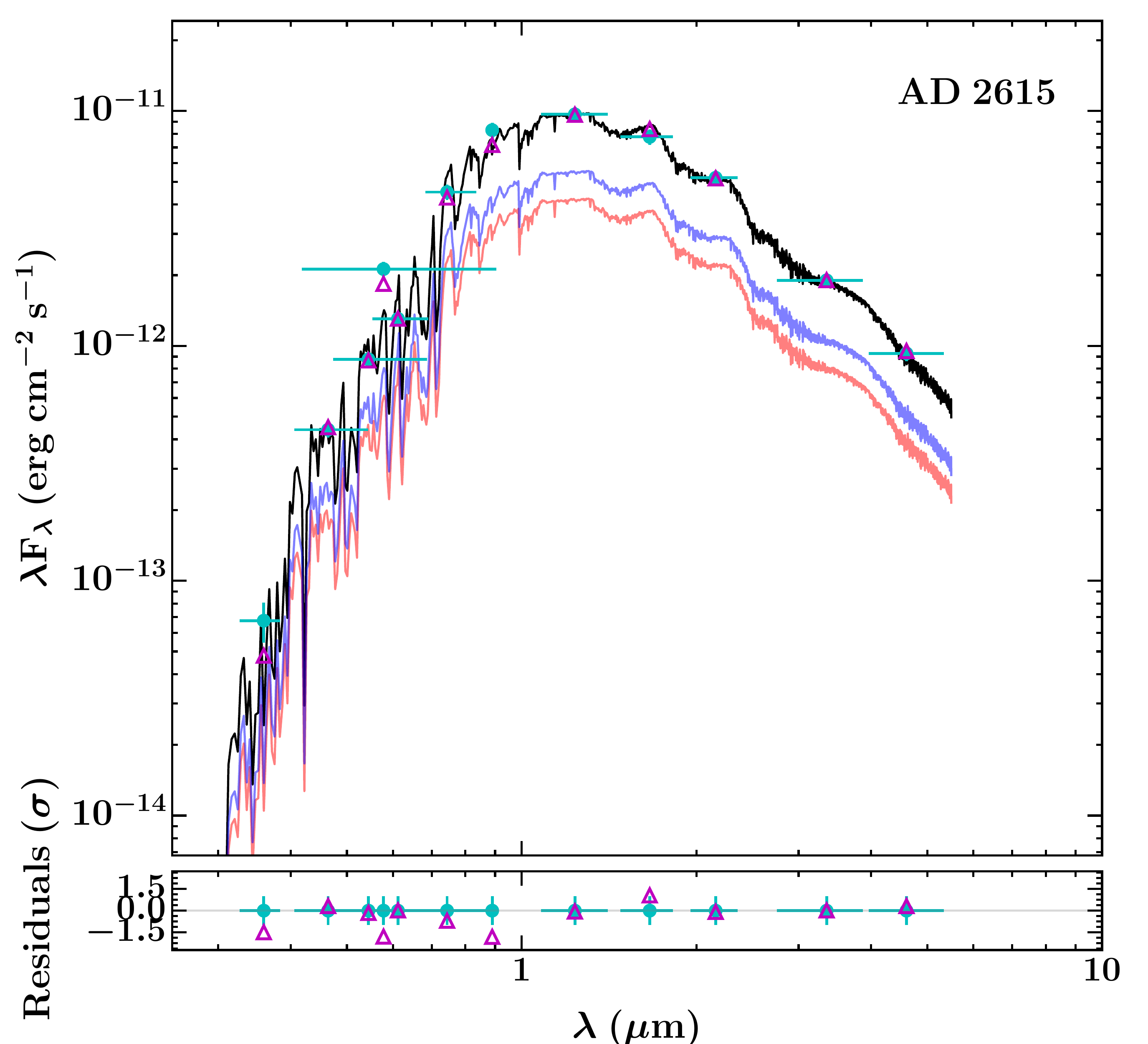} 
  \includegraphics[width=0.43\linewidth]{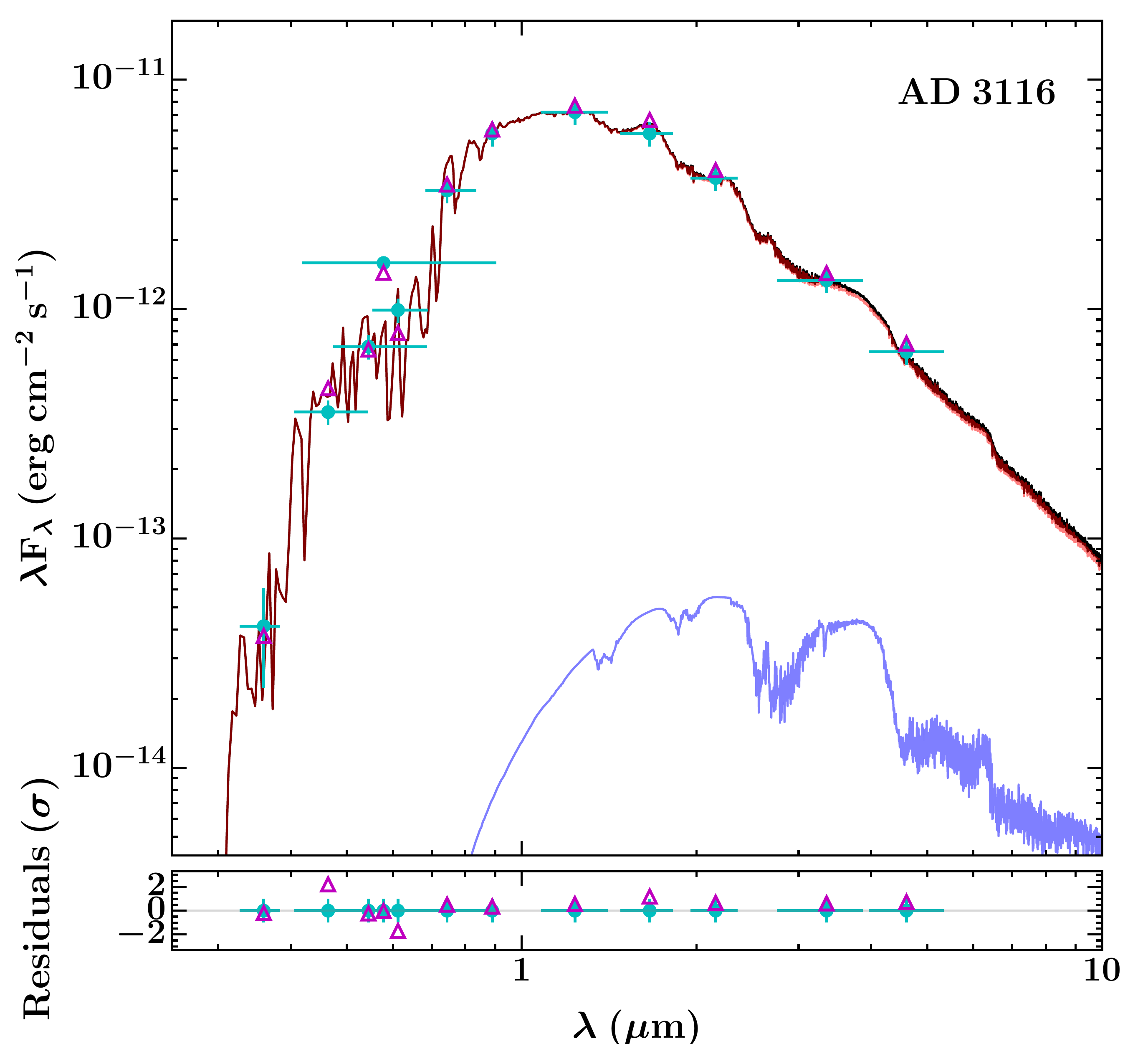} 
  \includegraphics[width=0.43\linewidth]{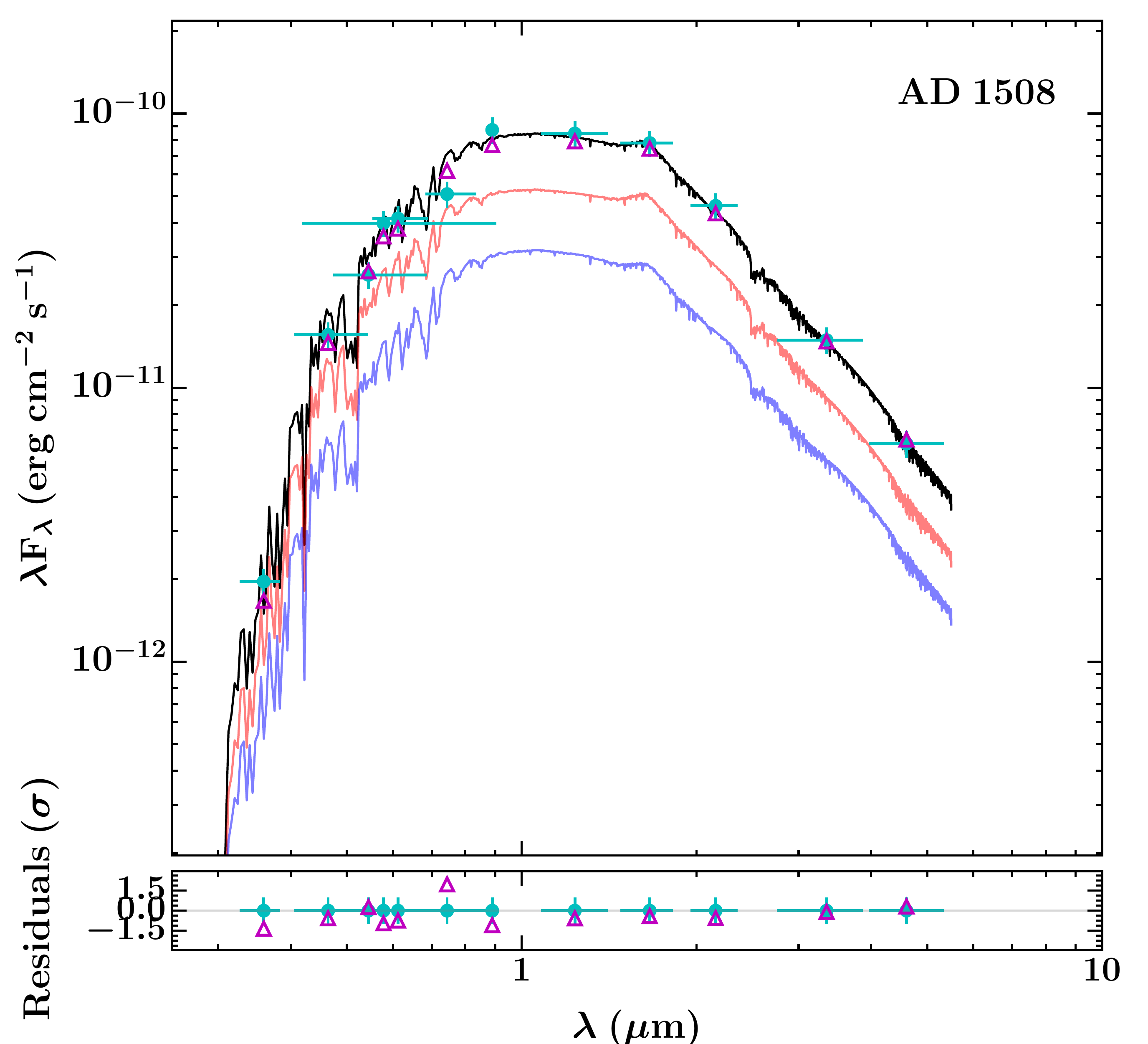} 
   \caption{Spectral energy distributions (SEDs) of the four new EBs. Clockwise from \emph{top left}: ADs 3814, 2615, 1508 and 3116. Cyan points represent the observed SED, which has been constructed from the broadband magnitudes reported in Table \ref{info_tab} (horizontal error bars indicate each band's spectral coverage). The primary and secondary star spectra are shown in red and blue, respectively. Their combined spectrum is shown in black and the hollow magenta triangles show the combined model convolved with the $ugriz$, $V$, $K_{p}$, $JHK$ and WISE 1\,\&\,2 bands. The models shown for ADs 3814, 2615 and 1508 are the PHOENIX v2 models are these produce a better fit to the data than the BT-SETTL models. However, we show the BT-SETTL models for AD 3116 as the PHOENIX models do not extend to low enough temperatures to explain the secondary component.}
   \label{SED}
\end{figure*}

The standard method of estimating \teff\ is the following: 1. estimate the primary star \teff\ from either system colors adopting empirical single-star relations, or use (typically) low resolution spectra to infer a combined spectral type (SpT) and convert this into a primary star \teff. 2. estimate the secondary star \teff\ from the primary \teff\ and the light curve surface brightness (and hence temperature) ratio. There are a number of issues with this approach: empirical color-\teff\ and SpT-\teff\ relations for single stars are not necessarily applicable for all binary systems and the temperature ratio estimated from the light curve is specific to that band, i.e. $(T_{\rm sec}/T_{\rm pri})_{\rm \,band}$; it is not a \teff\ ratio.

A more direct approach would be to model the system's spectra, but to do so would require high SNR (signal-to-noise) data, which would normally require the co-adding of spectra. While feasible for single star systems, this is not possible for binaries as there are two varying components. One approach would be to disentangle the spectra into their individual components and model these directly to estimate \teff\ of each star \citep[e.g.][]{Czekala15,Czekala17}. However, while powerful, this approach is both time and computationally intensive, and the distance to the system remains unknown (unless the spectra are also flux calibrated). 

A method of simultaneously determining \teff\ of both stars, and the distance to the system, is to model the system's spectral energy distribution (SED). This approach is not computationally intensive, does not rely on empirical single-star relations and readily incorporates priors from the joint light curve and RV modeling. Importantly, with respect to the last point, it correctly interprets the band-specific surface brightness ratio from the light curve modeling. Therefore, we simultaneously estimate \teff's and the distance to ADs 3814, 2615, 3116 and 1508 using the following method:

\begin{enumerate}
\item SEDs were constructed using broadband magnitudes readily available in the literature. We obtained SDSS $ugriz$ magnitudes from the Sloan Digital Sky Survey Data Release 13, and 2MASS JHK$_{\rm s}$ and WISE data from the NASA/IPAC Infrared Science Archive. These are reported in Table \ref{info_tab} along with their formal measurement uncertainties. 
\item Model grids of both BT-SETTL \citep{Allard12} and PHOENIX v2 model spectra \citep{Husser13} were convolved with commonly available bandpasses ($ugriz$, UBVRI, 2MASS JHK$_{s}$, \emph{Spitzer}/IRAC, WISE and Kepler) to create a model grid of bandpass fluxes.
\item Each SED was modeled by interpolating the model grids in \teff--$\log g$ space. We opted to fix the metallicity at Z=0.0 given the cluster [Fe/H] value but note it is possible to include in the interpolation.
\item The parameters of the fit were the \teff, radius and $\log g$ of each star, the distance to the system, the interstellar extinction and the uncertainty scale factor ($T_{\rm pri}$, $T_{\rm sec}$, $R_{\rm pri}$, $R_{\rm sec}$, $\log g_{\rm pri}$, $\log g_{\rm sec}$, $d_{\rm sys}$, $A_{\rm v}$ and $\sigma_{s}$). The radii and $\log g$'s have priors from the joint light curve and RV solution, $A_{\rm v}$ had a prior determined for the cluster \citep{Taylor06}, and the temperatures, distance and uncertainty scale factor had uninformative priors. The uncertainties on the magnitudes were initially set by adding the observed variability level to the formal measurement errors in quadrature and a further inflation term ($\sigma_{s}$) was then fit for.
\item The posterior parameter space was explored using {\tt emcee} with 50,000 steps and 196 `walkers'. Convergence was assessed using the Gelman–Rubin diagnostic plus examination of individual sections of the chains. A conservative burn-in was estimated comprising the first 25,000 steps for all systems and parameter distributions were derived from the remainder after thinning each chain based on the autocorrelation lengths of each parameter.
\item This method also gives the option of placing additional priors in the modeling. For example, one can place a prior, from the light curve modeling, on the surface brightness ratio between the two stars \emph{in the band observed}, rather than incorrectly placing a \teff\ ratio constraint. In the case of single-lined systems, radius ratio constraints and surface brightness upper limits can also be placed.

\end{enumerate}

Both BT-SETTL and PHOENIX v2 model spectra are able to reproduce the broadband magnitudes of ADs 3814, 2615, 3116 and 1508. We note, however, that the BT-SETTL models consistently underpredict the optical $r$ band fluxes, whereas the PHOENIX v2 models predict higher red-optical fluxes in agreement with the data for all sources. Accordingly, in Figure \ref{SED} we show the PHOENIX v2 model fits to the observed broadband magnitudes of ADs 3814, 2615 and 1508 reported in Table \ref{info_tab} (for AD 3116 we show the BT-SETTL fit as the PHOENIX models do not extend to low enough temperatures to explain the secondary brown dwarf component). The \teff\ and distance values derived from our SED fitting procedure with both the BT-SETTL and PHOENIX v2 models are reported in Table \ref{Td_comp_tab} along with the empirical relation predictions of \citet{Mann16} and \citet{David16}. We discuss the effective temperature and distance estimates in the following two sections.

\begin{table}  
 \centering  
 \caption{Effective temperatures and distance values for each EB estimated from SED modeling and the empirical relations of \citet{Mann16} and \citet{David16}. }  
 \label{Td_comp_tab}  
 \begin{tabular}{l l c c c }  
 \noalign{\smallskip} \noalign{\smallskip} \hline  \hline \noalign{\smallskip} 
  Method\,$^{*}$  &  Model\,$^{\dag}$  &  \multicolumn{2}{c}{\teff\,\,$^{\ddag}$}  &  Distance \\
       &       &  Primary  &  Secondary  &        \\
     &        &  (K)      &     (K)     &  ~(pc)  \\
\noalign{\smallskip} \hline \noalign{\smallskip} \noalign{\smallskip}

\multicolumn{5}{c}{............................................ AD 3814 ......................................} \\

\noalign{\smallskip} 

SED  &  PHOENIX  &  $3193 \pm 17$  &  $3085 \pm 21$  &  $168.8\,^{+6.1}_{-7.3}$ \\  [-0.3ex]
SED  &  BT-SETTL  &  $3230 \pm 36$  &  $3121 \pm 35$  &  $172.1 \pm 9.3$ \\  [-0.3ex]

ER   &  M15  &  $3241 \pm 76$  &  $3013 \pm 79$  &  $172 \pm 12$  \\  [-0.3ex]
ER   &  D16  &  3251  & 3023 &   \\  [-0.3ex]

 \noalign{\smallskip}
SED  &  Combined  & $3211\,^{+54}_{-36}$  &  $3103\,^{+53}_{-39}$  &  $170.4\,^{+11.0}_{-8.9}$ \\

 \noalign{\smallskip} \noalign{\smallskip} \noalign{\smallskip}
 
\multicolumn{5}{c}{............................................ AD 2615 ......................................} \\

\noalign{\smallskip} 

SED  &  PHOENIX  &  $3132 \pm 21$  &  $3112 \pm 20$  &  $177.2 \pm 7.9$ \\  [-0.3ex]
SED  &  BT-SETTL  &  $3172 \pm 37$  &  $3150 \pm 37$  &  $181 \pm 11$ \\  [-0.3ex]
ER   &  M15  &  $3187 \pm 75$  &  $3145 \pm 90$  &  $177 \pm 15$  \\  [-0.3ex]
ER   &  D16  &  3197 &  3156 &   \\  [-0.3ex]

 \noalign{\smallskip}
SED  &  Combined  & $3152\,^{+57}_{-40}$  &  $3131\,^{+56}_{-38}$  &  $179\,^{+13}_{-10}$ \\
      
 \noalign{\smallskip} \noalign{\smallskip} \noalign{\smallskip}
 
\multicolumn{5}{c}{............................................ AD 3116 ......................................} \\

\noalign{\smallskip} 

SED  &  BT-SETTL  &  $3184 \pm 29$  &  $1639 \pm 248$  &  --- \\  [-0.3ex]
ER   &  M15  &  $3237 \pm 74$  &  $880 \pm 217$  &  $183 \pm 14$  \\  [-0.3ex]
ER   &  D16  &  3236 &  880 &   \\

 \noalign{\smallskip} \noalign{\smallskip} \noalign{\smallskip}
 
\multicolumn{5}{c}{............................................ AD 1508 ......................................} \\

\noalign{\smallskip} 

SED  &  PHOENIX  &  $3754 \pm 78$  &  $3679 \pm 121$  &  $164 \pm 22$ \\  [-0.3ex]
SED  &  BT-SETTL  &  $3779 \pm 87$  &  $3706 \pm 117$  &  $167 \pm 23$ \\  [-0.3ex]
ER   &  M15  &  $3738 \pm 76$  &  $3639 \pm 284$  &  $156 \pm 28$  \\  [-0.3ex]
ER   &  D16  & 3746  &  3649 &   \\  [-0.3ex]

 \noalign{\smallskip}
SED  &  Combined  & $3767\,^{+99}_{-85}$  &  $3693\,^{+122}_{-135}$  &  $166\,^{+25}_{-23}$ \\
      
 \noalign{\smallskip} \noalign{\smallskip} 
 \hline  
 \end{tabular}  
\begin{list}{}{}  
\item[$^{*}$] SED = spectral energy distribution and ER = empirical relations.
\item[$^{\dag}$] M15 = empirical relations from \citet{Mann16}; D16 = \citet{David16} polynomial fit to the color and temperature data presented in \citet{Pecaut13}.
\item[$^{\ddag}$] For the two sets of empirical relations, the secondary \teff\ is estimated using the  \gpe\ temperature ratio in the \ktwo\ band as a proxy for the \teff\ ratio.
\end{list}  
 \end{table}

% =========================================================
\subsubsection{Effective temperatures}

We find that the BT-SETTL model temperatures are typically $\sim$40 K hotter than the PHOENIX v2 values, although both sets of temperatures agree to within 1$\sigma$. They are also both in agreement with the temperatures predicted by empirical relations. We note that both sets of empirical relations used the BT-SETTL models to calibrate their temperature scale and hence caution should be applied when interpreting the slightly closer agreement between the empirical relations and BT-SETTL SED temperatures than with the PHOENIX v2 values. 

Given the slight offset between the BT-SETTL and PHOENIX temperatures we opted to combine the two predictions for each star as our final \teff\ values. These are reported in Table \ref{Td_comp_tab} as the ``combined'' model and are:  $T_{\rm pri} = 3211\,^{+54}_{-36}$ K and $T_{\rm sec} = 3103\,^{+53}_{-39}$ K for AD 3814; $T_{\rm pri} = 3152\,^{+57}_{-40}$ K and $T_{\rm sec} = 3131\,^{+56}_{-38}$ K for AD 2615; and $T_{\rm pri} = 3767\,^{+99}_{-85}$ K and $T_{\rm sec} = 3693\,^{+122}_{-135}$ K for AD 1508. For AD 3116 we used only the BT-SETTL models given the expected temperature of the brown dwarf secondary.

While both SED modeling and empirical relations yield consistent results, the SED modeling constraints are significantly tighter (even combining both sets of results), which is perhaps unsurprising given they are system-specific and capitalize on the joint light curve and RV modeling constraints. Furthermore, interpreting the temperature ratio from the light curve modeling as a genuine \teff\ ratio is incorrect in all cases where the bandpass observed does not cover the majority of the integrated spectra of both EB components, and the system is not equal mass. For both ADs 3814 and 2615, using the \kepler\ bandpass temperature ratio as a \teff\ ratio (as required when using empirical relations) results in a steeper temperature scale than the light curve modeling results actually imply, i.e. the secondary is predicted to be cooler than expected relative to the primary temperature. This effect is most noticeable in AD 3814 given the larger mass ratio in this system.

% =========================================================
\subsubsection{Distance to \prae}

Literature distance estimates to \prae\ range from $\sim$160--190 pc with the more recent determinations clustering around $\sim$175--185 pc \citep{Mermilliod90,Reglero91,Gatewood94,Percival03,An07,vanLeeuwen09,vanLeeuwen17}. Gaia DR1 parallaxes imply a distance of $182.8 \pm 1.7 \pm 14$ \citep[the two uncertainties are the error on the cluster center determination and the observed spread of cluster members on the sky;][]{vanLeeuwen17}. Our distance estimates for ADs 3814, 2615 and 1508 are $170.4\,^{+11.0}_{-8.9}$, $179\,^{+13}_{-10}$ and $166\,^{+25}_{-23}$ pc, respectively, which are all in agreement with the Gaia parallax distance. As AD 3116 is single-lined, we do not have precise radii and surface gravities, so we placed a prior on the distance to the system of $d_{\rm sys} = 182.8\pm14$ pc, and hence do not quote a distance for this system as we essentially recover our prior.

Empirical bolometric corrections (BCs) are available for M-dwarfs \citep[e.g.][]{Mann16}. Combining these with our calculated radii gives the system bolometric flux, which can be converted to absolute bandpass magnitudes using the derived BCs and compared to apparent magnitudes to estimate the distance using the distance modulus (see M15 distances in Table \ref{Td_comp_tab}). We note that these are also in agreement with both our distances and the Gaia cluster value.

% =========================================================
\subsection{Comparison with stellar evolution models}
\label{model_comp}
\subsubsection{The Newly characterized EBs}
\label{model_comp_prae}

\begin{figure*}
\centering
   \includegraphics[width=0.32\linewidth]{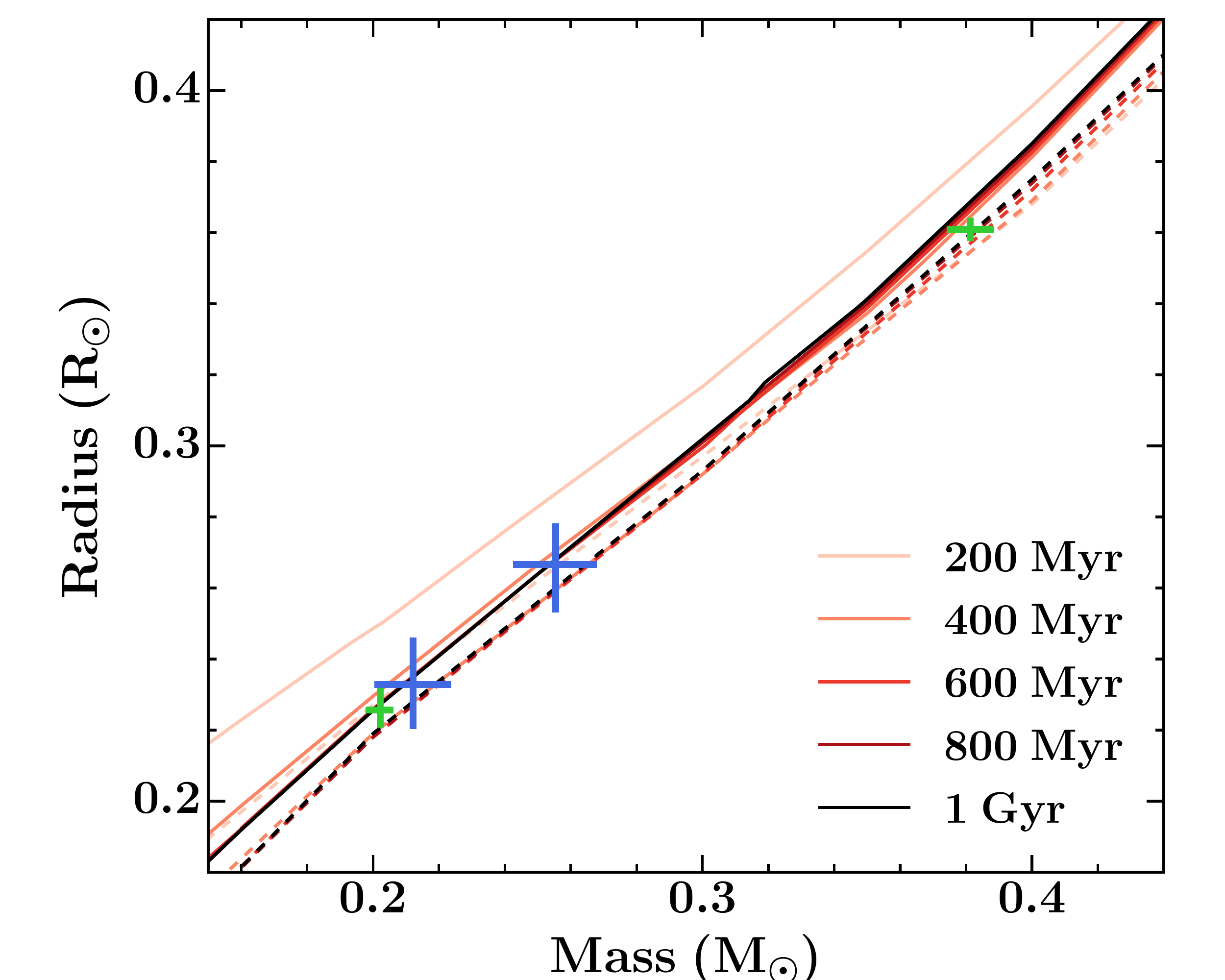} 
   \includegraphics[width=0.32\linewidth]{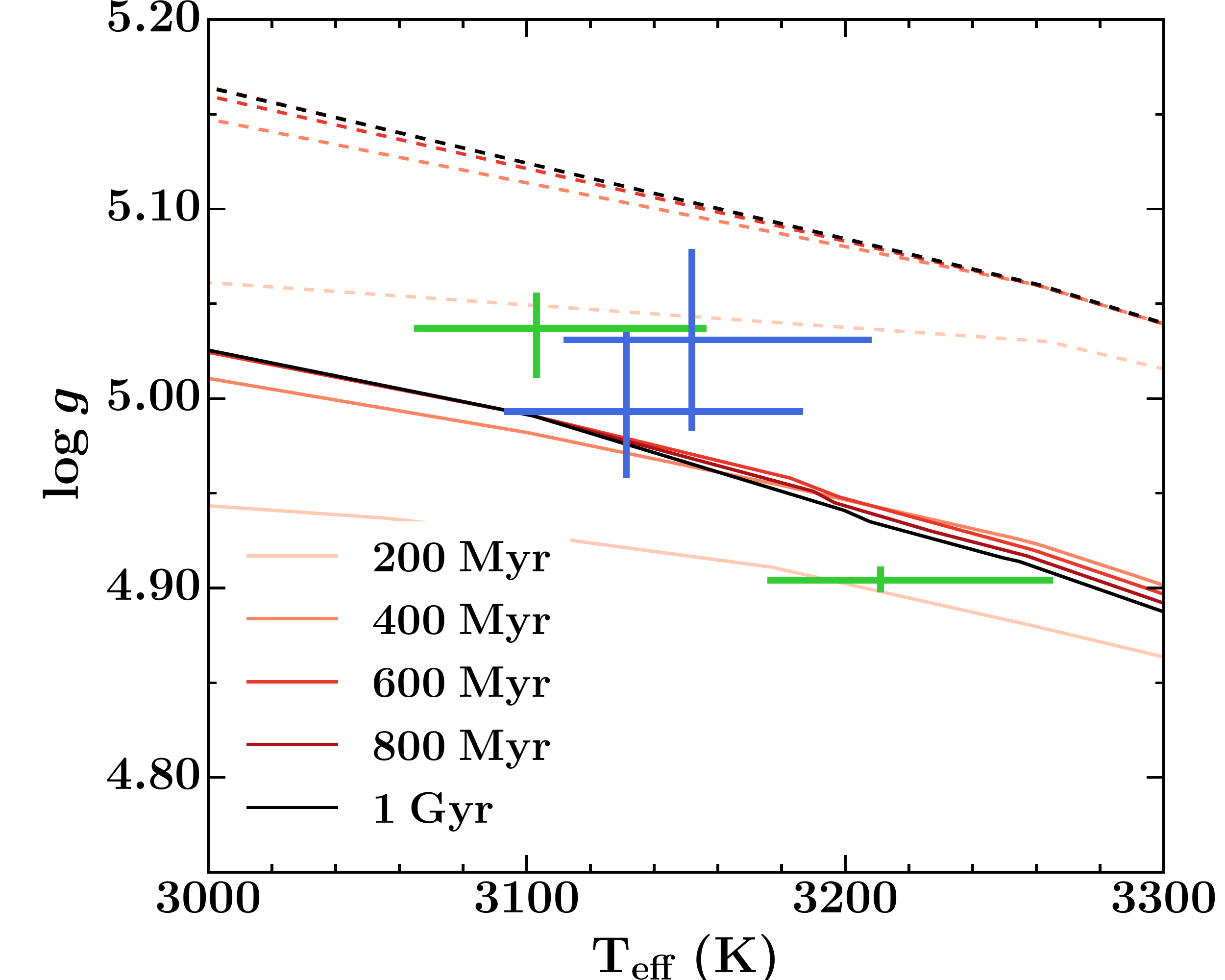} 
   \includegraphics[width=0.32\linewidth]{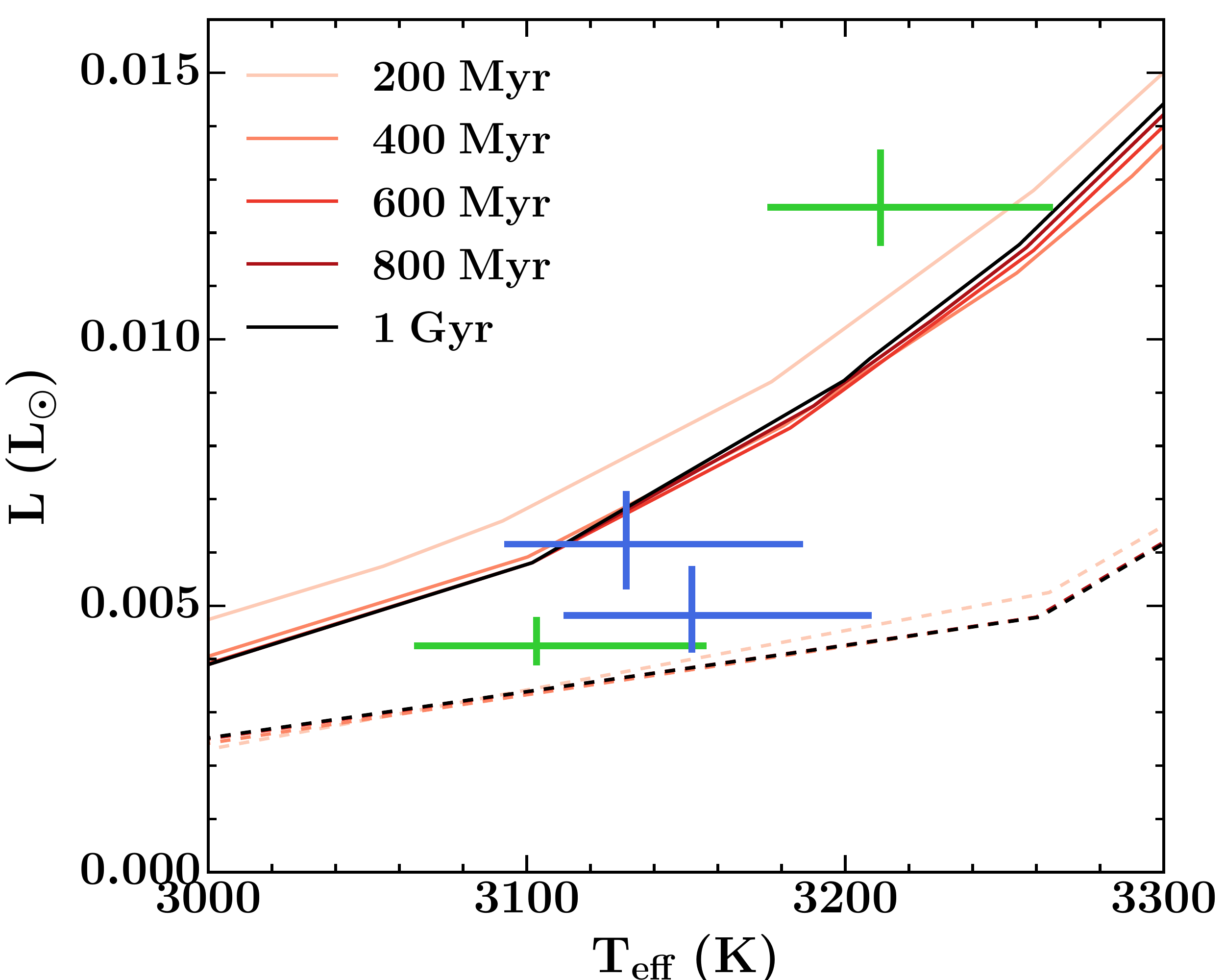} 
  
   \includegraphics[width=0.32\linewidth]{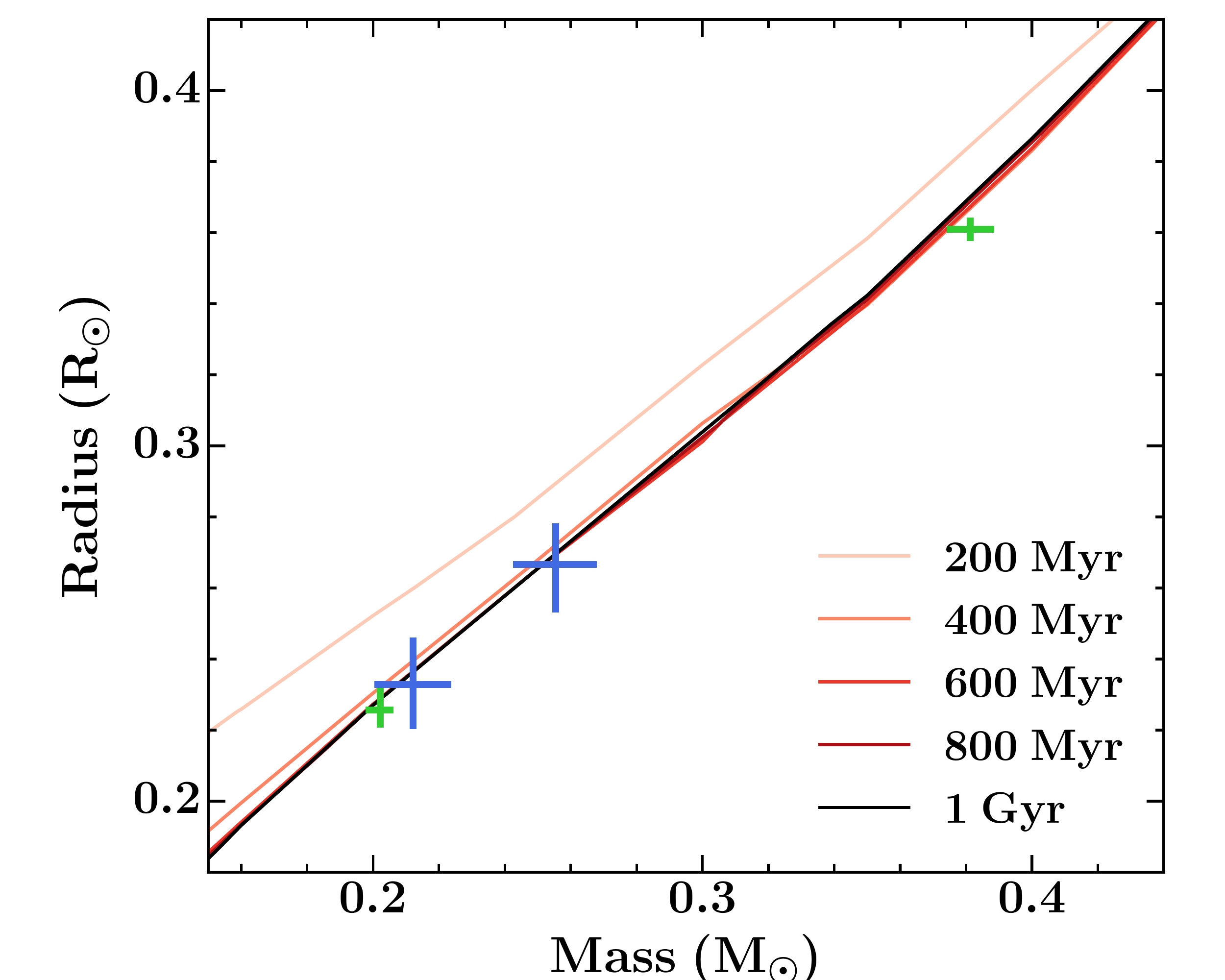} 
   \includegraphics[width=0.32\linewidth]{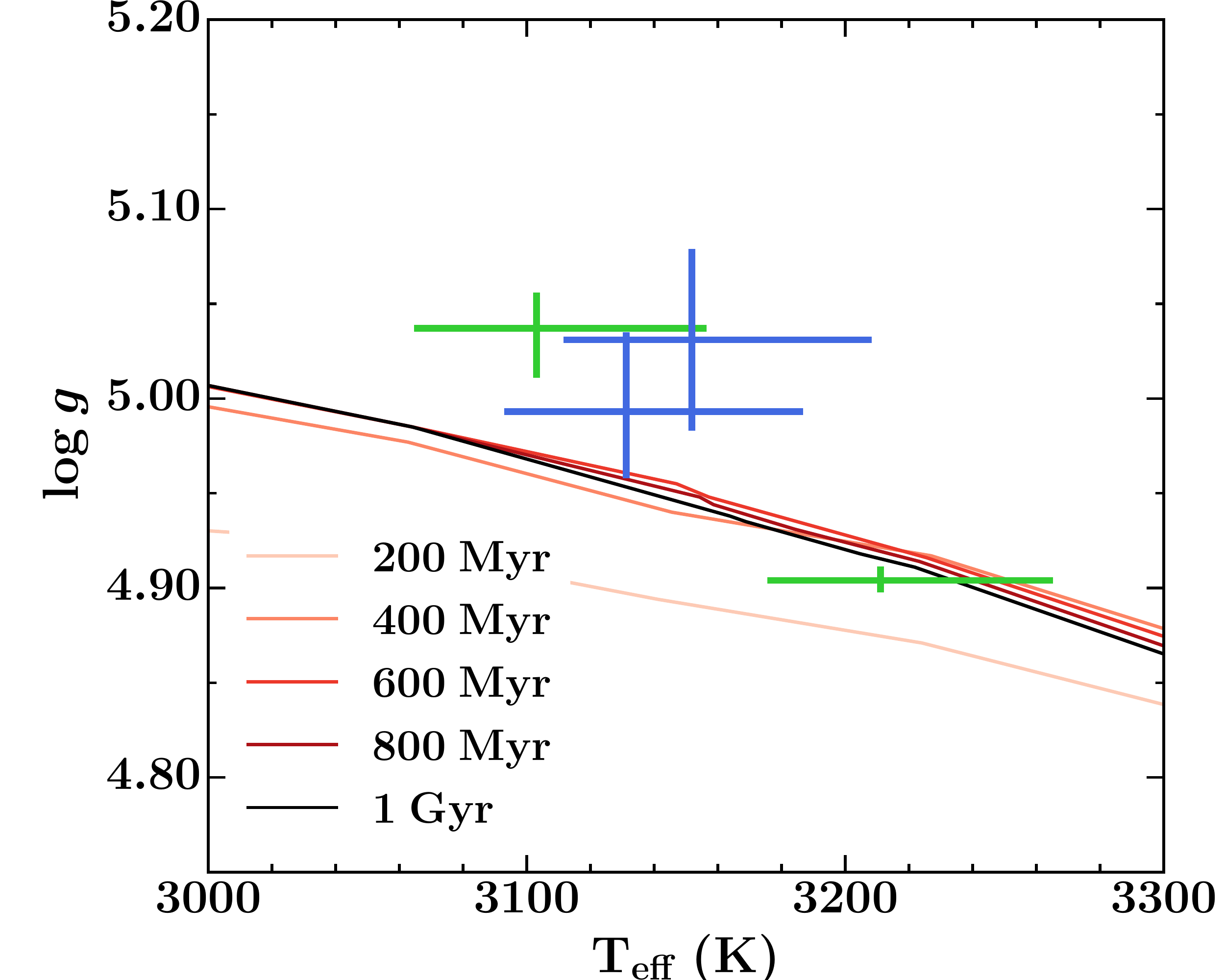} 
   \includegraphics[width=0.32\linewidth]{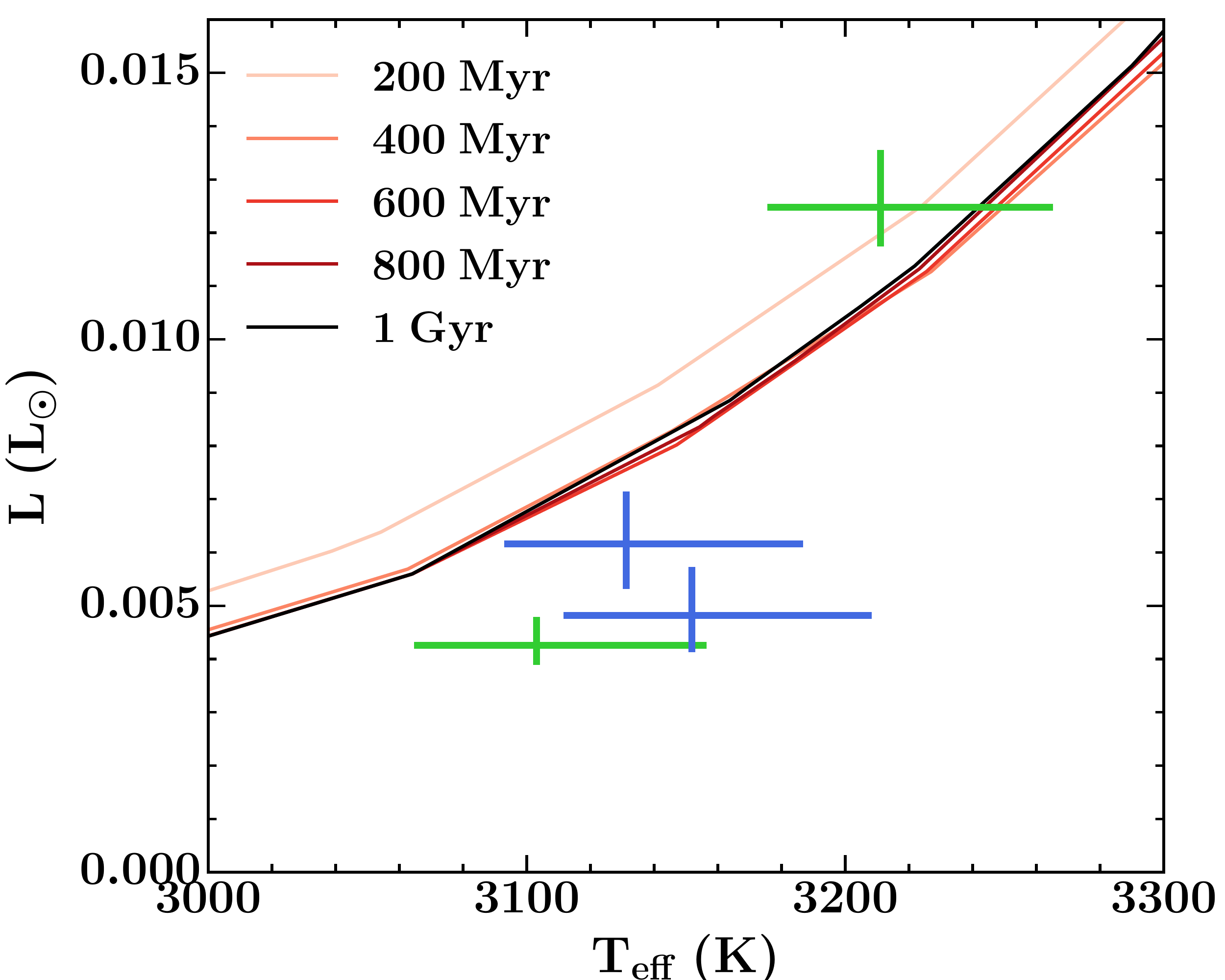} 
   
   \caption{Comparison of the fundamental parameters of ADs 3814 and 2615 (green and blue, respectively) to the PARSEC v1.2 and BHAC15 model isochrones (solid and dashed lines, respectively). The \emph{top row} shows the PARSEC (solid) and BHAC15 (dashed) models in the mass--radius, \teff--\logg\ and \teff--luminosity planes (\eleft\ to \eright) at solar metallicity. The \emph{bottom row} shows the same planes but for the PARSEC models at the metallicity of \prae\ ($Z=0.0174$). The model isochrones shown are common in all plots and range from 200 Myr (lightest) to 1 Gyr (darkest).}
   \label{MRT_comp}
\end{figure*}

With precise masses, radii and effective temperatures for both stars in ADs 3814 and 2615 we can test the predictions of stellar evolution theory for low mass stars at the beginning of the main sequence phase of evolution.
Figure \ref{MRT_comp} compares the fundamental parameters of ADs 3814 and 2615 to the PARSEC v1.2 \citep{Bressan12,Chen14} and BHAC15 \citep{Baraffe15} models. 
\prae\ is slightly metal-rich ([Fe/H]$\approx$0.14) but the closest BHAC15 models in metallicity are solar composition. Therefore, we compare our results with both the solar metallicity PARSEC and BHAC15 models (Figure \ref{MRT_comp}, \emph{top row}) and also compare to the PARSEC models at \prae\ metallicity (Figure \ref{MRT_comp}, \emph{bottom row}).   
In the mass-radius plane (\emph{left panels}) the PARSEC models (solid lines) predict slightly larger radii than BHAC15 (dashed lines) for a given mass, but both models are able to explain the two components of each system with a single isochrone at the 1$\sigma$ level (for PARSEC this is true for both solar and \prae\ metallicities). This agreement is encouraging as the masses of AD 3814 are constrained to 2\% for both components and the primary and secondary radii to 1\% and 3\%, respectively. The uncertainties on the masses and radii of AD 2615 are slightly larger, given the system is fainter, and there are fewer eclipses and RVs, but the masses and radii are still both constrained to 6\% for the primary and 5\% for the secondary. 

We note that both systems are young (sub-Gyr) and display modest H$\alpha$ emission. Therefore, compared to old M dwarfs these Praesepe stars are expected to have relatively strong magnetic fields and high spot coverage. Higher activity levels are thought to result in stars with lower effective temperatures and inflated radii \citep[e.g.][]{Chabrier07,Macdonald14}, and this is often seen in observations \citep[e.g.][]{Feiden12}. Stars in EBs with longer orbital periods appear to show better agreement with the models, but those that do show disagreement tend to be fully convective. This might suggest that for stars with radiative cores and convective outer envelopes, disagreements with models are driven by rotation and magnetic activity but comparisons for fully convective stars are subject to other errors \citep{Feiden15}. That these two fully (or almost fully) convective EB systems are active and have relatively short (6--12 day) periods yet agree well with the radius predictions of non-magnetic models presents a further challenge to stellar evolution theory.

While the masses and radii appear to be in agreement, including \teff\ complicates the picture. We next compare our results in the \teff--\logg\ plane. The surface gravity, \logg, combines the mass and radius information, which agree well for both models, and hence this parameter should also be well explained. In the middle column of Figure \ref{MRT_comp} we see significant discrepancies between the data and models, which points towards problems in the model \teff\ scales. The models substantially diverge in their \teff\ predictions, with the BHAC15 models being hotter by $\sim$200--250 K across the mid M-dwarf range, and the PARSEC models being perhaps 10--25K cooler than the data. We note that this is also seen in the mass--\teff\ and radius-\teff\ planes (not shown here). Both sets of models essentially predict the same \teff\ independent of age for $t \gtrsim$400 Myr out to 10 Gyr. Our SED analysis yields \teff\ values that are in closer agreement to the PARSEC models than BHAC15, but both models predict a steeper \teff\ scale than the data suggest (note that a steeper model \teff\ scale manifests as a shallower gradient in \teff\--\logg\ space, as observed). One option is that the model \teff\ scales are too steep for mid M-dwarfs but it could also be that additional phenomena, not included in the models, are responsible for the observed slope difference. Both ADs 3814 and 2615 display starpot modulation in the \ktwo\ light curves. As neither PARSEC nor BHAC15 include the effect of magnetic fields and starspots it could be that some of the discrepancy arises from these phenomena rather than the model \teff\ scale being too steep per se. 

Although the primary component of AD 3814 agrees with the PARSEC \prae\ metallicity models, the secondary lies above the relation. We can take the primary star as an example to explore the required spot coverage and contrast ratio needed to bring its computed \teff\ onto the same expected isochrone as the secondary component. We note that this scenario would require the PARSEC \teff\ scale to be underpredicting the true unspotted \teff\ but this is plausible so we continue with the exercise nonetheless. Assuming a spot-to-unspotted photospheric temperature ratio of 0.8 \citep[e.g.][]{Grankin98} would require $\sim$25\% spot coverage. To bring the primary and secondary components within 1$\sigma$ would only require a 10\% spot coverage on the primary. We note, however, that the radius posterior medians sit just below the zero age main sequence predicted by the PARSEC models and invoking starspots to redress the \teff\ slope differences would imply a corresponding decrease in the radii for these stars without spots.

To bring the primary and secondary components of AD 3814 into agreement with the BHAC15 models would require spot coverages of 30--40\% on each star. While high, this is consistent with observations of active late-type stars, especially those in close binaries \citep[e.g.][]{ONeal04}. We note that the BHAC15 models track a steeper path in \logg\--\teff\ space beyond 3400 K (corresponding to a shallower \teff\ scale). Simply shifting the BHAC15 models cooler by 250 K would bring them into agreement with all four stars. This is not possible with the PARSEC models, so it remains a valid option that the PARSEC model temperature scale is too steep over the mass range probed ($\sim$0.2--0.4 $M_{\odot}$). However, more precisely characterized M-dwarf binaries are required to confirm this tentative statement.

The radii and effective temperatures combine to determine the luminosity of a star. Stellar evolution models are typically found to underpredict the radii and overpredict the effective temperatures of active low-mass stars; however, these combine to essentially recover the correct luminosity. The \emph{right} column of Figure \ref{MRT_comp} shows the radiative \teff\--luminosity relation. As expected, the BHAC15 models appear to underpredict the luminosity because the model \teff\ is too high. The PARSEC models are in better agreement: they are able to follow the general trend of the data and explain the primary component of AD 3814 and the secondary of AD 2615, but the other components are slightly discrepant at the $\sim$1.5$\sigma$ level.

\begin{figure*}
  \centering
  \includegraphics[width=0.8\linewidth]{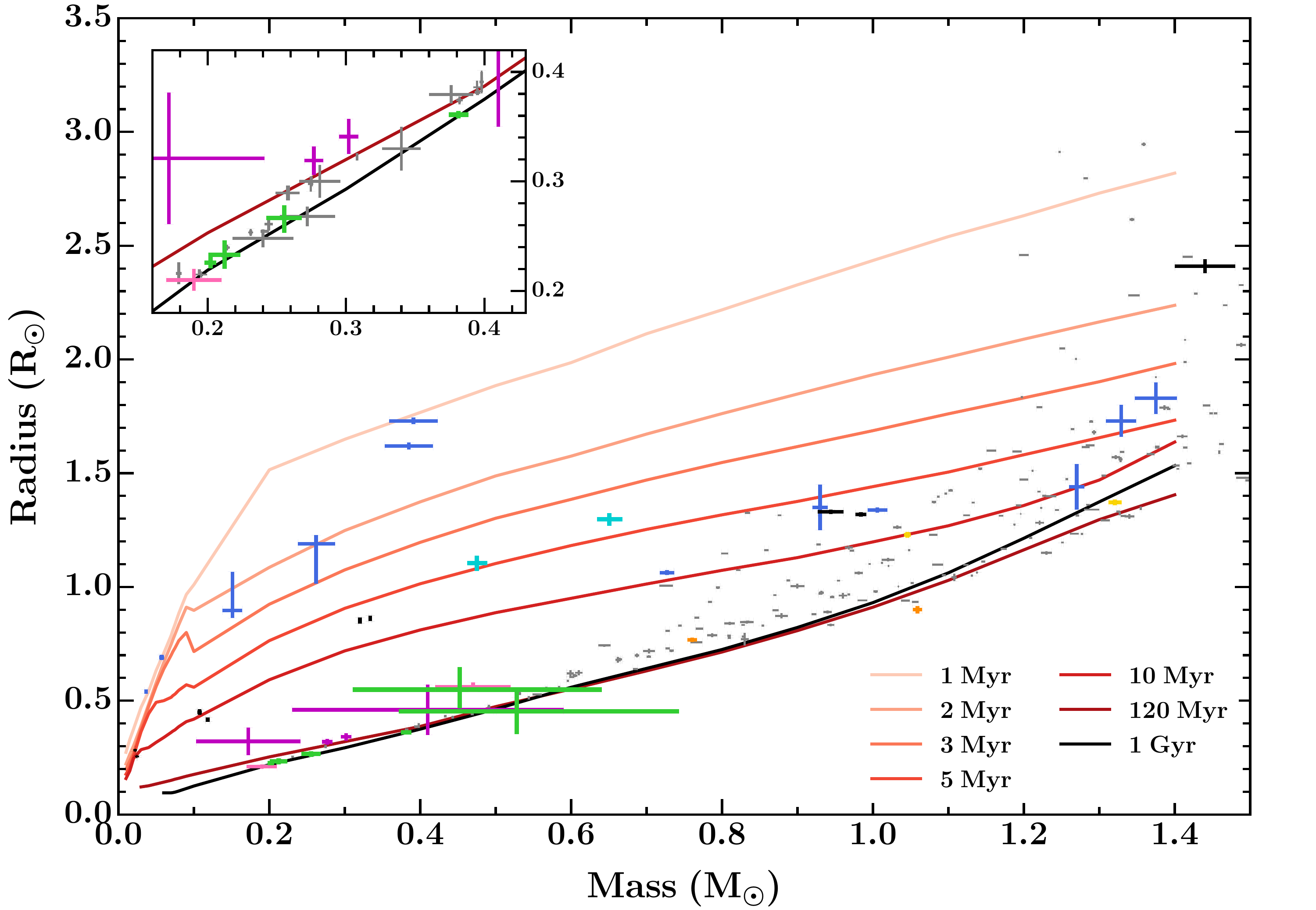}
  \caption{Mass-radius relation for detached double-lined eclipsing binaries (EBs) below 1.5 $M_{\odot}$. Data compiled from Table \ref{pms_ebs} and DEBCat (\href{http://www.astro.keele.ac.uk/~jkt/debdata/debs.html}{http://www.astro.keele.ac.uk/$\sim$jkt/debdata/debs.html}). EBs that are members of open clusters are colored while field EBs are shown in gray. The clusters containing well-characterized EBs are Orion (blue), Upper Scorpius (black), NGC\,2264 (cyan), Pleiades (magenta), Hyades (orange), NGC\,1647 (pink), Per OB2 (gold), and the new \prae\ EBs (green) presented here. The colored lines represent solar metallicity isochrones of \citet{Baraffe15} from 1 Myr to 1 Gyr (\etop\ to \ebot). Inset (\etop\ \eleft) is a zoom on the region containing ADs 3814 and 2615 to allow a closer comparison between the models and current observational constraints for low-mass stars. Here we also include the compilation of known low-mass EBs presented in \citet{Dittmann17}. }
  \label{MR_plot_full}
\end{figure*}

\begin{table*}
  \centering
  \caption{Published double-lined eclipsing binary systems in sub-Gyr open clusters where both components are below 1.5 M$_{\odot}$, ordered by ascending primary mass.}
  \label{pms_ebs}
  \resizebox{\textwidth}{!}{%
  \begin{tabular}{lccccllll}
    \hline
    \hline
    \noalign{\smallskip}
    Name & $M_{\rm{pri}}$ & $M_{\rm{sec}}$ & $R_{\rm{pri}}$ & $R_{\rm{sec}}$ & Cluster\,$^{a}$ & Age & Year & Refs. \\
    & ($M_\odot$) & ($M_\odot$) & ($R_\odot$) & ($R_\odot$) & & (Myr) & \\
    \noalign{\smallskip}
    \hline
    \noalign{\smallskip}

    EPIC 203868608        & $0.02216\pm0.00045$   & $0.02462\pm0.00055$   & $0.2823\pm0.0051$   & $0.2551\pm0.0036$     & Upper Sco   & 5--10     & 2016   & 1  \\ %[-0.7ex]
    2MJ0535-05            & $0.0572\pm0.0033$     & $0.0366\pm0.0022$     & $0.690\pm0.011$     & $0.540\pm0.009$       & ONC         & 1--2      & 2006   & 2,3 \\ %[-0.7ex]
    EPIC 203710387        & $0.1183\pm0.0028$     & $0.1076\pm0.0031$     & $0.417\pm0.010$     & $0.450\pm0.012$       & Upper Sco   & 5--10     & 2015   & 4,1 \\ %[-0.7ex]
    JW\,380               & $0.262\pm0.025$       & $0.151\pm0.013$       & $1.189\pm0.175$     & $0.897\pm0.170$       & ONC         & 1--2      & 2007   & 5 \\ %[-0.7ex]
    HCG 76                & $0.2768\pm0.0072$     & $0.3020\pm0.0073$     & $0.319\pm0.036$     & $0.34\pm0.11$         & Pleiades    & 125       & 2016   & 6  \\ %[-0.7ex]
    UScoCTIO 5            & $0.3336\pm0.0022$     & $0.3200\pm0.0022$     & $0.862\pm0.012$     & $0.852\pm0.013$       & Upper Sco   & 5--10     & 2015   & 7,1 \\ %[-0.7ex]
    Par\,1802             & $0.391\pm0.032$       & $0.385\pm0.032$       & $1.73\pm0.015$      & $1.62\pm0.015$        & ONC         & 1--2      & 2008   & 8,9 \\ %[-0.7ex]
    MHO 9                 & $0.41\pm0.18$         & $0.172\pm0.069$       & $0.46\pm0.11$       & $0.321\pm0.060$       & Pleiades    & 125       & 2016   & 6 \\ %[-0.7ex]
    2MJ0446+19            & $0.47\pm0.05$         & $0.19\pm0.02$         & $0.56\pm0.02$       & $0.21\pm0.01$         & NGC 1647    &  150      & 2006   & 10 \\ %[-0.7ex]
    CoRoT\,223992193      & $0.668\pm0.012$       & $0.4953\pm0.0073$     & $1.295\pm0.040$     & $1.107\pm0.044$       & NGC 2264    & 3--6      & 2014   & 11 \\ %[-0.7ex]
    MML\,53               & $0.994\pm0.030$       & $0.857\pm0.026$       & \multicolumn{2}{c}{$2.201\pm0.071$\,$^b$}   & UCL         & 15        & 2010   & 12,13 \\ %[-0.7ex]
    HD144548              & $0.984\pm0.007$       & $0.944\pm0.017$       & $1.319\pm0.010$     & $1.330\pm0.010$       & Upper Sco   & 5--10     & 2015   & 14 \\ %[-0.7ex]
                          & \multicolumn{2}{c}{$1.44\pm0.04$\,$^c$}       & \multicolumn{2}{c}{$2.41\pm0.03$\,$^c$}     &             &           &        &   \\ %[-0.7ex]
	V1174\,Ori            & $1.006\pm0.013$       & $0.7271\pm0.0096$     & $1.338\pm0.011$     & $1.063\pm0.011$       & Ori OB 1c   & 5--10     & 2004   & 15 \\  %[-0.7ex]
    V818 Tau              & $1.06\pm0.01$         & $0.90\pm0.02$         & $0.76\pm0.01$       & $0.77\pm0.01$         & Hyades      & 600--800  & 2002   & 16 \\ %[-0.7ex]
    RXJ\,0529.4$+$0041A   & $1.27\pm0.01$         &  $0.93\pm0.01$        & $1.44\pm0.10$       & $1.35\pm0.10$         & Ori OB 1a   & 7--13     & 2000   & 17,18,19 \\ %[-0.7ex]
    NP Per                & $1.3207\pm0.0087$     & $1.0456\pm0.0046$     & $1.372\pm0.013$     & $1.229\pm0.013$       & Per OB 2    & 6--15     & 2016   & 20 \\ %[-0.7ex]
    ASAS\,J0528$+$03      & $1.375\pm0.028$       &  $1.329\pm0.020$      & $1.83\pm0.07$       & $1.73\pm0.07$         & Ori OB 1a   & 7--13     & 2008   & 21 \\ %[-0.7ex]

    \noalign{\smallskip}  \noalign{\smallskip}                                        
     \multicolumn{9}{c}{\emph{Praesepe systems published in this paper}} \\
    \noalign{\smallskip}
    
    AD 3814               & $0.3813\pm0.0074$     & $0.2022\pm0.0045$     & $0.3610\pm0.0033$   & $0.2256\pm0.0063$     & Praesepe    & 600--800  & 2017   & this work \\ %[-0.7ex]
    AD 2615               & $0.212\pm0.012$       & $0.255\pm0.013$       & $0.233\pm0.013$     & $0.267\pm0.014$       & Praesepe    & 600--800  & 2017   & this work \\ %[-0.7ex]
    AD 1508               & $0.45\pm0.19$         & $0.53\pm0.21$         & $0.548\pm0.099$     & $0.45\pm0.10$         & Praesepe    & 600--800  & 2017   & this work \\ %[-0.7ex]
    
    \noalign{\smallskip}
    \hline
  \end{tabular} }
  \begin{list}{}{}
  \item[\textbf{Notes.}]Where asymmetric error bars were reported in the original papers we quote the larger of the two here. 
  \item[$^a$]ONC = Orion Nebula Cluster; UCL = Upper Centaurus Lupus; Upper Sco = Upper Scorpius.
  \item[$^b$]Radius sum (individual radii have not been determined).
  \item[$^c$]Tertiary component that is also eclipsed.
  \item[\textbf{References.}] 
  1. \citet{David16}; 2. \citet{Stassun06}; 3. \citet{Stassun07}; 4. \citet{Lodieu15}; 5. \citet{Irwin07}; 
  6. \citet{David16a}; 7. \citet{Kraus15}; 8. \citet{Cargile08}; 9. \citet{Stassun08}; 10. \citet{Hebb06}; 11. \citet{Gillen14}
  12. \citet{Hebb10}; 13. \citet{Hebb11}; 14. \citet{Alonso15}; 15. \citet{Stassun04}; 16. \citet{Torres02}; 17. \citet{Covino00}; 18. \citet{Covino01}; 19. \citet{Covino04}; 
  20. \citet{Lacy16}; 21. \citet{Stempels08}.
  \end{list}
\end{table*}

\subsubsection{Updated mass--radius relation for low-mass EBs}

Figure \ref{MR_plot_full} shows the mass-radius relation for detached double-lined eclipsing binaries below $M<1.5$ $M_{\odot}$. Field EBs are shown in gray while members of young open clusters -- including our newly discovered systems reported here -- are colored by cluster (see figure caption for color scheme). The fundamental parameters of the known cluster EBs with both components below 1.5 $M_{\odot}$ are reported in Table \ref{pms_ebs}. The three double-lined \prae\ systems reported here make a significant contribution to known cluster EBs, increasing the total number below 1.5 $M_{\odot}$ by almost 20\% (and increasing the known double-lined M-dwarf EB population by 30\%). Furthermore, ADs 3814 and 2615 add precise constraints for stellar evolution models at the zero-to-early age main sequence for low-mass stars.

% =========================================================
\subsubsection{Age of \prae}
\label{age_discussion}

As briefly discussed in the introduction, the age of \prae\ has been debated in recent years. It has typically been estimated at $\sim$600--650 Myr by isochrone fitting, often through association with the Hyades \citep[e.g.][]{Perryman98,Salaris04,Fossati08}. However, \citet{Brandt15} found that including rotation in stellar models implied an age of $790\pm60$ Myr (2$\sigma$ uncertainty) for \prae, which is in agreement with their Hyades age of $\sim$750--800 Myr \citep{Brandt15a}. This older age estimate arises from the fact that rotation results in longer main sequence lifetimes and hence older ages for post-turnoff populations. This result was corroborated by \cite{David15}, who also include the effect of stellar rotation in their comparison between stellar atmospheric parameters (derived from Str\"{o}mgren photometry) and theoretical isochrones.

Somewhat orthogonal to the ages inferred from radiative properties such as $L$ and $T$, the ages of EB systems can be determined through comparison of their masses and radii with stellar evolution models (see section \ref{model_comp_prae}). Unfortunately, over the mass range probed by our EBs, the several hundred Myr \prae\ sits roughly at the zero age main sequence. As M-dwarf evolution is slow, their increase in radius as they evolve through their first several Gyr on the main sequence is correspondingly small. Therefore, using our masses and radii to independently estimate the age of \prae\ would carry significant uncertainty and would not provide useful input to the current 600 vs. 800 Myr age discussion.

%%%%%%%%%%%%%%%%%%%%%%%%%%%%%%%%%%%%%%%%%%%%%%%%%
\subsection{Circularization and synchronization}
\label{sync}

\subsubsection{Tidal circularization}

In this section we compare our findings for the new EB systems to the expectations for tidal circularization and spin-orbit synchronization at the age of Praesepe.
The binaries presented here are particularly valuable benchmarks for studies of tidal dissipation timescales in close binaries, as they are at or near the beginning of their main sequence evolution. \citet{Zahn89} posited that essentially all tidal circularization should occur during the PMS phase, when stars are larger and have deeper convective envelopes. If this theory were correct, all late-type main sequence binaries with periods less than $\sim$8 days should be circularized. Binaries with longer orbital periods would retain their primordial eccentricities and experience negligible tidal circularization after the PMS phase.

However, \citet{Meibom05} used observations of binaries in coeval stellar populations to clearly show that tidal dissipation proceeds to circularize orbits well after the PMS stage (see their Fig. 9). While standard equilibrium tide theory \citep{Zahn89a, ClaretCunha97} and dynamical tide theory \citep{Witte02} do predict exactly this trend, binaries are generally observed to circularize more quickly than theory predicts (i.e. tidal dissipation is a more efficient process than expected). The binary population of Praesepe and the Hyades is a conspicuous outlier to this trend, indicating agreement with theory but significant tension with observations of all other well-characterized clusters. However, \citet{Zahn89} cautioned that two short-period eccentric binaries in Praesepe and Hyades (KW 181 and VB 121) are single-lined systems, in which the secondaries could possibly be white dwarfs, meaning that the standard theory of tidal dissipation would not apply. Ignoring these two systems, those authors estimated binaries with periods below 8.5--11.9 days should be circularized by the age of the Hyades, and by extension Praesepe. Our findings for AD 3814 and AD 2615 corroborate the notion that the circularization period for Praesepe is larger than previously measured, and to our knowledge AD 2615 is the longest period circular binary in either Praesepe or the Hyades. Revisiting the analysis of \citet{Meibom05} including these two systems would bring the observations for Praesepe into better agreement with those of other clusters, in the sense that binaries of a given age are observed to be circular out to longer periods than theory predicts.

As for AD 3116, tidal dissipation proceeds differently for extreme mass ratio systems \citep{Ogilvie14}, and so we caution against drawing conclusions based on its relatively high eccentricity ($e=0.15$) given its short orbital period of $<2$ days. In fact, the recently discovered transiting brown dwarf in the significantly older Ruprecht 147 cluster similarly exhibits a relatively high eccentricity and short orbital period \citep{Nowak16}.

Finally, we note that the transition between circular and eccentric binaries in a coeval stellar population (as demarcated by either the ``cutoff period'', i.e. the longest period circular binary, or preferably by the ``tidal circularization period'') can in principle be used to estimate the age of the stellar population \citep{MathieuMazeh88}. Given sufficient data and a well-calibrated relation amongst clusters, the method could also be extended to close binaries in the field to provide an upper limit in age if the binary is eccentric, or a lower limit if it is circular.

%SYNCHRONIZATION

\subsubsection{Spin-orbit synchronization}

The theoretical outcome of tidal evolution within a binary system is a circular orbit and a state of double synchronous rotation with spin axes aligned to the orbital angular momentum vector. However, as noted by \citet{Ogilvie14}, this theoretical prediction has never been observationally verified for a binary star system. This is in part due to the difficulty of measuring stellar rotation, particularly for both components of a binary, and the need for an eclipsing system to precisely measure obliquities. 

Binaries for which the rotation period of one or more component can be measured, particularly within coeval stellar populations, are thus critical benchmarks for tidal synchronization studies. For the four binaries discussed here, one appears to be nearly synchronized (AD 1508) while the other three appear to be rotating subsynchronously (i.e. at a frequency lower than the orbital frequency). This observation is based on the measured $P_\mathrm{spot}/P_\mathrm{orb}$ ratios of 1.25, 1.08, and 1.14 for ADs 3814, 2615, and 3116, respectively. On the surface, this is surprising given that 1) the expected synchronization timescales are much smaller than the cluster age, and 2) tidal synchronization is expected to occur more quickly than circularization in close binaries \citep{Zahn77, Hut81} and two of the subsynchronous binaries are on nearly circular orbits (ADs 3814 and 2615). 

It is important to note that photometric variations indicate a star's surface rotation rate, but the spin of interior layers is not measured and known only to the extent to which there is reason to believe the interior is coupled to the surface. For the binaries with mass ratios near unity, both stars are contributing to the observed brightness modulations, but in the absence of multiple distinct peaks in a periodogram we infer the modulation period to indicate the rotation of the primary. Notably, surface differential rotation can lead to configurations in which the spin of equatorial regions is synchronized with the orbit while higher latitudes may be rotating more slowly. Such a scenario was suggested to explain observations of the late-type EB HII 2407 in the Pleiades \citep{David15b}. Indeed, there is observational evidence \citep{Barnes15} and theoretical motivation \citep{Schuessler92, Granzer00} for polar spots on rapidly rotating, fully convective stars. 

However, unlike the Pleiades EB, the binaries presented here exhibit much larger discrepancies between the rotation and orbital periods. The measured rates of differential rotation have been observed to decrease strongly with stellar temperature \citep{Barnes05,CollierCameron07}. Using the empirical formula of \citet{CollierCameron07}, the expected rates of differential rotation for the stars considered here are all below $10^{-3}$ rad~d$^{-1}$. If we assume the orbits are synchronized at the equator and that polar spots are responsible for the measured rotation periods, then the implied rates of differential rotation for ADs 3814, 2615, and 3116 would be 0.21, 0.04, and 0.38 rad~d$^{-1}$, respectively. These values are significantly higher than the differential rotation rates measured for fully convective stars \citep{Morin08,Reinhold13,Davenport15}. Our observations therefore indicate either: 1) tidal synchronization proceeds more slowly in fully convective stars than the theory of equilibrium tides predicts, 2) magnetic braking is currently playing a more important role in the spin evolution of these binaries than tidal forces, or 3) differential rotation in fully convective stars can be much more important than previously appreciated. We consider the last explanation to be the least plausible.

Subsynchronous rotation has previously been observed for short period binaries in the younger M35 and M34 clusters, aged $\sim$150 Myr and $\sim$250 Myr, respectively \citep{Meibom06}. As those authors noted, this result is in direct contradiction with expectations of tidal evolution on the main sequence which predicts binaries with periods near or below the circularization period (which AD 3814 and AD 2615 apparently are) to be rotating pseudosynchronously (synchronized with the instantaneous orbital angular velocity at periastron) or slightly \textit{supersynchronous}.  

We conclude by noting that current theories of tidal evolution carry significant and under-explored uncertainties. In particular, theory for solar-type and early-type stars is more developed than that for fully convective stars. Tidal dissipation is expected to be more efficient, and thus circularization more rapid, in stars with convective outer layers \citep{Zahn75}, which is supported observationally \citep{vanEylen16}.

%%%%%%%%%%%%%%%%%%%%%%%%%%%%%%%%%%%%%%%%%%%%%%%%%

\section{Conclusions}
\label{conclusions}

We have presented photometric timeseries data from \kepler/\ktwo\ and follow-up high dispersion spectroscopy from Keck/HIRES in order to characterize four new EB systems in the sub-Gyr old \prae\ cluster. These new discoveries increase the number of characterized EBs below 1.5 $M_{\odot}$ in sub-Gyr open clusters by 20\%, and add 40\% of the cluster EB population with masses $M \lesssim 0.6 \, M_\odot$. 

We analyze these low-mass EBs with \gpe, a new multi-purpose Gaussian process eclipsing binary and transiting exoplanet model, to determine model-independent stellar masses and radii. We present an updated method of simultaneously determining the effective temperatures of both stars as well as the distance to an EB by modeling the system's spectral energy distribution. This approach capitalizes on the posterior constraints from the joint light curve and RV modeling to break existing degeneracies and also correctly interprets the light curve model's band-specific surface brightness ratio, rather than using it to approximate an effective temperature ratio.

We determine the masses of AD 3814 to 2\% precision and the primary and secondary radii to 1\% and 3\%, respectively. The masses and radii of AD 2615 are both determined to 6\% precision for the primary and to 5\% for the secondary. Together with effective temperatures determined to a typical precision of $\pm$50 K, we test the PARSEC v1.2 and BHAC15 stellar evolution models. Overall, the EB parameters are most consistent with the PARSEC models, primarily because the BHAC15 temperature scale is too hot over the mass--age range probed. 
Both the PARSEC and BHAC15 models are able to explain the masses and radii of ADs 3814 and 2615 with a single isochrone in the range $\sim$400--1000 Myr, but predicting \teff\ proves more challenging. Our SED-derived \teff\ values, which are consistent with those derived from empirical M-dwarf relations, are better matched to the PARSEC models. We find that the BHAC15 models predict temperatures \teff\ $\sim$100--300 K hotter than our data, whereas the PARSEC models lie in the correct \teff\ range. However, both models predict a steeper \teff\ track over the mass range $M \sim 0.2-0.4$ $M_{\odot}$ than our data suggest. More M-dwarf EBs with precise \teff\ values on the main sequence are required to confirm this tentative statement. Our luminosities are in agreement with the PARSEC model predictions but we find that the BHAC15 models overpredict this parameter primarily due to their high \teff\ values. 
While both ADs 3814 and 2615 possess precise solutions, we note that AD 3814 would benefit from a more detailed modeling of the individual eclipses (especially incorporating a full starspot model), and AD 2615 would benefit from additional RVs to tighten the existing solution.

We present a preliminary solution for a third detached double-lined system, AD 1508. The \ktwo\ light curve displays clear, but shallow, eclipses on both stars and the three Keck/HIRES RVs we obtained show the two stars not to be rapid rotators. This system is therefore amenable to precise characterization but would require further RV measurements throughout the orbital phase and may also benefit from targeted eclipse monitoring with moderate-aperture ground-based telescopes.

The final system, AD 3116, comprises a mid M-dwarf primary star with a transiting brown dwarf companion ($M$$\sim$54 $M_{\rm Jup}$). There are only $\sim$20 transiting brown dwarf systems known: AD 3116 is one of only three systems where the primary is an M-dwarf, and is only the second transiting brown dwarf system discovered in an open cluster (and the first younger than a Gyr). It will therefore be a favorable target for future transiting brown dwarf studies.

Finally, we find that ADs 3814 and 2615, which have orbital periods of 6.0 and 11.6 days, are circularized but not synchronized, with at least one component rotating sub-synchronously. This contradicts the expectations of tidal evolution, which would predict synchronization to proceed faster than circularization in these systems and for it to have been achieved by the age of \prae. Our observations therefore suggest that either tidal synchronization proceeds more slowly in fully convective stars than the theory of equilibrium tides predicts, or magnetic braking is currently playing a more important role in the spin evolution of these binaries than tidal forces.

%%%%%%%%%%%%%%%%%%%%%%%%%%%%%%%%%%%%%%%%%%%%%%%%%%%%%%%%%%%%%%%%%%%%%%
%%%%%%%%%%%%%%%%%%%%       Parameter table      %%%%%%%%%%%%%%%%%%%%%%
%%%%%%%%%%%%%%%%%%%%%%%%%%%%%%%%%%%%%%%%%%%%%%%%%%%%%%%%%%%%%%%%%%%%%%

\begin{table*}  
 \centering  
 \footnotesize
 \caption[Model parameters]{Fitted and derived parameters of the models applied to AD~3814, AD~2615, AD~3116 and AD~1508.}  
 % \emph{ 20170203 14.55 20170203 02.06 20170423 22.31 20170328 12.19} 
 \label{lc_model_tab}  
 \resizebox{\textwidth}{!}{%
 \begin{tabular}{l l l c c c c }  
 \noalign{\smallskip} \noalign{\smallskip} \hline  \hline \noalign{\smallskip}  
 Parameter  &   Symbol  &  Unit  & \multicolumn{4}{c}{Value} \\   
     &        &            &    \,\,\,\,\,AD~3814    &    \,\,\,\,\,AD~2615    &    \,\,\,\,\,AD~3116    &    \,\,\,\,\,AD~1508 \\ 
 \noalign{\smallskip} \hline \noalign{\smallskip} \noalign{\smallskip}  

\multicolumn{7}{c}{\emph{Eclipse parameters}} \\   
 \noalign{\smallskip} \noalign{\smallskip}   
 Sum of radii    &    $(R_{\rm{pri}} + R_{\rm{sec}})/ a$    &        &    $\,\,\,\,\,0.05044\,^{+0.00069}_{-0.00055}$    &    $\,\,\,\,\,0.02979 \pm 0.00034$    &    $\,\,\,\,\,0.0845\,^{+0.0066}_{-0.0053}$    &    $\,\,\,\,\,0.1774\,^{+0.0066}_{-0.0076}$    \\  [\linejump ex] 
 Radius ratio    &    $R_{\rm{sec}} / R_{\rm{pri}}$    &       &    $\,\,\,\,\,0.624\,^{+0.017}_{-0.010}$    &    $\,\,\,\,\,1.15 \pm 0.11$    &    $\,\,\,\,\,0.3599\,^{+0.0094}_{-0.0128}$    &    $\,\,\,\,\,0.83 \pm 0.24$             \\  [\linejump ex]
 Orbital inclination & $i$ & $^{\circ}$        &    $\,\,\,\,\,89.177\,^{+0.051}_{-0.064}$    &    $\,\,\,\,\,88.996 \pm 0.013$    &    $\,\,\,\,\,88.41\,^{+0.49}_{-0.42}$    &    $\,\,\,\,\,80.54\,^{+0.46}_{-0.39}$    \\  [\linejump ex]
 Orbital period    &    $P$    &    days         &    $\,\,\,\,\,6.015717 \pm 0.000013$    &    $\,\,\,\,\,11.615254 \pm 0.000073$    &    $\,\,\,\,\,1.9827960 \pm 0.0000060$    &    $\,\,\,\,\,1.5568370\,^{+0.0000100}_{-0.0000090}$    \\  [\linejump ex]
 Time of eclipse center    &    $T_{\rm{prim}}$    &    BJD         &    $\,\,\,\,\,2457178.982842 \pm 0.000059$    &    $\,\,\,\,\,2457176.63998 \pm 0.00019$    &    $\,\,\,\,\,2457178.817792 \pm 0.000080$    &    $\,\,\,\,\,2457147.26784 \pm 0.00026$    \\  [\linejump ex]
     &    $\sqrt{e} \cos \omega$    &         &    $-0.0301\,^{+0.0103}_{-0.0057}$    &    $\,\,\,\,\,0.0337\,^{+0.0067}_{-0.0128}$    &    $\,\,\,\,\,0.364\,^{+0.016}_{-0.026}$    &    $-0.0081\,^{+0.0069}_{-0.0094}$    \\  [\linejump ex]
     &    $\sqrt{e} \sin \omega$    &         &    $\,\,\,\,\,0.031 \pm 0.034$    &    $\,\,\,\,\,0.020 \pm 0.052$    &    $\,\,\,\,\,0.04 \pm 0.14$    &    $\,\,\,\,\,0.010\,^{+0.049}_{-0.041}$    \\  [\linejump ex]
 Central surface brightness ratio    &    $J_{\rm{K2}}$    &         &    $\,\,\,\,\,0.748 \pm 0.034$    &    $\,\,\,\,\,0.950 \pm 0.060$    &    $\,\,\,\,\,0.0051\,^{+0.0049}_{-0.0036}$    &    $\,\,\,\,\,0.90\,^{+0.30}_{-0.21}$    \\  [\linejump ex]
 \noalign{\smallskip} \noalign{\smallskip}
 Primary linear LDC\,*    &    $u_{\rm{pri~K2}}$    &         &    $\,\,\,\,\,0.54 \pm 0.20$    &    $\,\,\,\,\,0.51 \pm 0.13$    &    $\,\,\,\,\,0.66 \pm 0.16$    &    $\,\,\,\,\,0.48 \pm 0.14$    \\  % [\linejump ex]
 Primary non-linear LDC\,*    &    $u'_{\rm{pri~K2}}$    &         &    $\,\,\,\,\,0.24 \pm 0.29$    &    $\,\,\,\,\,0.31 \pm 0.22$    &    $\,\,\,\,\,0.03 \pm 0.23$    &    $\,\,\,\,\,0.22 \pm 0.24$    \\ % [\linejump ex] 
 Secondary linear LDC\,*    &    $u_{\rm{sec~K2}}$    &         &    $\,\,\,\,\,0.39 \pm 0.11$    &    $\,\,\,\,\,0.41 \pm 0.13$    &    $\,\,\,\,\,0.43 \pm 0.13$    &    $\,\,\,\,\,0.44 \pm 0.14$    \\  % [\linejump ex]  
 Secondary non-linear LDC\,*  &    $u'_{\rm{sec~K2}}$  &       &    $\,\,\,\,\,0.12 \pm 0.21$    &    $\,\,\,\,\,0.04\,^{+0.25}_{-0.18}$    &    $\,\,\,\,\,0.12 \pm 0.26$    &    $\,\,\,\,\,0.12 \pm 0.23$    \\  % [\linejump ex]
 \noalign{\smallskip} \noalign{\smallskip} \noalign{\smallskip} \noalign{\smallskip}  \noalign{\smallskip}  
 \multicolumn{7}{c}{\emph{Out-of-eclipse variability parameters}} \\  
 \noalign{\smallskip} \noalign{\smallskip}  
 Amplitude                        &    $A_{\rm{K2}}$               &    \%      &   $\,\,\,\,\,0.00785\,^{+0.00077}_{-0.00057}$    &   $\,\,\,\,\,0.0179 \pm 0.0023$             &   $\,\,\,\,\,0.0103\,^{+0.0033}_{-0.0020}$    &   $\,\,\,\,\,0.136\,^{+0.085}_{-0.136}$    \\ [\linejump ex]
 Timescale of SqExp term          &    $l_{\rm{SE ~ K2}}$          &    days    &   $\,\,\,\,\,8.55\,^{+0.43}_{-0.36}$             &   $\,\,\,\,\,16.97 \pm 0.97$                &   $\,\,\,\,\,7.32\,^{+0.90}_{-0.67}$          &  ---  \\  [\linejump ex]
 Scale factor of ExpSine2 term    &    $\Gamma_{\rm{ESS ~ K2}}$    &    days    &   $\,\,\,\,\,11.5 \pm 4.0$                       &   $\,\,\,\,\,9.56\,^{+0.96}_{-0.82}$        &   $\,\,\,\,\,0.55^{+0.27}_{-0.20}$            &  ---  \\  [\linejump ex]
  Period of ExpSine2 term         &    $P_{\rm{ESS ~ K2}}$         &    days    &   $\,\,\,\,\,7.375\,^{+0.059}_{-0.069}$          &   $\,\,\,\,\,12.150\,^{+0.074}_{-0.062}$    &   $\,\,\,\,\,2.252 \pm 0.020$                 &  ---  \\  [\linejump ex]
 Timescale of Matern32 term       &    $l_{\rm{M32 ~ K2}}$         &    days        &  ---  &  ---  &  ---  &   $\,\,\,\,\,223.6 \pm 2.3$ \\  [\linejump ex]
 White noise scale factor                 &    $\sigma_{\rm{K2}}$          &            &   $\,\,\,\,\,1.901 \pm 0.039$                    &   $\,\,\,\,\,1.448 \pm 0.02$                &   $\,\,\,\,\,1.363 \pm 0.019$                 &   $\,\,\,\,\,1.329\,^{+0.031}_{-0.023}$    \\  [\linejump ex]
 \noalign{\smallskip} \noalign{\smallskip} \noalign{\smallskip} \noalign{\smallskip} \noalign{\smallskip}  
\multicolumn{7}{c}{\emph{Radial velocity parameters}} \\  
 \noalign{\smallskip} \noalign{\smallskip}  
 Systemic velocity    &    $V_{\rm{sys}}$    &    km\,s$^{-1}$        &    $\,\,\,\,\,33.60 \pm 0.24$    &    $\,\,\,\,\,34.91 \pm 0.39$    &    $\,\,\,\,\,34.93\,^{+0.61}_{-0.53}$    &    $\,\,\,\,\,33.1 \pm 1.7$    \\  [\linejump ex]
 Primary RV semi-amplitude    &    $K_{\rm{pri}}$    &    km\,s$^{-1}$        &    $\,\,\,\,\,33.90 \pm 0.39$    &    $\,\,\,\,\,39.86\,^{+0.80}_{-0.88}$    &    $\,\,\,\,\,18.66\,^{+0.95}_{-1.00}$    &    $\,\,\,\,\,98 \pm 15$    \\  [\linejump ex]
 Secondary RV semi-amplitude    &    $K_{\rm{sec}}$    &    km\,s$^{-1}$        &    $\,\,\,\,\,63.93 \pm 0.49$    &    $\,\,\,\,\,33.12\,^{+0.83}_{-0.89}$    &  ---  &    $\,\,\,\,\,84 \pm 14$    \\  [\linejump ex]
 HIRES jitter term    &    $\sigma_{\rm{HIRES}}$    &    km\,s$^{-1}$        &    $\,\,\,\,\,0.50\,^{+0.37}_{-0.30}$    &    $\,\,\,\,\,0.95\,^{+0.48}_{-0.35}$    &    $\,\,\,\,\,0.93\,^{+1.18}_{-0.62}$    &    $\,\,\,\,\,1.6\,^{+2.5}_{-1.2}$    \\  [\linejump ex]
 \noalign{\smallskip} \noalign{\smallskip} \noalign{\smallskip} \noalign{\smallskip} \noalign{\smallskip} 
\multicolumn{7}{c}{\emph{Fundamental parameters}} \\  
 \noalign{\smallskip} \noalign{\smallskip}  
 Primary mass      &    $M_{\rm pri}$    &    M$_{\odot}$        &    $\,\,\,\,\,0.3813 \pm 0.0074$    &    $\,\,\,\,\,0.212 \pm 0.012$    &   $\,\,\,\,\,0.276\pm0.020$\,$^{a}$    &    $\,\,\,\,\,0.45^{+0.19}_{-0.14}$    \\  [\linejump ex]
 Secondary mass    &    $M_{\rm sec}$    &    M$_{\odot}$        &    $\,\,\,\,\,0.2022 \pm 0.0045$    &    $\,\,\,\,\,0.255 \pm 0.013$    &   $\,\,\,\,\,0.0517\pm0.0041 ~~ (54.2\pm4.3)\,^{b,c}$     &    $\,\,\,\,\,0.53^{+0.22}_{-0.16}$    \\  [\linejump ex]
 Primary radius    &    $R_{\rm pri}$    &    R$_{\odot}$        &    $\,\,\,\,\,0.3610 \pm 0.0033$    &    $\,\,\,\,\,0.233 \pm 0.013$    &   $0.29 \pm 0.08$\,$^{d}$                                        &    $\,\,\,\,\,0.549\,^{+0.099}_{-0.082}$    \\ [\linejump ex]
 Secondary radius  &    $R_{\rm sec}$    &    R$_{\odot}$        &    $\,\,\,\,\,0.2256\,^{+0.0063}_{-0.0049}$    &    $\,\,\,\,\,0.267 \pm 0.014$    &    $0.10\pm0.03 ~~ (1.02\pm0.28)\,^{c,e}$                               &    $\,\,\,\,\,0.454\,^{+0.094}_{-0.101}$    \\ [\linejump ex]
 Primary effective temperature    &    $T_{\rm pri}$    &    K       &    $\,\,\,\,\,3211\,^{+54}_{-36}$    &    $\,\,\,\,\, 3152\,^{+57}_{-40}$    &    $3191 \pm 27$    &   $3767\,^{+99}_{-85}$   \\  [\linejump ex]
 Secondary effective temperature    &    $T_{\rm sec}$    &    K       &    $\,\,\,\,\,3103\,^{+53}_{-39}$    &    $\,\,\,\,\,3131\,^{+56}_{-38}$    &    $1669\,^{+244}_{-258}$   &  $3693\,^{+122}_{-135}$   \\ [\linejump ex]
 Mass sum     &    $M_{\rm pri} + M_{\rm sec}$    &    M$_{\odot}$        &    $\,\,\,\,\,0.583 \pm 0.011$    &    $\,\,\,\,\,0.468 \pm 0.023$    &  ---  &    $\,\,\,\,\,0.98\,^{+0.38}_{-0.29}$    \\  [\linejump ex]
 Radius sum    &    $R_{\rm pri} + R_{\rm sec}$    &    R$_{\odot}$        &    $\,\,\,\,\,0.5868\,^{+0.0084}_{-0.0073}$    &    $\,\,\,\,\,0.4991\,^{+0.0096}_{-0.0102}$    &  ---  &    $\,\,\,\,\,1.00 \pm 0.13$    \\  [\linejump ex]
 Semi-major axis    &    $a$    &    R$_{\odot}$        &    $\,\,\,\,\,11.630 \pm 0.073$    &    $\,\,\,\,\,16.75 \pm 0.28$    &  ---  &    $\,\,\,\,\,5.67 \pm 0.65$    \\  [\linejump ex]
 Eccentricity    &    $e$    &         &    $\,\,\,\,\,0.00194\,^{+0.00253}_{-0.00057}$    &    $\,\,\,\,\,0.00254\,^{+0.00406}_{-0.00078}$    &    $\,\,\,\,\,0.146\,^{+0.024}_{-0.016}$    &    $\,\,\,\,\,0.00108\,^{+0.00347}_{-0.00078}$    \\  [\linejump ex]
 Longitude of periastron    &    $\omega$    &    $^{\circ}$         &    $\,\,\,\,\,116.0 \pm 39.0$    &    $\,\,\,\,\,27.0 \pm 69.0$    &    $\,\,\,\,\,5 \pm 20$    &    $\,\,\,\,\,91.0 \pm 29.0$    \\  [\linejump ex]
 Primary surface gravity    &    $\log g_{\rm pri}$    &    (cm\,s$^{-2}$)        &    $\,\,\,\,\,4.9040\,^{+0.0073}_{-0.0064}$    &    $\,\,\,\,\,5.031 \pm 0.048$    &  ---  &    $\,\,\,\,\,4.61 \pm 0.13$    \\  [\linejump ex]
 Secondary surface gravity    &    $\log g_{\rm sec}$    &    (cm\,s$^{-2}$)        &    $\,\,\,\,\,5.037\,^{+0.019}_{-0.026}$    &    $\,\,\,\,\,4.993\,^{+0.042}_{-0.035}$    &  ---  &    $\,\,\,\,\,4.84 \pm 0.20$    \\  [\linejump ex]
 Primary synchronized velocity      &    $V_{\rm pri ~ sync}$    &    km\,s$^{-1}$       &    $\,\,\,\,\,3.036 \pm 0.028$    &    $\,\,\,\,\,1.014 \pm 0.058$    &  ---  &    $\,\,\,\,\,17.8 \pm 3.2$    \\  [\linejump ex]
 Secondary synchronized velocity    &    $V_{\rm sec ~ sync}$    &    km\,s$^{-1}$       &    $\,\,\,\,\,1.898\,^{+0.053}_{-0.041}$    &    $\,\,\,\,\,1.162\,^{+0.050}_{-0.059}$    &  ---  &    $\,\,\,\,\,14.7 \pm 3.3$    \\  [\linejump ex]
 Synchronization timescale      &    $t_{\rm sync}$    &    Myr       &    $\,\,\,\,\,27.29 \pm 0.49$    &    $\,\,\,\,\,152.6 \pm 4.7$    &  ---  &    $\,\,\,\,\,0.0510\,^{+0.0120}_{-0.0090}$    \\  [\linejump ex]
 Circularization timescale    &    $t_{\rm circ}$    &    Gyr       &    $\,\,\,\,\,17.30 \pm 0.10$    &    $\,\,\,\,\,467.6 \pm 1.5$    &  ---  &    $\,\,\,\,\,0.01040\,^{+0.00033}_{-0.00013}$   \\  [\linejump ex]
 \noalign{\smallskip} \noalign{\smallskip} \noalign{\smallskip}  
 \hline  
 \end{tabular}%
 } 
  \begin{list}{}{}  
 \item[* LDC = limb darkening coefficient]  
 \item[$^{a}$ Derived from the empirical relations of \citet{Benedict16}.] 
 \item[$^{b}$ Derived from the system mass function.]
 \item[$^{c}$ Units in brackets are relative to Jupiter.] 
 \item[$^{d}$ Derived from the empirical relations of \citet{Mann16}.] 
 \item[$^{e}$ Derived from the light curve radius ratio and the empirically determined primary radius.] 
 \end{list}   
 \end{table*}

%%%%%%%%%%%%%%%%%%%%%%%%%%%%%%%%%%%%%%%%%%%%%%%%%%%%%%%%%%%%%%%%%%%%%%
%%%%%%%%%%%%%%%%%%%%%%%%%%%      3814     %%%%%%%%%%%%%%%%%%%%%%%%%%%%
%%%%%%%%%%%%%%%%%%%%%%%%%%%%%%%%%%%%%%%%%%%%%%%%%%%%%%%%%%%%%%%%%%%%%%

%\pagebreak 

\begin{figure*}
\centering
   \includegraphics[width=0.84\linewidth]{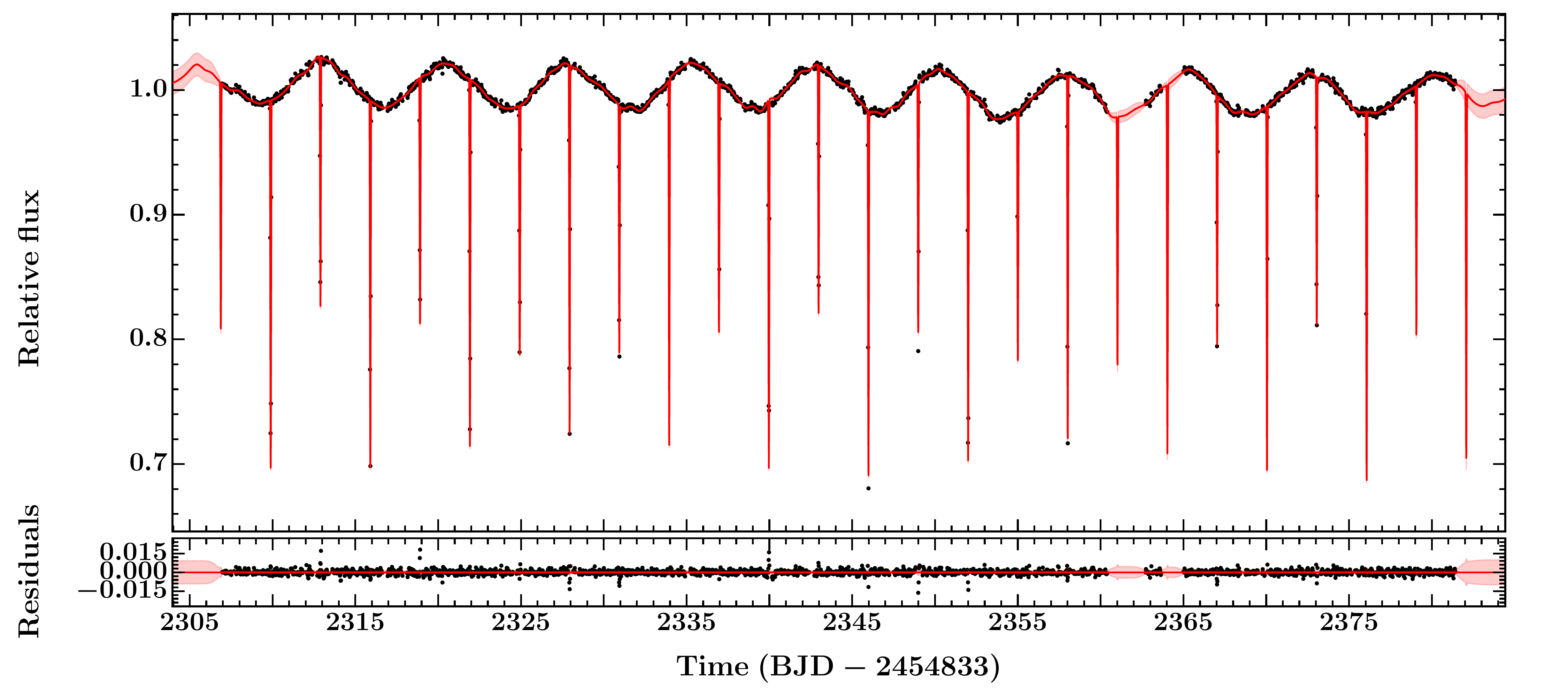} 
   \caption{Systematics-corrected \ktwo\ light curve of AD 3814 (black points) with the \gpe\ model in red. The red line and pink shaded region represent the mean and 2$\sigma$ uncertainty of the model's predictive posterior distribution. }
   \label{3814_LC}
\end{figure*}  

\begin{figure}
\centering
   \vspace{-0.0cm}
   \includegraphics[width=0.8\linewidth]{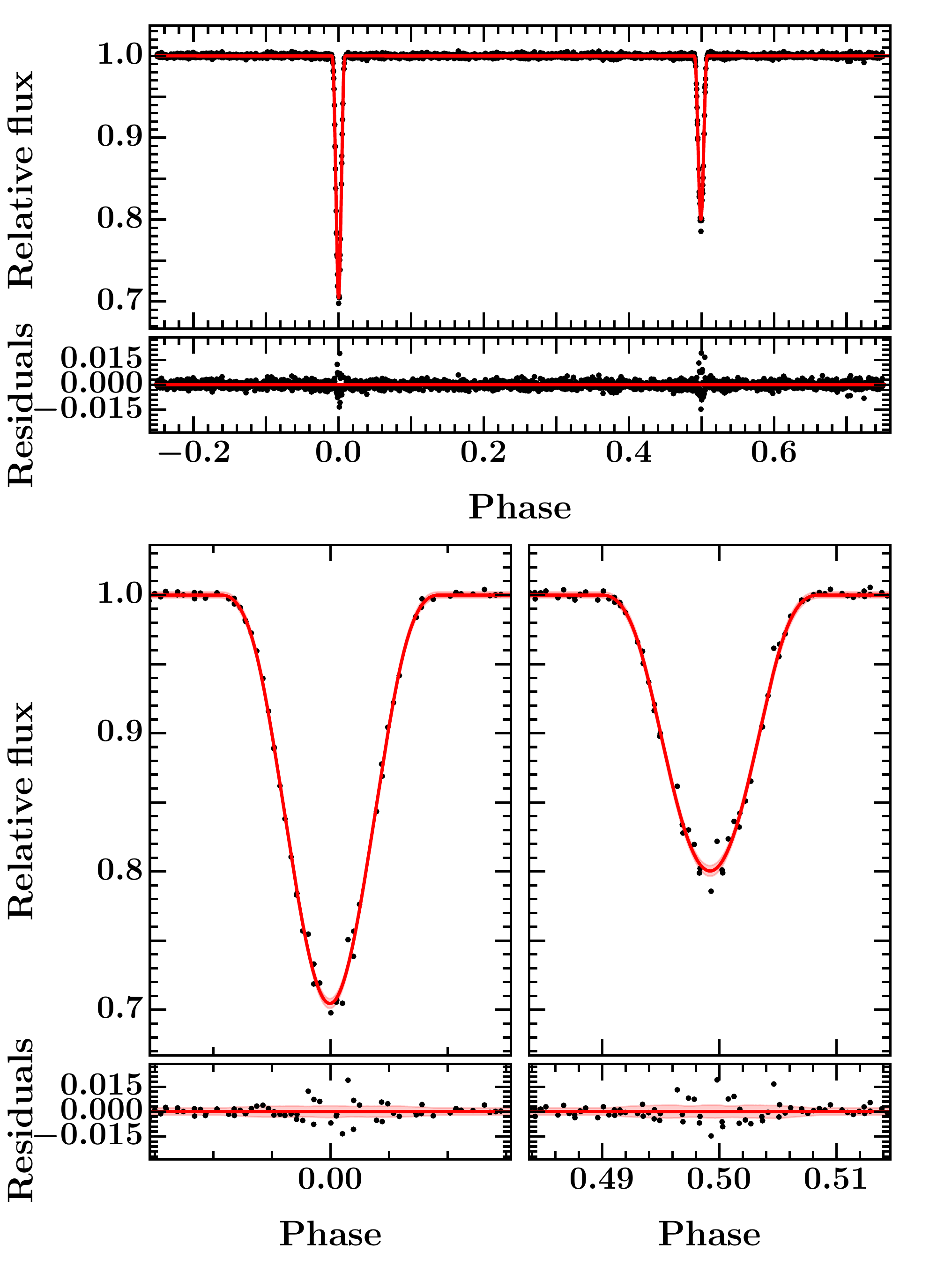}
   \vspace{-0.1cm}
   \caption{\eTop\ panels: phase-folded \ktwo\ light curve of AD 3814 (black), which has been detrended with respect to the Gaussian process model. The red line indicates the median EB model derived from the posterior distribution, i.e. individual draws are calculated across phase space and the median of their paths plotted. Phase zero marks the center of the primary eclipse. Immediately below are the residuals of the fit. \eBot\ panels: zooms on primary and secondary eclipses (\eleft\ and \eright\ respectively) with the median model and 2$\sigma$ uncertainties shown (red line and pink shaded region, respectively). Residuals are shown immediately below. }
   \label{3814_eclipses}
\end{figure}  

\begin{figure}
\centering
   \vspace{-0.0cm}
   \includegraphics[width=0.8\linewidth]{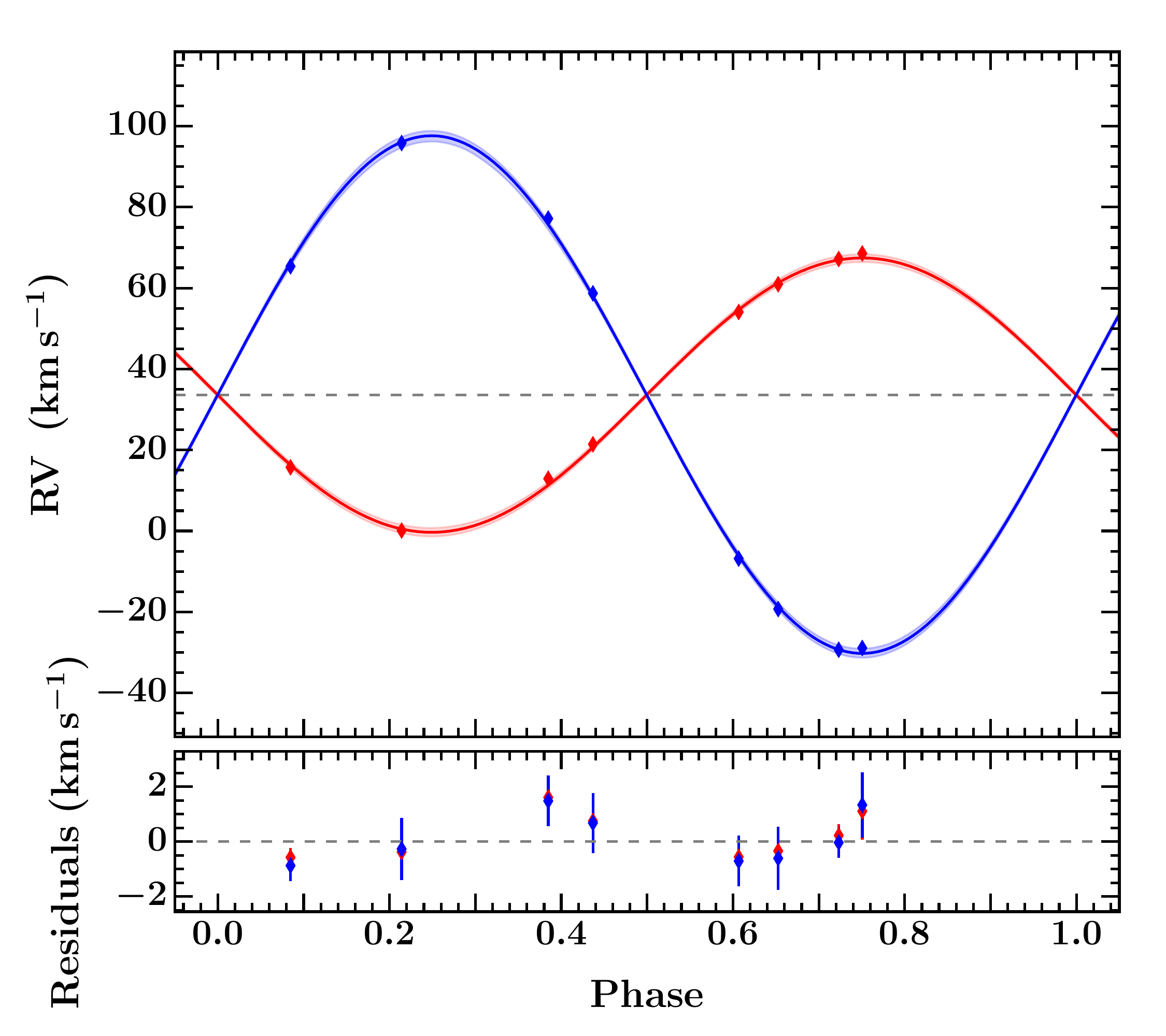}
    \vspace{-0.2cm}
   \caption{\eTop: phase-folded RV orbit of AD 3814 with Keck/HIRES RVs for the primary and secondary stars (red and blue, respectively). The lines and shaded regions indicate the median and 2$\sigma$ uncertainties on the posterior distributions of the RV orbits. The gray horizontal dotted line shows the systemic velocity. \eBot: Residuals of the fit. }
   \label{3814_RV}
\end{figure}
  
\begin{figure}
\centering
   \vspace{-0.05cm}
   \includegraphics[width=0.7\linewidth]{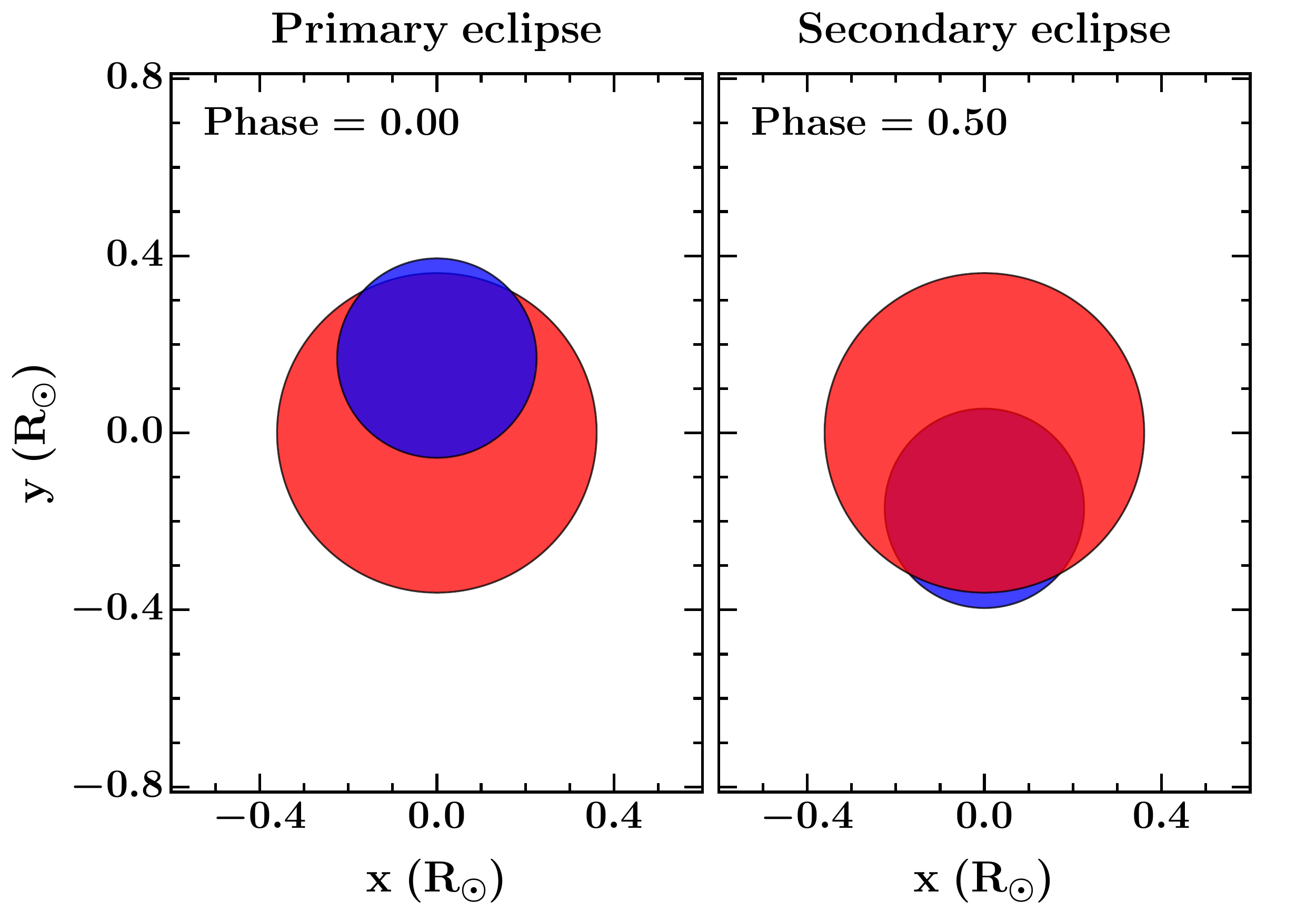}
   \vspace{-0.1cm}
   \caption{Geometry of AD 3814, to scale, as observed at primary and secondary eclipse. The primary star is shown in red and the secondary in blue. }
   \label{3814_geom}
\end{figure}

%%%%%%%%%%%%%%%%%%%%%%%%%%%%%%%%%%%%%%%%%%%%%%%%%%%%%%%%%%%%%%%%%%%%%%
%%%%%%%%%%%%%%%%%%%%%%%%%%%      2615     %%%%%%%%%%%%%%%%%%%%%%%%%%%%
%%%%%%%%%%%%%%%%%%%%%%%%%%%%%%%%%%%%%%%%%%%%%%%%%%%%%%%%%%%%%%%%%%%%%%

%\pagebreak 

\begin{figure*}
\centering
   \includegraphics[width=0.84\linewidth]{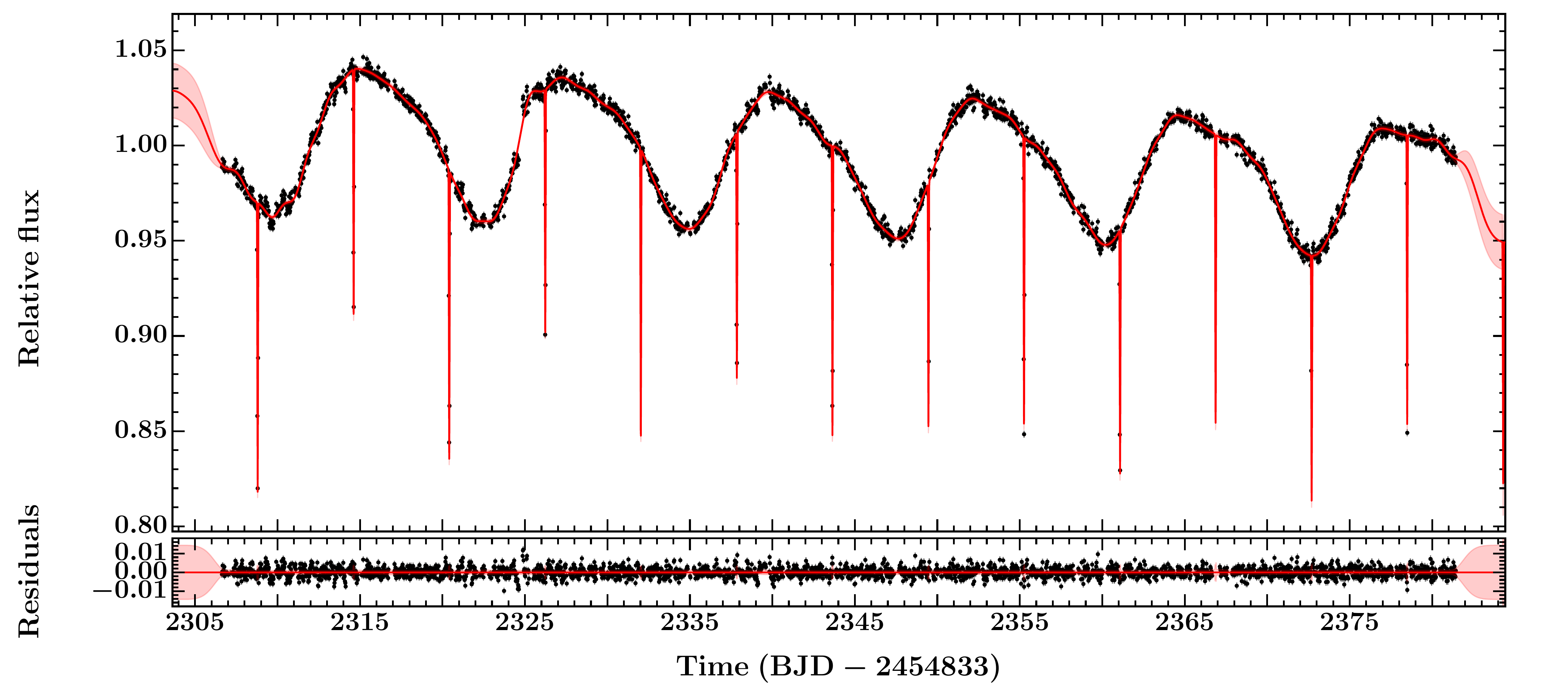}
   \caption{Systematics-corrected \ktwo\ light curve of AD 2615 (black points) with the \gpe\ model in red. The red line and pink shaded region represent the mean and 2$\sigma$ uncertainty of the model's predictive posterior distribution. }
   \label{2615_LC}
\end{figure*}  

\begin{figure}
\centering
   \vspace{0.0cm}
   \includegraphics[width=0.8\linewidth]{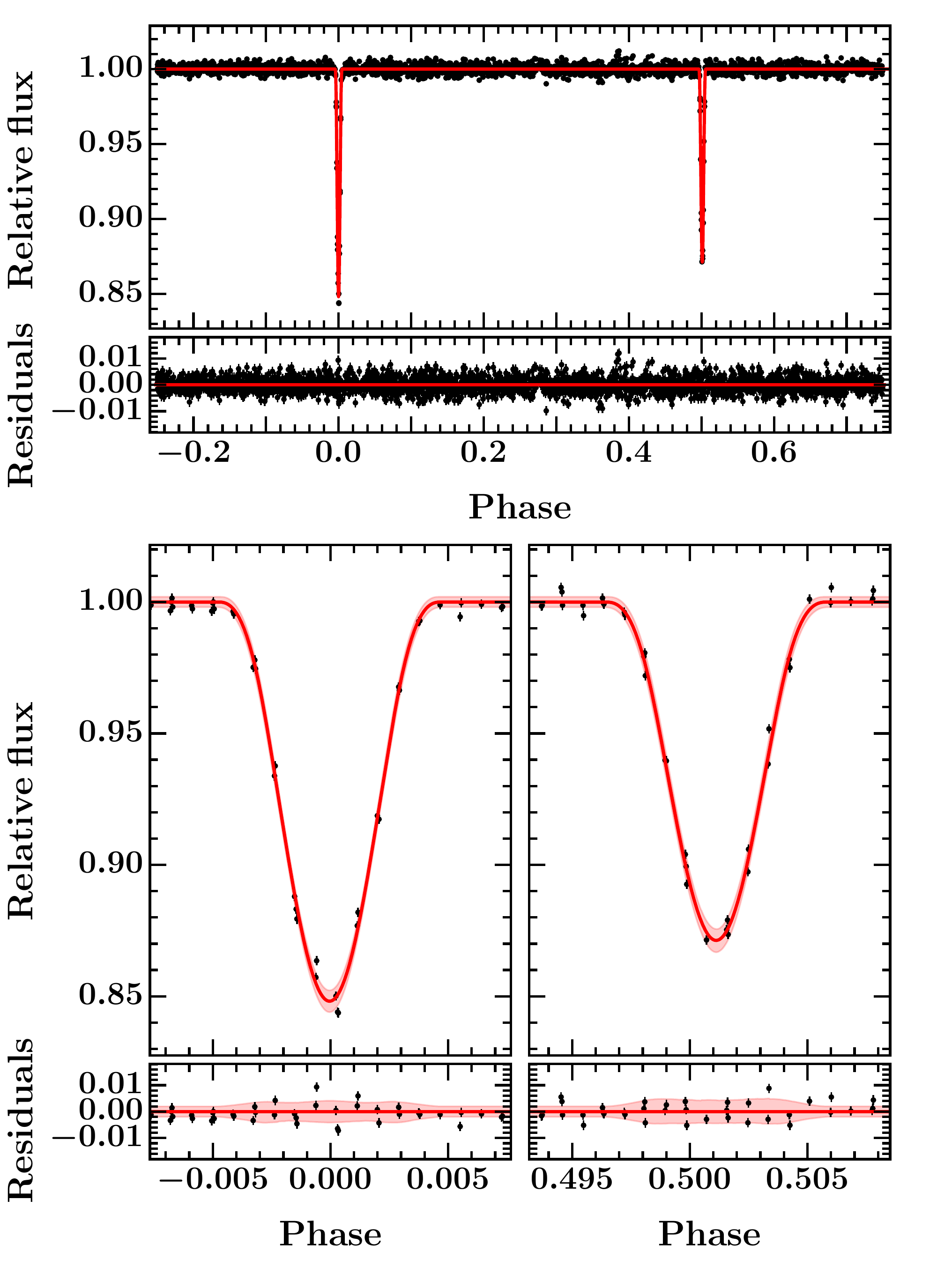}
   \vspace{-0.1cm}
   \caption{\eTop\ panels: phase-folded \ktwo\ light curve of AD 2615 (black), which has been detrended with respect to the Gaussian process model. The red line indicates the median EB model derived from the posterior distribution, i.e. individual draws are calculated across phase space and the median of their paths plotted. Phase zero marks the center of the primary eclipse. Immediately below are the residuals of the fit. \eBot\ panels: zooms on primary and secondary eclipses (\eleft\ and \eright\ respectively) with the median model and 2$\sigma$ uncertainties shown (red line and pink shaded region, respectively). Residuals are shown immediately below. }
   \label{2615_eclipses}
\end{figure}  

\begin{figure}
\centering
   \vspace{0.0cm}
   \includegraphics[width=0.8\linewidth]{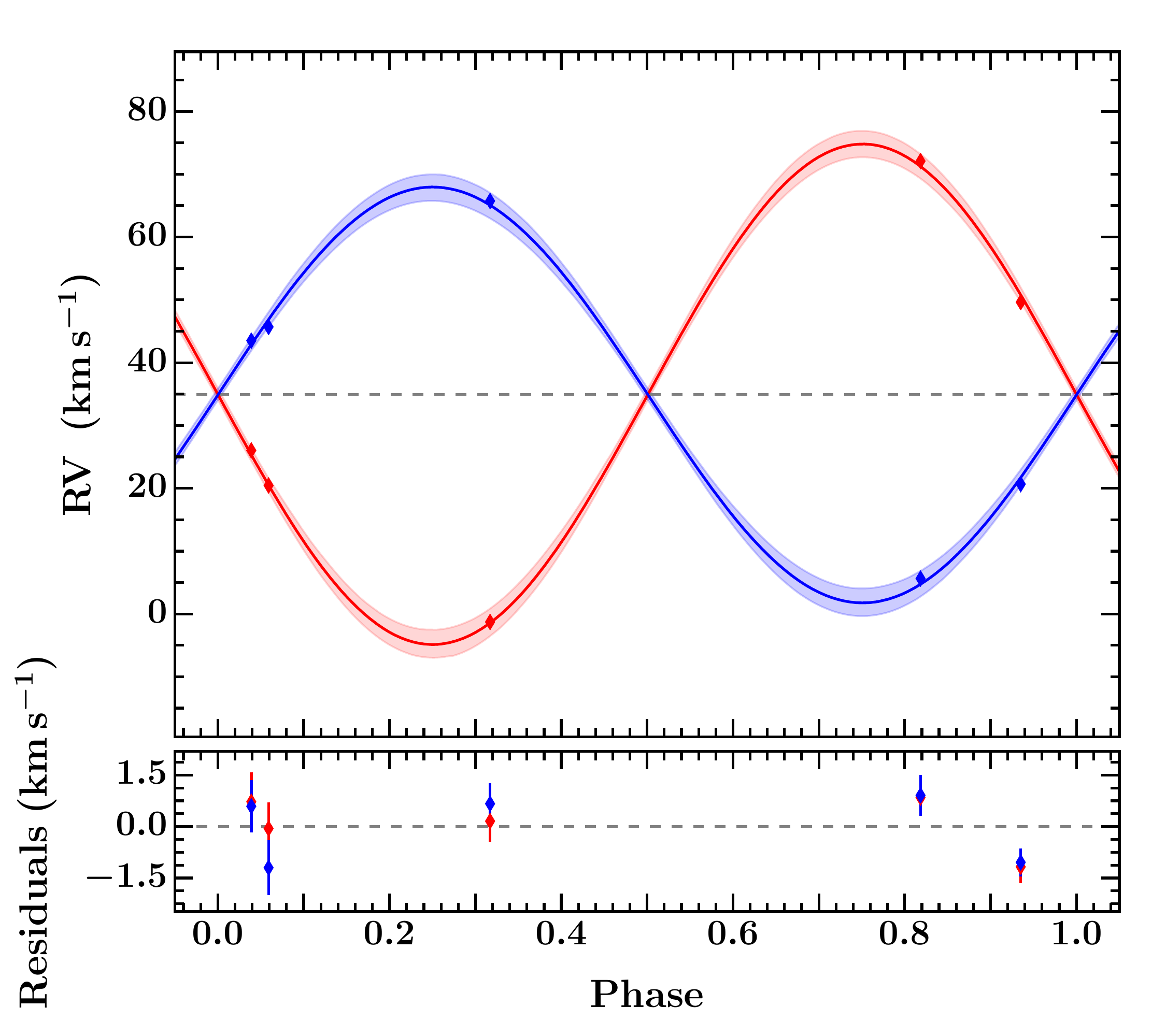}
    \vspace{-0.2cm}
   \caption{\eTop: phase-folded RV orbit of AD 2615 with Keck/HIRES RVs for the primary and secondary stars (red and blue, respectively). The lines and shaded regions indicate the median and 2$\sigma$ uncertainty on the posterior distribution of the RV orbits. The gray horizontal dotted line shows the systemic velocity. \eBot: Residuals of the fit. }
   \label{2615_RV}
\end{figure}
  
\begin{figure}
\centering
   \vspace{-0.05cm}
   \includegraphics[width=0.7\linewidth]{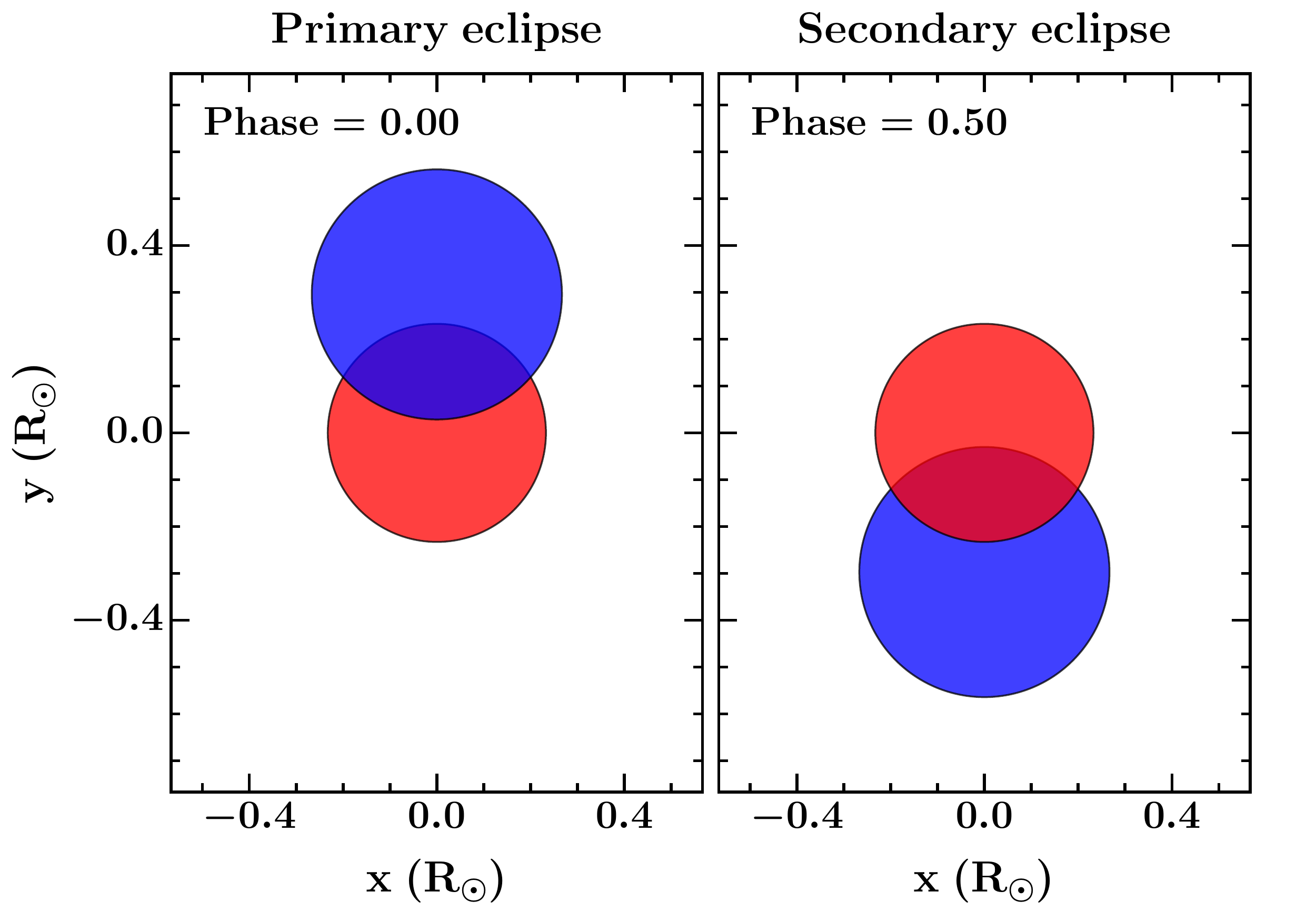}
   \vspace{-0.1cm}
   \caption{Geometry of AD 2615, to scale, as observed at primary and secondary eclipse. The primary star is shown in red and the secondary in blue. }
   \label{2615_geom}
\end{figure}

%%%%%%%%%%%%%%%%%%%%%%%%%%%%%%%%%%%%%%%%%%%%%%%%%%%%%%%%%%%%%%%%%%%%%%
%%%%%%%%%%%%%%%%%%%%%%%%%%%      3116     %%%%%%%%%%%%%%%%%%%%%%%%%%%%
%%%%%%%%%%%%%%%%%%%%%%%%%%%%%%%%%%%%%%%%%%%%%%%%%%%%%%%%%%%%%%%%%%%%%%

%\pagebreak 

%\begin{minipage}{\textwidth}

\begin{figure*}
\centering
   \includegraphics[width=0.84\linewidth]{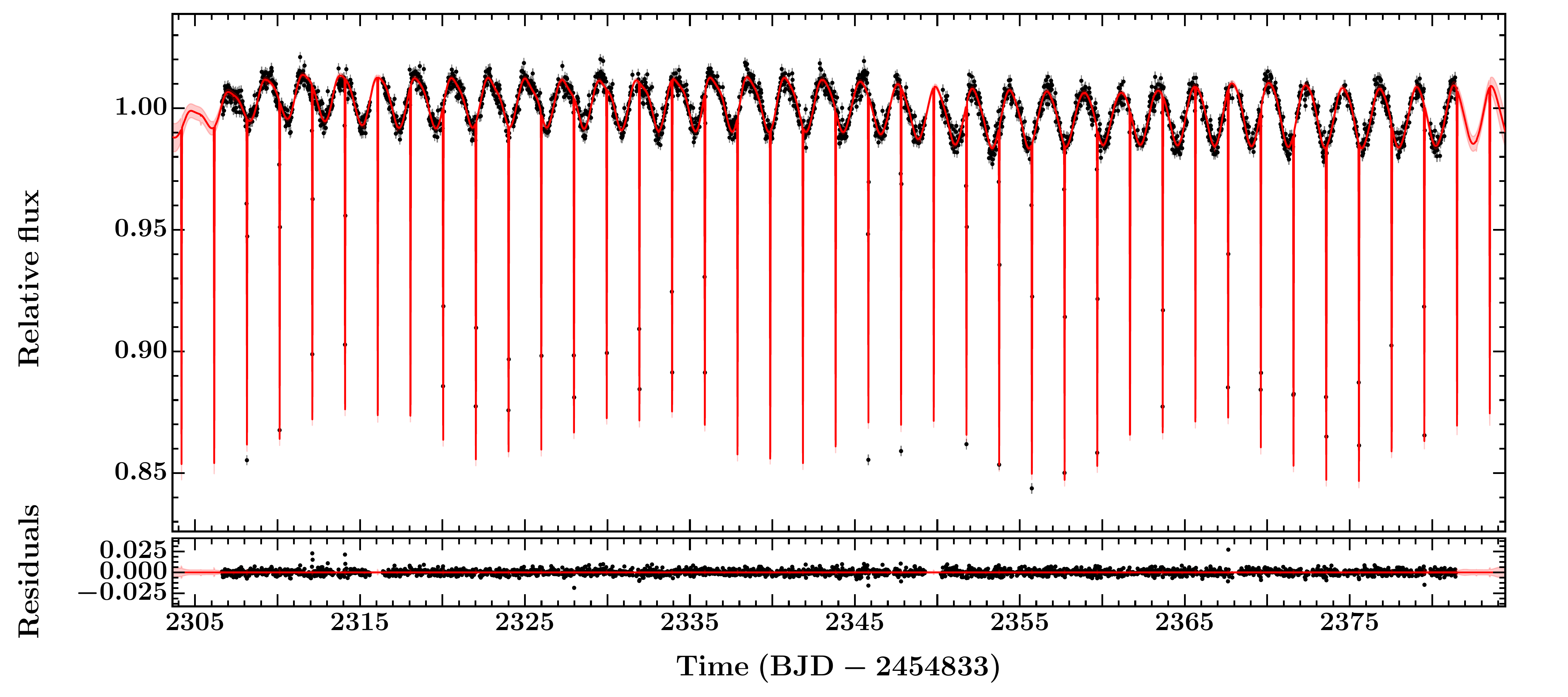}
   \caption{Systematics-corrected \ktwo\ light curve of AD 3116 (black points) with the \gpe\ model in red. The red line and pink shaded region represent the mean and 2$\sigma$ uncertainty of the model's predictive posterior distribution. }
   \label{3116_LC}
\end{figure*}  

\begin{figure}
\centering
   \vspace{0.0cm}
   \includegraphics[width=0.8\linewidth]{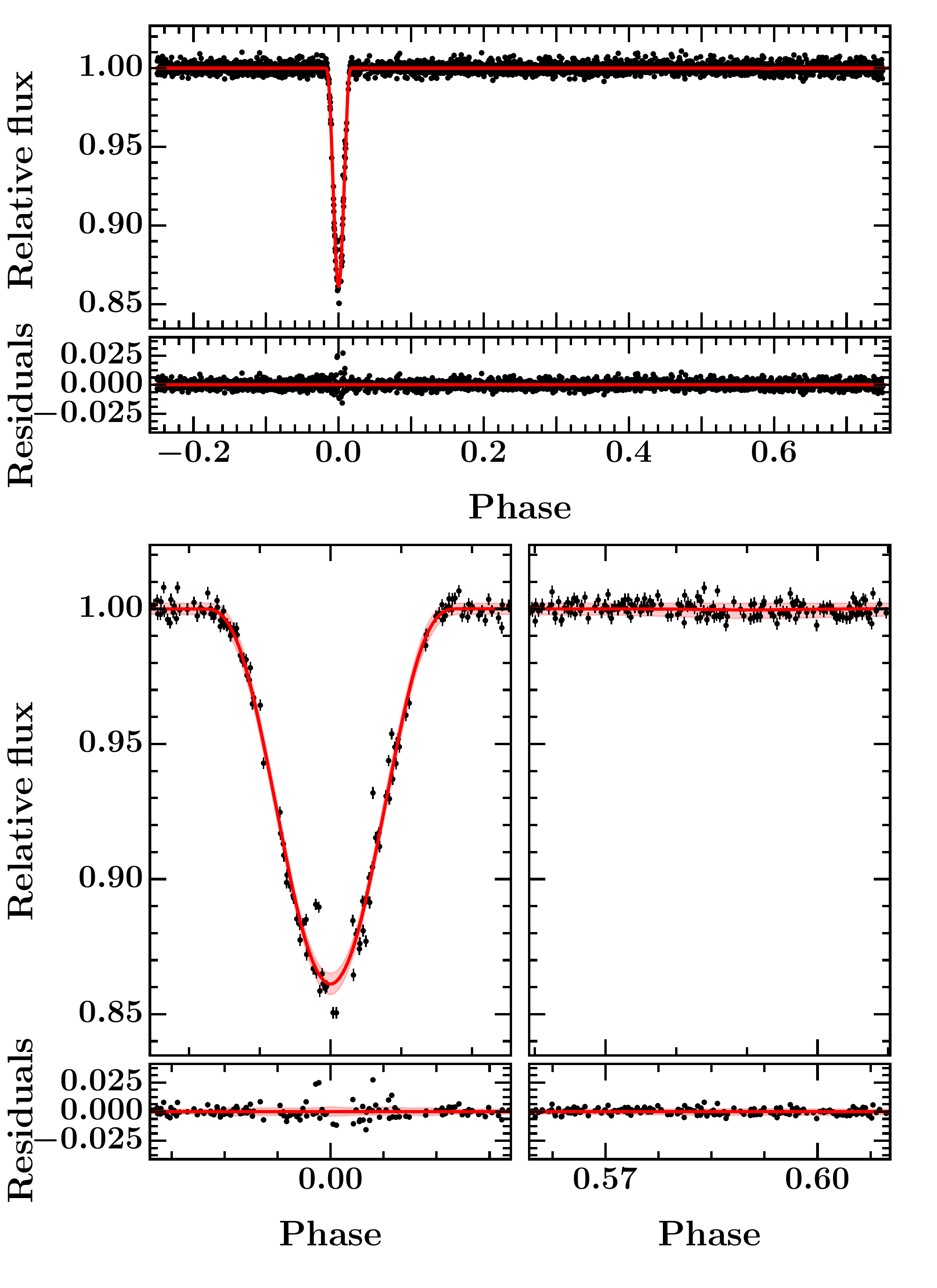}
   \vspace{-0.1cm}
   \caption{\eTop\ panels: phase-folded \ktwo\ light curve of AD 3116 (black), which has been detrended with respect to the Gaussian process model. The red line indicates the median EB model derived from the posterior distribution, i.e. individual draws are calculated across phase space and the median of their paths plotted. Phase zero marks the center of the primary eclipse. Immediately below are the residuals of the fit. \eBot\ panels: zooms on primary and secondary eclipses (\eleft\ and \eright\ respectively) with the median model and 2$\sigma$ uncertainties shown (red line and pink shaded region, respectively). Residuals are shown immediately below. }
   \label{3116_eclipses}
\end{figure}  

\begin{figure}
\centering
   \vspace{0.0cm}
   \includegraphics[width=0.8\linewidth]{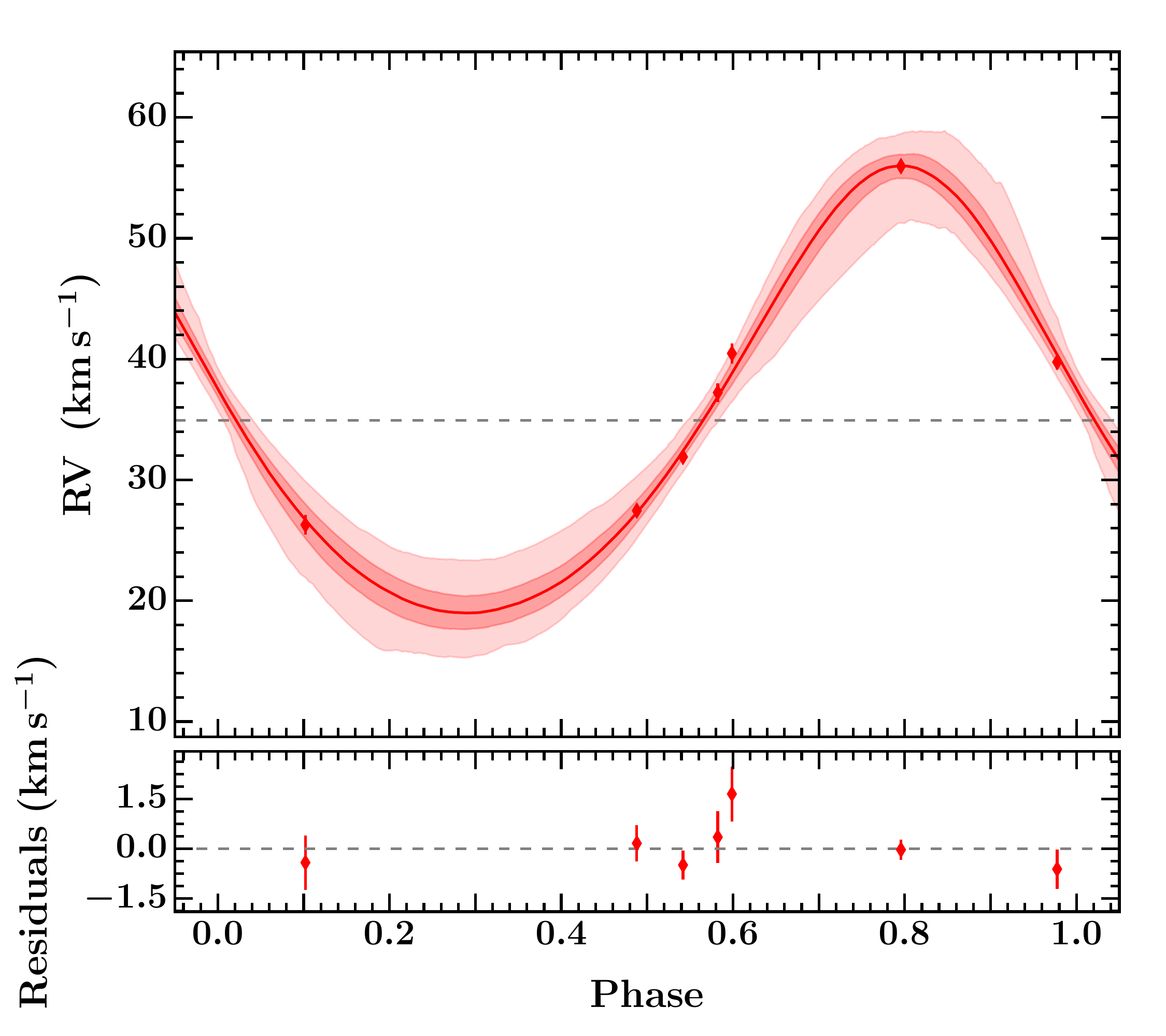}
    \vspace{-0.2cm}
   \caption{\eTop: phase-folded RV orbit of AD 3116 with Keck/HIRES RVs for the primary and secondary stars (red and blue, respectively). The line and shaded regions indicate the median and 1 and 2$\sigma$ uncertainties on the posterior distribution of the primary RV orbit. The gray horizontal dotted line shows the systemic velocity. \eBot: Residuals of the fit. }
   \label{3116_RV}
\end{figure}
  
\begin{figure}
\centering
   \vspace{-0.05cm}
   \includegraphics[width=0.7\linewidth]{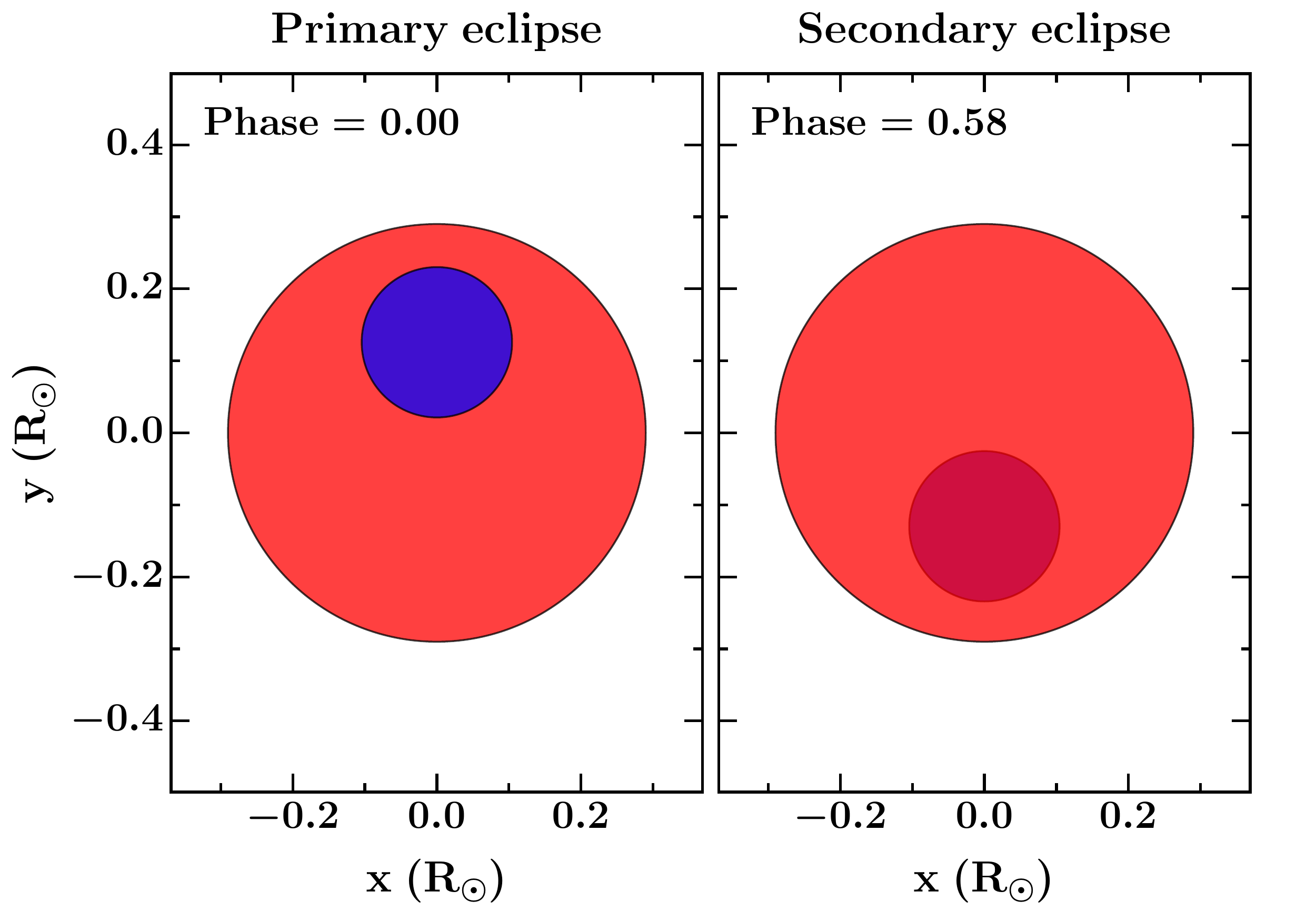}
   \vspace{-0.1cm}
   \caption{Geometry of AD 3116, to scale, as observed at primary and secondary eclipse. The primary star is shown in red and the secondary in blue. }
   \label{3116_geom}
\end{figure}

%\end{minipage}

%%%%%%%%%%%%%%%%%%%%%%%%%%%%%%%%%%%%%%%%%%%%%%%%%%%%%%%%%%%%%%%%%%%%%%
%%%%%%%%%%%%%%%%%%%%%%%%%%%      1508     %%%%%%%%%%%%%%%%%%%%%%%%%%%%
%%%%%%%%%%%%%%%%%%%%%%%%%%%%%%%%%%%%%%%%%%%%%%%%%%%%%%%%%%%%%%%%%%%%%%

%\pagebreak 

\begin{figure*}
\centering
   \includegraphics[width=0.84\linewidth]{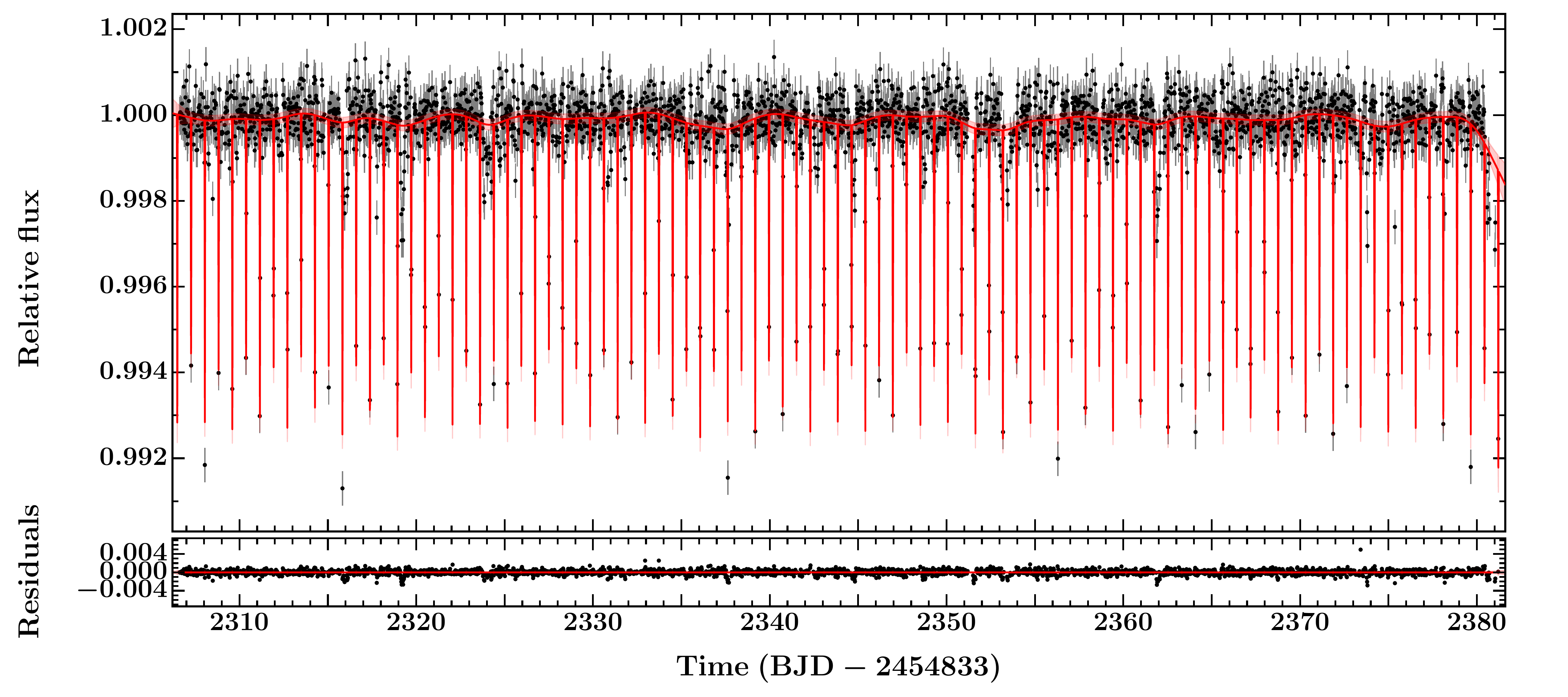}
   \caption{Systematics-corrected \ktwo\ light curve of AD 1508 (black points) with the \gpe\ model in red. The red line and pink shaded region represent the mean and 2$\sigma$ uncertainty of the model's predictive posterior distribution. }
   \label{1508_LC}
\end{figure*}  

\begin{figure}
\centering
   \vspace{0.0cm}
   \includegraphics[width=0.8\linewidth]{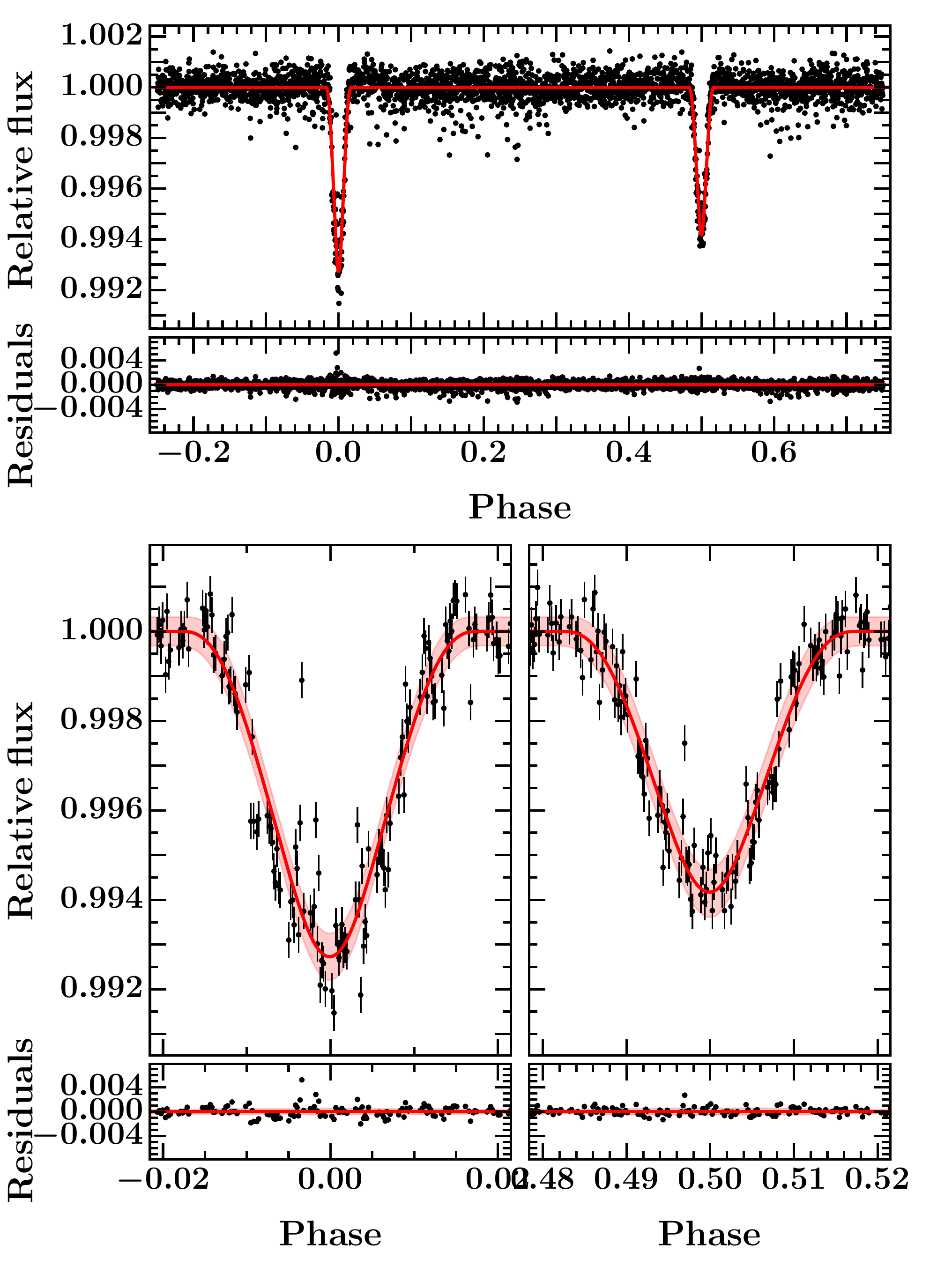}
   \vspace{-0.1cm}
   \caption{\eTop\ panels: phase-folded \ktwo\ light curve of AD 1508 (black), which has been detrended with respect to the Gaussian process model. The red line indicates the median EB model derived from the posterior distribution, i.e. individual draws are calculated across phase space and the median of their paths plotted. Phase zero marks the center of the primary eclipse. Immediately below are the residuals of the fit. \eBot\ panels: zooms on primary and secondary eclipses (\eleft\ and \eright\ respectively) with the median model and 2$\sigma$ uncertainties shown (red line and pink shaded region, respectively). Residuals are shown immediately below. }
   \label{1508_eclipses}
\end{figure}  

\begin{figure}
\centering
   \vspace{0.0cm}
   \includegraphics[width=0.8\linewidth]{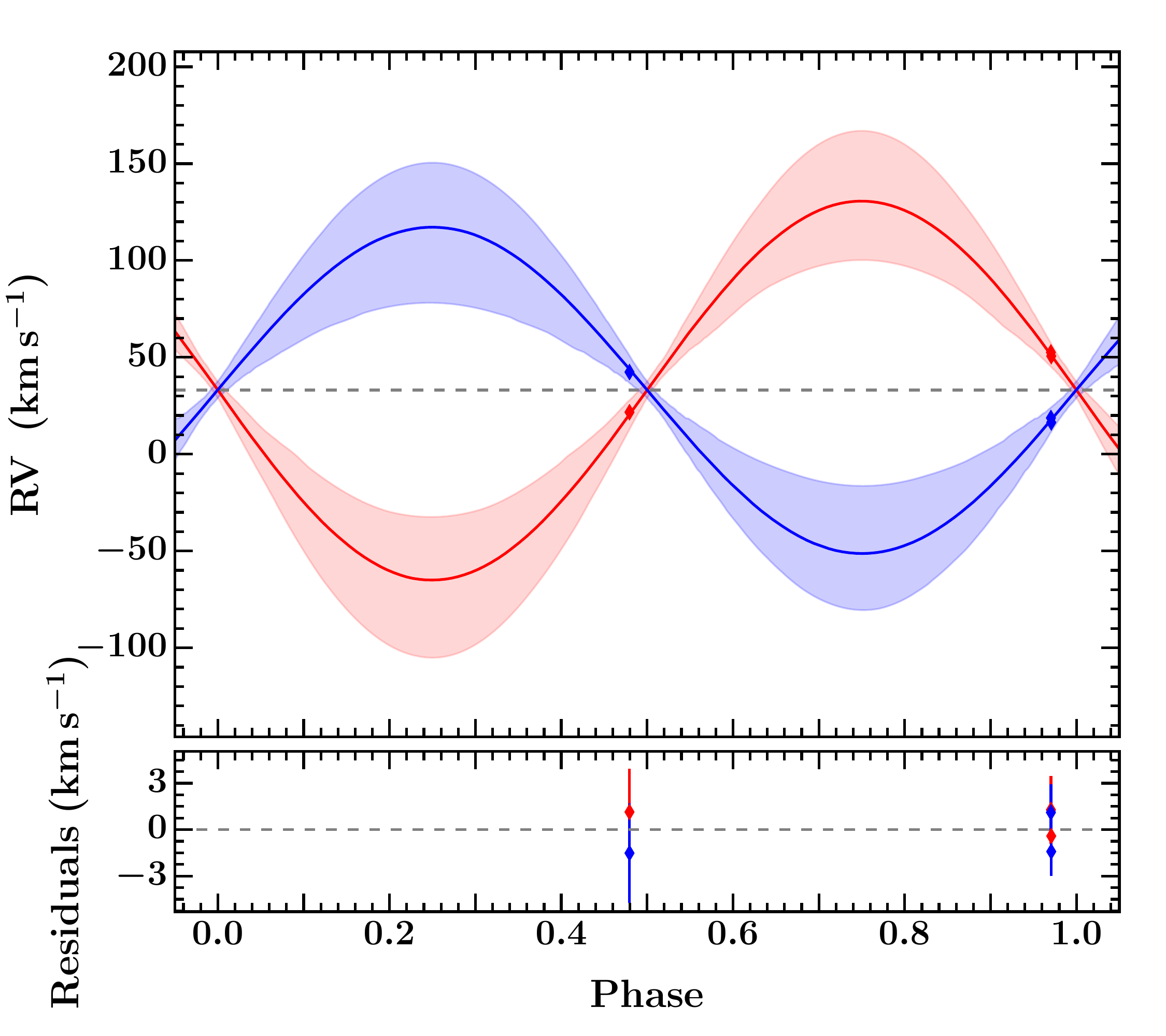}
    \vspace{-0.2cm}
   \caption{\eTop: phase-folded RV orbit of AD 1508 with Keck/HIRES RVs for the primary and secondary stars (red and blue, respectively). The lines and shaded regions indicate the median and 2$\sigma$ uncertainty on the posterior distribution of the RV orbits. The gray horizontal dotted line shows the systemic velocity. \eBot: Residuals of the fit. }
   \label{1508_RV}
\end{figure}
  
\begin{figure}
\centering
   \vspace{-0.05cm}
   \includegraphics[width=0.7\linewidth]{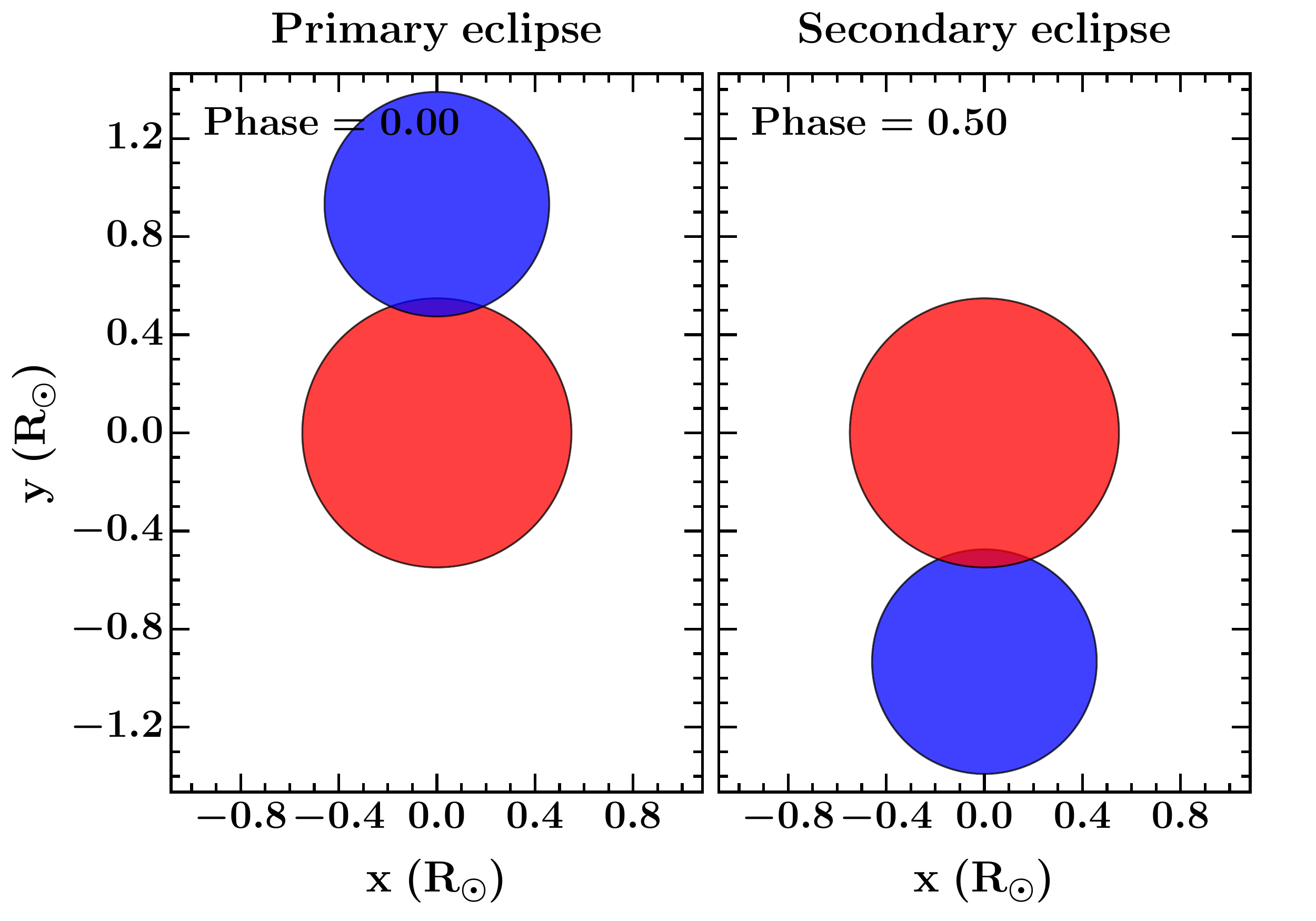}
   \vspace{-0.1cm}
   \caption{Geometry of AD 1508, to scale, as observed at primary and secondary eclipse. The primary star is shown in red and the secondary in blue. }
   \label{1508_geom}
\end{figure}

%%%%%%%%%%%%%%%%%%%%%%%%%%%%%%%%%%%%%%

\acknowledgments

We thank Pierre Maxted for interesting discussions and John Southworth for help compiling Table \ref{pms_ebs}.
This paper includes data collected by the Kepler/K2 mission. Funding for the K2 mission of Kepler is provided by the NASA Science Mission directorate.
Some of the data presented in this paper were obtained from the Mikulski Archive for Space Telescopes (MAST) under support by the NASA Office of Space Science via grant NNX09AF08G and by other grants and contracts. 
Some of the data presented herein were obtained at the W.M. Keck Observatory, which is operated as a scientific partnership among the California Institute of Technology, the University of California and the National Aeronautics and Space Administration. The Observatory was made possible by the generous financial support of the W.M. Keck Foundation.
The authors wish to recognize and acknowledge the very significant cultural role and reverence that the summit of Mauna Kea has always had within the indigenous Hawaiian community.  We are most fortunate to have the opportunity to conduct observations from this mountain.
Finally, we would like to thank the anonymous referee for their careful reading of the manuscript and helpful suggestions for improvement.


\vspace{5mm}
\facilities{\kepler/\ktwo, Keck/HIRES, SDSS, 2MASS, WISE.}


\software{astropy \citep{astropy13}, {\tt emcee} \citep{Foreman-Mackey13}, {\tt george} \citep{Ambikasaran14}.
          }

\bibliographystyle{aasjournal}
\bibliography{ref}

\begin{thebibliography}{}
\expandafter\ifx\csname natexlab\endcsname\relax\def\natexlab#1{#1}\fi

\bibitem[{{Adams} {et~al.}(2002){Adams}, {Stauffer}, {Skrutskie}, {Monet},
  {Portegies Zwart}, {Janes}, \& {Beichman}}]{Adams02}
{Adams}, J.~D., {Stauffer}, J.~R., {Skrutskie}, M.~F., {et~al.} 2002, \aj, 124,
  1570

\bibitem[{{Aigrain} {et~al.}(2007){Aigrain}, {Hodgkin}, {Irwin}, {Hebb},
  {Irwin}, {Favata}, {Moraux}, \& {Pont}}]{Aigrain07}
{Aigrain}, S., {Hodgkin}, S., {Irwin}, J., {et~al.} 2007, \mnras, 375, 29

\bibitem[{{Aigrain} {et~al.}(2016){Aigrain}, {Parviainen}, \&
  {Pope}}]{Aigrain16}
{Aigrain}, S., {Parviainen}, H., \& {Pope}, B.~J.~S. 2016, \mnras, 459, 2408

\bibitem[{{Aigrain} {et~al.}(2012){Aigrain}, {Pont}, \& {Zucker}}]{Aigrain12}
{Aigrain}, S., {Pont}, F., \& {Zucker}, S. 2012, \mnras, 419, 3147

\bibitem[{{Alencar} \& {Vaz}(1997)}]{Alencar97}
{Alencar}, S.~H.~P., \& {Vaz}, L.~P.~R. 1997, \aap, 326, 257

\bibitem[{{Allard} {et~al.}(2012){Allard}, {Homeier}, \& {Freytag}}]{Allard12}
{Allard}, F., {Homeier}, D., \& {Freytag}, B. 2012, Philosophical Transactions
  of the Royal Society of London Series A, 370, 2765

\bibitem[{{Alonso} {et~al.}(2015){Alonso}, {Deeg}, {Hoyer}, {Lodieu}, {Palle},
  \& {Sanchis-Ojeda}}]{Alonso15}
{Alonso}, R., {Deeg}, H.~J., {Hoyer}, S., {et~al.} 2015, \aap, 584, L8

\bibitem[{{Ambikasaran} {et~al.}(2014){Ambikasaran}, {Foreman-Mackey},
  {Greengard}, {Hogg}, \& {O'Neil}}]{Ambikasaran14}
{Ambikasaran}, S., {Foreman-Mackey}, D., {Greengard}, L., {Hogg}, D.~W., \&
  {O'Neil}, M. 2014, ArXiv e-prints, arXiv:1403.6015

\bibitem[{{An} {et~al.}(2007){An}, {Terndrup}, {Pinsonneault}, {Paulson},
  {Hanson}, \& {Stauffer}}]{An07}
{An}, D., {Terndrup}, D.~M., {Pinsonneault}, M.~H., {et~al.} 2007, \apj, 655,
  233

\bibitem[{{Astropy Collaboration} {et~al.}(2013){Astropy Collaboration},
  {Robitaille}, {Tollerud}, {Greenfield}, {Droettboom}, {Bray}, {Aldcroft},
  {Davis}, {Ginsburg}, {Price-Whelan}, {Kerzendorf}, {Conley}, {Crighton},
  {Barbary}, {Muna}, {Ferguson}, {Grollier}, {Parikh}, {Nair}, {Unther},
  {Deil}, {Woillez}, {Conseil}, {Kramer}, {Turner}, {Singer}, {Fox}, {Weaver},
  {Zabalza}, {Edwards}, {Azalee Bostroem}, {Burke}, {Casey}, {Crawford},
  {Dencheva}, {Ely}, {Jenness}, {Labrie}, {Lim}, {Pierfederici}, {Pontzen},
  {Ptak}, {Refsdal}, {Servillat}, \& {Streicher}}]{astropy13}
{Astropy Collaboration}, {Robitaille}, T.~P., {Tollerud}, E.~J., {et~al.} 2013,
  \aap, 558, A33

\bibitem[{{Baker} {et~al.}(2010){Baker}, {Jameson}, {Casewell}, {Deacon},
  {Lodieu}, \& {Hambly}}]{Baker10}
{Baker}, D.~E.~A., {Jameson}, R.~F., {Casewell}, S.~L., {et~al.} 2010, \mnras,
  408, 2457

\bibitem[{{Baraffe} {et~al.}(2015){Baraffe}, {Homeier}, {Allard}, \&
  {Chabrier}}]{Baraffe15}
{Baraffe}, I., {Homeier}, D., {Allard}, F., \& {Chabrier}, G. 2015, \aap, 577,
  A42

\bibitem[{{Barnes} {et~al.}(2005){Barnes}, {Collier Cameron}, {Donati},
  {James}, {Marsden}, \& {Petit}}]{Barnes05}
{Barnes}, J.~R., {Collier Cameron}, A., {Donati}, J.-F., {et~al.} 2005, \mnras,
  357, L1

\bibitem[{{Barnes} {et~al.}(2015){Barnes}, {Jeffers}, {Jones}, {Pavlenko},
  {Jenkins}, {Haswell}, \& {Lohr}}]{Barnes15}
{Barnes}, J.~R., {Jeffers}, S.~V., {Jones}, H.~R.~A., {et~al.} 2015, \apj, 812,
  42

\bibitem[{{Bate}(2009)}]{Bate09}
{Bate}, M.~R. 2009, \mnras, 392, 590

\bibitem[{{Bayliss} {et~al.}(2017){Bayliss}, {Hojjatpanah}, {Santerne},
  {Dragomir}, {Zhou}, {Shporer}, {Col{\'o}n}, {Almenara}, {Armstrong},
  {Barrado}, {Barros}, {Bento}, {Boisse}, {Bouchy}, {Brown}, {Brown},
  {Cameron}, {Cochran}, {Demangeon}, {Deleuil}, {D{\'{\i}}az}, {Fulton},
  {Horne}, {H{\'e}brard}, {Lillo-Box}, {Lovis}, {Mawet}, {Ngo}, {Osborn},
  {Palle}, {Petigura}, {Pollacco}, {Santos}, {Sefako}, {Siverd}, {Sousa}, \&
  {Tsantaki}}]{Bayliss16}
{Bayliss}, D., {Hojjatpanah}, S., {Santerne}, A., {et~al.} 2017, \aj, 153, 15

\bibitem[{{Benedict} {et~al.}(2016){Benedict}, {Henry}, {Franz}, {McArthur},
  {Wasserman}, {Jao}, {Cargile}, {Dieterich}, {Bradley}, {Nelan}, \&
  {Whipple}}]{Benedict16}
{Benedict}, G.~F., {Henry}, T.~J., {Franz}, O.~G., {et~al.} 2016, \aj, 152, 141

\bibitem[{{Boesgaard} {et~al.}(2013){Boesgaard}, {Roper}, \&
  {Lum}}]{Boesgaard13}
{Boesgaard}, A.~M., {Roper}, B.~W., \& {Lum}, M.~G. 2013, \apj, 775, 58

\bibitem[{{Boudreault} {et~al.}(2012){Boudreault}, {Lodieu}, {Deacon}, \&
  {Hambly}}]{Boudreault12}
{Boudreault}, S., {Lodieu}, N., {Deacon}, N.~R., \& {Hambly}, N.~C. 2012,
  \mnras, 426, 3419

\bibitem[{{Brandt} \& {Huang}(2015{\natexlab{a}})}]{Brandt15a}
{Brandt}, T.~D., \& {Huang}, C.~X. 2015{\natexlab{a}}, \apj, 807, 58

\bibitem[{{Brandt} \& {Huang}(2015{\natexlab{b}})}]{Brandt15}
---. 2015{\natexlab{b}}, \apj, 807, 24

\bibitem[{{Bressan} {et~al.}(2012){Bressan}, {Marigo}, {Girardi}, {Salasnich},
  {Dal Cero}, {Rubele}, \& {Nanni}}]{Bressan12}
{Bressan}, A., {Marigo}, P., {Girardi}, L., {et~al.} 2012, \mnras, 427, 127

\bibitem[{{Cargile} {et~al.}(2008){Cargile}, {Stassun}, \&
  {Mathieu}}]{Cargile08}
{Cargile}, P.~A., {Stassun}, K.~G., \& {Mathieu}, R.~D. 2008, \apj, 674, 329

\bibitem[{{Chabrier} {et~al.}(2007){Chabrier}, {Gallardo}, \&
  {Baraffe}}]{Chabrier07}
{Chabrier}, G., {Gallardo}, J., \& {Baraffe}, I. 2007, \aap, 472, L17

\bibitem[{{Chen} {et~al.}(2014){Chen}, {Girardi}, {Bressan}, {Marigo},
  {Barbieri}, \& {Kong}}]{Chen14}
{Chen}, Y., {Girardi}, L., {Bressan}, A., {et~al.} 2014, \mnras, 444, 2525

\bibitem[{{Claret} \& {Cunha}(1997)}]{ClaretCunha97}
{Claret}, A., \& {Cunha}, N.~C.~S. 1997, \aap, 318, 187

\bibitem[{{Claret} {et~al.}(2012){Claret}, {Hauschildt}, \& {Witte}}]{Claret12}
{Claret}, A., {Hauschildt}, P.~H., \& {Witte}, S. 2012, \aap, 546, A14

\bibitem[{{Collier Cameron}(2007)}]{CollierCameron07}
{Collier Cameron}, A. 2007, Astronomische Nachrichten, 328, 1030

\bibitem[{{Covino} {et~al.}(2004){Covino}, {Frasca}, {Alcal{\'a}}, {Paladino},
  \& {Sterzik}}]{Covino04}
{Covino}, E., {Frasca}, A., {Alcal{\'a}}, J.~M., {Paladino}, R., \& {Sterzik},
  M.~F. 2004, \aap, 427, 637

\bibitem[{{Covino} {et~al.}(2001){Covino}, {Melo}, {Alcal{\'a}}, {Torres},
  {Fern{\'a}ndez}, {Frasca}, \& {Paladino}}]{Covino01}
{Covino}, E., {Melo}, C., {Alcal{\'a}}, J.~M., {et~al.} 2001, \aap, 375, 130

\bibitem[{{Covino} {et~al.}(2000){Covino}, {Catalano}, {Frasca}, {Marilli},
  {Fern{\'a}ndez}, {Alcal{\'a}}, {Melo}, {Paladino}, {Sterzik}, \&
  {Stelzer}}]{Covino00}
{Covino}, E., {Catalano}, S., {Frasca}, A., {et~al.} 2000, \aap, 361, L49

\bibitem[{{Csizmadia}(2016)}]{Csizmadia16}
{Csizmadia}, S. 2016, {III.6 Exploration of the brown dwarf regime around
  solar-like stars by CoRoT}, ed. {CoRot Team}, 143

\bibitem[{{Czekala} {et~al.}(2015){Czekala}, {Andrews}, {Mandel}, {Hogg}, \&
  {Green}}]{Czekala15}
{Czekala}, I., {Andrews}, S.~M., {Mandel}, K.~S., {Hogg}, D.~W., \& {Green},
  G.~M. 2015, \apj, 812, 128

\bibitem[{{Czekala} {et~al.}(2017){Czekala}, {Mandel}, {Andrews}, {Dittmann},
  {Ghosh}, {Montet}, \& {Newton}}]{Czekala17}
{Czekala}, I., {Mandel}, K.~S., {Andrews}, S.~M., {et~al.} 2017, ArXiv
  e-prints, arXiv:1702.05652

\bibitem[{{Davenport} {et~al.}(2015){Davenport}, {Hebb}, \&
  {Hawley}}]{Davenport15}
{Davenport}, J.~R.~A., {Hebb}, L., \& {Hawley}, S.~L. 2015, \apj, 806, 212

\bibitem[{{David} \& {Hillenbrand}(2015)}]{David15}
{David}, T.~J., \& {Hillenbrand}, L.~A. 2015, \apj, 804, 146

\bibitem[{{David} {et~al.}(2016{\natexlab{a}}){David}, {Hillenbrand}, {Cody},
  {Carpenter}, \& {Howard}}]{David16}
{David}, T.~J., {Hillenbrand}, L.~A., {Cody}, A.~M., {Carpenter}, J.~M., \&
  {Howard}, A.~W. 2016{\natexlab{a}}, \apj, 816, 21

\bibitem[{{David} {et~al.}(2015){David}, {Stauffer}, {Hillenbrand}, {Cody},
  {Conroy}, {Stassun}, {Pope}, {Aigrain}, {Gillen}, {Collier Cameron},
  {Barrado}, {Rebull}, {Isaacson}, {Marcy}, {Zhang}, {Riddle}, {Ziegler},
  {Law}, \& {Baranec}}]{David15b}
{David}, T.~J., {Stauffer}, J., {Hillenbrand}, L.~A., {et~al.} 2015, \apj, 814,
  62

\bibitem[{{David} {et~al.}(2016{\natexlab{b}}){David}, {Conroy}, {Hillenbrand},
  {Stassun}, {Stauffer}, {Rebull}, {Cody}, {Isaacson}, {Howard}, \&
  {Aigrain}}]{David16a}
{David}, T.~J., {Conroy}, K.~E., {Hillenbrand}, L.~A., {et~al.}
  2016{\natexlab{b}}, \aj, 151, 112

\bibitem[{{Dittmann} {et~al.}(2017){Dittmann}, {Irwin}, {Charbonneau},
  {Berta-Thompson}, {Newton}, {Latham}, {Latham}, {Esquerdo}, {Berlind}, \&
  {Calkins}}]{Dittmann17}
{Dittmann}, J.~A., {Irwin}, J.~M., {Charbonneau}, D., {et~al.} 2017, \apj, 836,
  124

\bibitem[{{Feiden}(2015)}]{Feiden15}
{Feiden}, G.~A. 2015, in Astronomical Society of the Pacific Conference Series,
  Vol. 496, Astronomical Society of the Pacific Conference Series, ed. S.~M.
  {Rucinski}, G.~{Torres}, \& M.~{Zejda}, 137

\bibitem[{{Feiden} \& {Chaboyer}(2012)}]{Feiden12}
{Feiden}, G.~A., \& {Chaboyer}, B. 2012, \apj, 757, 42

\bibitem[{{Foreman-Mackey} {et~al.}(2013){Foreman-Mackey}, {Hogg}, {Lang}, \&
  {Goodman}}]{Foreman-Mackey13}
{Foreman-Mackey}, D., {Hogg}, D.~W., {Lang}, D., \& {Goodman}, J. 2013, \pasp,
  125, 306

\bibitem[{{Fossati} {et~al.}(2008){Fossati}, {Bagnulo}, {Landstreet}, {Wade},
  {Kochukhov}, {Monier}, {Weiss}, \& {Gebran}}]{Fossati08}
{Fossati}, L., {Bagnulo}, S., {Landstreet}, J., {et~al.} 2008, \aap, 483, 891

\bibitem[{{Gaia Collaboration} {et~al.}(2017){Gaia Collaboration}, {van
  Leeuwen}, {Vallenari}, {Jordi}, {Lindegren}, {Bastian}, {Prusti}, {de
  Bruijne}, {Brown}, {Babusiaux}, \& et~al.}]{vanLeeuwen17}
{Gaia Collaboration}, {van Leeuwen}, F., {Vallenari}, A., {et~al.} 2017, ArXiv
  e-prints, arXiv:1703.01131

\bibitem[{{Gatewood} \& {de Jonge}(1994)}]{Gatewood94}
{Gatewood}, G., \& {de Jonge}, J.~K. 1994, \apj, 428, 166

\bibitem[{{Gillen} {et~al.}(2014){Gillen}, {Aigrain}, {McQuillan}, {Bouvier},
  {Hodgkin}, {Alencar}, {Terquem}, {Southworth}, {Gibson}, {Cody}, {Lendl},
  {Morales-Calder{\'o}n}, {Favata}, {Stauffer}, \& {Micela}}]{Gillen14}
{Gillen}, E., {Aigrain}, S., {McQuillan}, A., {et~al.} 2014, \aap, 562, A50

\bibitem[{{Grankin}(1998)}]{Grankin98}
{Grankin}, K.~N. 1998, Astronomy Letters, 24, 497

\bibitem[{{Granzer} {et~al.}(2000){Granzer}, {Sch{\"u}ssler}, {Caligari}, \&
  {Strassmeier}}]{Granzer00}
{Granzer}, T., {Sch{\"u}ssler}, M., {Caligari}, P., \& {Strassmeier}, K.~G.
  2000, \aap, 355, 1087

\bibitem[{{Hambly} {et~al.}(1995){Hambly}, {Steele}, {Hawkins}, \&
  {Jameson}}]{Hambly95}
{Hambly}, N.~C., {Steele}, I.~A., {Hawkins}, M.~R.~S., \& {Jameson}, R.~F.
  1995, \aaps, 109

\bibitem[{{Hebb} {et~al.}(2011){Hebb}, {Cegla}, {Stassun}, {Stempels},
  {Cargile}, \& {Palladino}}]{Hebb11}
{Hebb}, L., {Cegla}, H.~M., {Stassun}, K.~G., {et~al.} 2011, \aap, 531, A61

\bibitem[{{Hebb} {et~al.}(2006){Hebb}, {Wyse}, {Gilmore}, \&
  {Holtzman}}]{Hebb06}
{Hebb}, L., {Wyse}, R.~F.~G., {Gilmore}, G., \& {Holtzman}, J. 2006, \aj, 131,
  555

\bibitem[{{Hebb} {et~al.}(2010){Hebb}, {Stempels}, {Aigrain},
  {Collier-Cameron}, {Hodgkin}, {Irwin}, {Maxted}, {Pollacco}, {Street},
  {Wilson}, \& {Stassun}}]{Hebb10}
{Hebb}, L., {Stempels}, H.~C., {Aigrain}, S., {et~al.} 2010, \aap, 522, A37

\bibitem[{{Howell} {et~al.}(2014){Howell}, {Sobeck}, {Haas}, {Still},
  {Barclay}, {Mullally}, {Troeltzsch}, {Aigrain}, {Bryson}, {Caldwell},
  {Chaplin}, {Cochran}, {Huber}, {Marcy}, {Miglio}, {Najita}, {Smith},
  {Twicken}, \& {Fortney}}]{Howell14}
{Howell}, S.~B., {Sobeck}, C., {Haas}, M., {et~al.} 2014, \pasp, 126, 398

\bibitem[{{Husser} {et~al.}(2013){Husser}, {Wende-von Berg}, {Dreizler},
  {Homeier}, {Reiners}, {Barman}, \& {Hauschildt}}]{Husser13}
{Husser}, T.-O., {Wende-von Berg}, S., {Dreizler}, S., {et~al.} 2013, \aap,
  553, A6

\bibitem[{{Hut}(1981)}]{Hut81}
{Hut}, P. 1981, \aap, 99, 126

\bibitem[{{Irwin} {et~al.}(2007){Irwin}, {Aigrain}, {Hodgkin}, {Stassun},
  {Hebb}, {Irwin}, {Moraux}, {Bouvier}, {Alapini}, {Alexander}, {Bramich},
  {Holtzman}, {Mart{\'{\i}}n}, {McCaughrean}, {Pont}, {Verrier}, \& {Zapatero
  Osorio}}]{Irwin07}
{Irwin}, J., {Aigrain}, S., {Hodgkin}, S., {et~al.} 2007, \mnras, 380, 541

\bibitem[{{Irwin} {et~al.}(2011){Irwin}, {Quinn}, {Berta}, {Latham}, {Torres},
  {Burke}, {Charbonneau}, {Dittmann}, {Esquerdo}, {Stefanik}, {Oksanen},
  {Buchhave}, {Nutzman}, {Berlind}, {Calkins}, \& {Falco}}]{Irwin11}
{Irwin}, J.~M., {Quinn}, S.~N., {Berta}, Z.~K., {et~al.} 2011, \apj, 742, 123

\bibitem[{{Jones} \& {Cudworth}(1983)}]{Jones83}
{Jones}, B.~F., \& {Cudworth}, K. 1983, \aj, 88, 215

\bibitem[{{Jones} \& {Stauffer}(1991)}]{Jones91}
{Jones}, B.~F., \& {Stauffer}, J.~R. 1991, \aj, 102, 1080

\bibitem[{{Kafka} \& {Honeycutt}(2006)}]{Kafka06}
{Kafka}, S., \& {Honeycutt}, R.~K. 2006, \aj, 132, 1517

\bibitem[{{Khalaj} \& {Baumgardt}(2013)}]{Khalaj13}
{Khalaj}, P., \& {Baumgardt}, H. 2013, \mnras, 434, 3236

\bibitem[{{Kipping}(2013)}]{Kipping13}
{Kipping}, D.~M. 2013, \mnras, 435, 2152

\bibitem[{{Klein Wassink}(1927)}]{Klein-Wassink27}
{Klein Wassink}, W.~J. 1927, Publications of the Kapteyn Astronomical
  Laboratory Groningen, 41, 1

\bibitem[{{Kraus} {et~al.}(2015){Kraus}, {Cody}, {Covey}, {Rizzuto}, {Mann}, \&
  {Ireland}}]{Kraus15}
{Kraus}, A.~L., {Cody}, A.~M., {Covey}, K.~R., {et~al.} 2015, ArXiv e-prints,
  arXiv:1505.02446

\bibitem[{{Kraus} \& {Hillenbrand}(2007)}]{Kraus07}
{Kraus}, A.~L., \& {Hillenbrand}, L.~A. 2007, \aj, 134, 2340

\bibitem[{{Lacy} {et~al.}(2016){Lacy}, {Fekel}, {Pavlovski}, {Torres}, \&
  {Muterspaugh}}]{Lacy16}
{Lacy}, C.~H.~S., {Fekel}, F.~C., {Pavlovski}, K., {Torres}, G., \&
  {Muterspaugh}, M.~W. 2016, \aj, 152, 2

\bibitem[{{Lodieu} {et~al.}(2015){Lodieu}, {Alonso}, {Gonz{\'a}lez
  Hern{\'a}ndez}, {Sanchis-Ojeda}, {Narita}, {Kawashima}, {Kawauchi},
  {Su{\'a}rez Mascare{\~n}o}, {Deeg}, {Prieto Arranz}, {Rebolo}, {Pall{\'e}},
  {B{\'e}jar}, {Ferragamo}, \& {Rubi{\~n}o-Mart{\'{\i}}n}}]{Lodieu15}
{Lodieu}, N., {Alonso}, R., {Gonz{\'a}lez Hern{\'a}ndez}, J.~I., {et~al.} 2015,
  \aap, 584, A128

\bibitem[{{Lucy}(1967)}]{Lucy67}
{Lucy}, L.~B. 1967, \zap, 65, 89

\bibitem[{{MacDonald} \& {Mullan}(2014)}]{Macdonald14}
{MacDonald}, J., \& {Mullan}, D.~J. 2014, \apj, 787, 70

\bibitem[{{Mandel} \& {Agol}(2002)}]{Mandel02}
{Mandel}, K., \& {Agol}, E. 2002, \apjl, 580, L171

\bibitem[{{Mann} {et~al.}(2015){Mann}, {Feiden}, {Gaidos}, {Boyajian}, \& {von
  Braun}}]{Mann16}
{Mann}, A.~W., {Feiden}, G.~A., {Gaidos}, E., {Boyajian}, T., \& {von Braun},
  K. 2015, \apj, 804, 64

\bibitem[{{Mathieu} \& {Mazeh}(1988)}]{MathieuMazeh88}
{Mathieu}, R.~D., \& {Mazeh}, T. 1988, \apj, 326, 256

\bibitem[{{Meibom} \& {Mathieu}(2005)}]{Meibom05}
{Meibom}, S., \& {Mathieu}, R.~D. 2005, \apj, 620, 970

\bibitem[{{Meibom} {et~al.}(2006){Meibom}, {Mathieu}, \& {Stassun}}]{Meibom06}
{Meibom}, S., {Mathieu}, R.~D., \& {Stassun}, K.~G. 2006, \apj, 653, 621

\bibitem[{{Mermilliod} {et~al.}(1990){Mermilliod}, {Weis}, {Duquennoy}, \&
  {Mayor}}]{Mermilliod90}
{Mermilliod}, J.-C., {Weis}, E.~W., {Duquennoy}, A., \& {Mayor}, M. 1990, \aap,
  235, 114

\bibitem[{{Morales-Calder{\'o}n} {et~al.}(2012){Morales-Calder{\'o}n},
  {Stauffer}, {Stassun}, {Vrba}, {Prato}, {Hillenbrand}, {Terebey}, {Covey},
  {Rebull}, {Terndrup}, {Gutermuth}, {Song}, {Plavchan}, {Carpenter},
  {Marchis}, {Garc{\'{\i}}a}, {Margheim}, {Luhman}, {Angione}, \&
  {Irwin}}]{Morales-Calderon12}
{Morales-Calder{\'o}n}, M., {Stauffer}, J.~R., {Stassun}, K.~G., {et~al.} 2012,
  \apj, 753, 149

\bibitem[{{Morin} {et~al.}(2008){Morin}, {Donati}, {Petit}, {Delfosse},
  {Forveille}, {Albert}, {Auri{\`e}re}, {Cabanac}, {Dintrans}, {Fares},
  {Gastine}, {Jardine}, {Ligni{\`e}res}, {Paletou}, {Ramirez Velez}, \&
  {Th{\'e}ado}}]{Morin08}
{Morin}, J., {Donati}, J.-F., {Petit}, P., {et~al.} 2008, \mnras, 390, 567

\bibitem[{{Netopil} {et~al.}(2016){Netopil}, {Paunzen}, {Heiter}, \&
  {Soubiran}}]{Netopil16}
{Netopil}, M., {Paunzen}, E., {Heiter}, U., \& {Soubiran}, C. 2016, \aap, 585,
  A150

\bibitem[{{Neuh{\"a}user} {et~al.}(2011){Neuh{\"a}user}, {Errmann}, {Berndt},
  {Maciejewski}, {Takahashi}, {Chen}, {Dimitrov}, {Pribulla}, {Nikogossian},
  {Jensen}, {Marschall}, {Wu}, {Kellerer}, {Walter}, {Brice{\~n}o}, {Chini},
  {Fernandez}, {Raetz}, {Torres}, {Latham}, {Quinn}, {Niedzielski},
  {Bukowiecki}, {Nowak}, {Tomov}, {Tachihara}, {Hu}, {Hung}, {Kjurkchieva},
  {Radeva}, {Mihov}, {Slavcheva-Mihova}, {Bozhinova}, {Budaj}, {Va{\v n}ko},
  {Kundra}, {Hamb{\'a}lek}, {Krushevska}, {Movsessian}, {Harutyunyan},
  {Downes}, {Hernandez}, {Hoffmeister}, {Cohen}, {Abel}, {Ahmad}, {Chapman},
  {Eckert}, {Goodman}, {Guerard}, {Kim}, {Koontharana}, {Sokol}, {Trinh},
  {Wang}, {Zhou}, {Redmer}, {Kramm}, {Nettelmann}, {Mugrauer}, {Schmidt},
  {Moualla}, {Ginski}, {Marka}, {Adam}, {Seeliger}, {Baar}, {Roell}, {Schmidt},
  {Trepl}, {Eisenbei{\ss}}, {Fiedler}, {Tetzlaff}, {Schmidt}, {Hohle}, {Kitze},
  {Chakrova}, {Gr{\"a}fe}, {Schreyer}, {Hambaryan}, {Broeg}, {Koppenhoefer}, \&
  {Pandey}}]{Neuhauser11}
{Neuh{\"a}user}, R., {Errmann}, R., {Berndt}, A., {et~al.} 2011, Astronomische
  Nachrichten, 332, 547

\bibitem[{{Nidever} {et~al.}(2002){Nidever}, {Marcy}, {Butler}, {Fischer}, \&
  {Vogt}}]{Nidever02}
{Nidever}, D.~L., {Marcy}, G.~W., {Butler}, R.~P., {Fischer}, D.~A., \& {Vogt},
  S.~S. 2002, \apjs, 141, 503

\bibitem[{{Nowak} {et~al.}(2016){Nowak}, {Palle}, {Gandolfi}, {Dai}, {Lanza},
  {Hirano}, {Barrag{\'a}n}, {Fukui}, {Bruntt}, {Endl}, {Cochran},
  {Prieto-Arranz}, {Kiilerich}, {Nespral}, {Hatzes}, {Albrecht}, {Deeg},
  {Winn}, {Yu}, {Kuzuhara}, {Grziwa}, {Smith}, {Prada Moroni}, {Guenther}, {Van
  Eylen}, {Csizmadia}, {Fridlund}, {Cabrera}, {Eigm{\"u}ller}, {Erikson},
  {Korth}, {Narita}, {P{\"a}tzold}, {Rauer}, \& {Ribas}}]{Nowak16}
{Nowak}, G., {Palle}, E., {Gandolfi}, D., {et~al.} 2016, ArXiv e-prints,
  arXiv:1610.08571

\bibitem[{{Ogilvie}(2014)}]{Ogilvie14}
{Ogilvie}, G.~I. 2014, \araa, 52, 171

\bibitem[{{O'Neal} {et~al.}(2004){O'Neal}, {Neff}, {Saar}, \&
  {Cuntz}}]{ONeal04}
{O'Neal}, D., {Neff}, J.~E., {Saar}, S.~H., \& {Cuntz}, M. 2004, \aj, 128, 1802

\bibitem[{{Pace} {et~al.}(2008){Pace}, {Pasquini}, \& {Fran{\c c}ois}}]{Pace08}
{Pace}, G., {Pasquini}, L., \& {Fran{\c c}ois}, P. 2008, \aap, 489, 403

\bibitem[{{Parviainen} \& {Aigrain}(2015)}]{Parviainen15}
{Parviainen}, H., \& {Aigrain}, S. 2015, \mnras, 453, 3821

\bibitem[{{Pecaut} \& {Mamajek}(2013)}]{Pecaut13}
{Pecaut}, M.~J., \& {Mamajek}, E.~E. 2013, \apjs, 208, 9

\bibitem[{{Pepper} {et~al.}(2017){Pepper}, {Gillen}, {Parviainen},
  {Hillenbrand}, {Cody}, {Aigrain}, {Stauffer}, {Vrba}, {David}, {Lillo-Box},
  {Stassun}, {Conroy}, {Pope}, \& {Barrado}}]{Pepper17}
{Pepper}, J., {Gillen}, E., {Parviainen}, H., {et~al.} 2017, \aj, 153, 177

\bibitem[{{Percival} {et~al.}(2003){Percival}, {Salaris}, \&
  {Kilkenny}}]{Percival03}
{Percival}, S.~M., {Salaris}, M., \& {Kilkenny}, D. 2003, \aap, 400, 541

\bibitem[{{Perryman} {et~al.}(1998){Perryman}, {Brown}, {Lebreton}, {Gomez},
  {Turon}, {Cayrel de Strobel}, {Mermilliod}, {Robichon}, {Kovalevsky}, \&
  {Crifo}}]{Perryman98}
{Perryman}, M.~A.~C., {Brown}, A.~G.~A., {Lebreton}, Y., {et~al.} 1998, \aap,
  331, 81

\bibitem[{{Pinfield} {et~al.}(2003){Pinfield}, {Dobbie}, {Jameson}, {Steele},
  {Jones}, \& {Katsiyannis}}]{Pinfield03}
{Pinfield}, D.~J., {Dobbie}, P.~D., {Jameson}, R.~F., {et~al.} 2003, \mnras,
  342, 1241

\bibitem[{{Quinn} {et~al.}(2012){Quinn}, {White}, {Latham}, {Buchhave},
  {Cantrell}, {Dahm}, {F{\H u}r{\'e}sz}, {Szentgyorgyi}, {Geary}, {Torres},
  {Bieryla}, {Berlind}, {Calkins}, {Esquerdo}, \& {Stefanik}}]{Quinn12}
{Quinn}, S.~N., {White}, R.~J., {Latham}, D.~W., {et~al.} 2012, \apjl, 756, L33

\bibitem[{{Rasmussen} \& {Williams}(2006)}]{Rasmussen06}
{Rasmussen}, C.~E., \& {Williams}, C.~K.~I. 2006, Gaussian Processes for
  Machine Learning, MIT Press

\bibitem[{{Rebull} {et~al.}(2017){Rebull}, {Stauffer}, {Hillenbrand}, {Cody},
  {Bouvier}, {Soderblom}, {Pinsonneault}, \& {Hebb}}]{Rebull17}
{Rebull}, L.~M., {Stauffer}, J.~R., {Hillenbrand}, L.~A., {et~al.} 2017, AAS,
  refereed

\bibitem[{{Reglero} \& {Fabregat}(1991)}]{Reglero91}
{Reglero}, V., \& {Fabregat}, J. 1991, \aaps, 90, 25

\bibitem[{{Reinhold} {et~al.}(2013){Reinhold}, {Reiners}, \&
  {Basri}}]{Reinhold13}
{Reinhold}, T., {Reiners}, A., \& {Basri}, G. 2013, \aap, 560, A4

\bibitem[{Roberts {et~al.}(2012)Roberts, Osborne, Ebden, Reece, Gibson, \&
  Aigrain}]{Roberts12}
Roberts, S., Osborne, M., Ebden, M., {et~al.} 2012, Philosophical Transactions
  of the Royal Society of London A: Mathematical, Physical and Engineering
  Sciences, 371,
  http://rsta.royalsocietypublishing.org/content/371/1984/20110550.full.pdf

\bibitem[{{Salaris} {et~al.}(2004){Salaris}, {Weiss}, \&
  {Percival}}]{Salaris04}
{Salaris}, M., {Weiss}, A., \& {Percival}, S.~M. 2004, \aap, 414, 163

\bibitem[{{Schuessler} \& {Solanki}(1992)}]{Schuessler92}
{Schuessler}, M., \& {Solanki}, S.~K. 1992, \aap, 264, L13

\bibitem[{{Sing}(2010)}]{Sing10}
{Sing}, D.~K. 2010, \aap, 510, A21

\bibitem[{{Stassun} {et~al.}(2014){Stassun}, {Feiden}, \& {Torres}}]{Stassun14}
{Stassun}, K.~G., {Feiden}, G.~A., \& {Torres}, G. 2014, \nar, 60, 1

\bibitem[{{Stassun} {et~al.}(2008){Stassun}, {Mathieu}, {Cargile}, {Aarnio},
  {Stempels}, \& {Geller}}]{Stassun08}
{Stassun}, K.~G., {Mathieu}, R.~D., {Cargile}, P.~A., {et~al.} 2008, \nat, 453,
  1079

\bibitem[{{Stassun} {et~al.}(2006){Stassun}, {Mathieu}, \&
  {Valenti}}]{Stassun06}
{Stassun}, K.~G., {Mathieu}, R.~D., \& {Valenti}, J.~A. 2006, \nat, 440, 311

\bibitem[{{Stassun} {et~al.}(2007){Stassun}, {Mathieu}, \&
  {Valenti}}]{Stassun07}
---. 2007, \apj, 664, 1154

\bibitem[{{Stassun} {et~al.}(2004){Stassun}, {Mathieu}, {Vaz}, {Stroud}, \&
  {Vrba}}]{Stassun04}
{Stassun}, K.~G., {Mathieu}, R.~D., {Vaz}, L.~P.~R., {Stroud}, N., \& {Vrba},
  F.~J. 2004, \apjs, 151, 357

\bibitem[{{Stauffer}(1982)}]{Stauffer82}
{Stauffer}, J. 1982, \pasp, 94, 678

\bibitem[{{Stempels} {et~al.}(2008){Stempels}, {Hebb}, {Stassun}, {Holtzman},
  {Dunstone}, {Glowienka}, \& {Frandsen}}]{Stempels08}
{Stempels}, H.~C., {Hebb}, L., {Stassun}, K.~G., {et~al.} 2008, \aap, 481, 747

\bibitem[{{Taylor}(2006)}]{Taylor06}
{Taylor}, B.~J. 2006, \aj, 132, 2453

\bibitem[{{Torres} {et~al.}(2010){Torres}, {Andersen}, \&
  {Gim{\'e}nez}}]{Torres10}
{Torres}, G., {Andersen}, J., \& {Gim{\'e}nez}, A. 2010, \aapr, 18, 67

\bibitem[{{Torres} \& {Ribas}(2002)}]{Torres02}
{Torres}, G., \& {Ribas}, I. 2002, \apj, 567, 1140

\bibitem[{{van Eyken} {et~al.}(2011){van Eyken}, {Ciardi}, {Rebull},
  {Stauffer}, {Akeson}, {Beichman}, {Boden}, {von Braun}, {Gelino}, {Hoard},
  {Howell}, {Kane}, {Plavchan}, {Ram{\'{\i}}rez}, {Bloom}, {Cenko}, {Kasliwal},
  {Kulkarni}, {Law}, {Nugent}, {Ofek}, {Poznanski}, {Quimby}, {Grillmair},
  {Laher}, {Levitan}, {Mattingly}, \& {Surace}}]{vanEyken11}
{van Eyken}, J.~C., {Ciardi}, D.~R., {Rebull}, L.~M., {et~al.} 2011, \aj, 142,
  60

\bibitem[{{Van Eylen} {et~al.}(2016){Van Eylen}, {Winn}, \&
  {Albrecht}}]{vanEylen16}
{Van Eylen}, V., {Winn}, J.~N., \& {Albrecht}, S. 2016, \apj, 824, 15

\bibitem[{{van Leeuwen}(2009)}]{vanLeeuwen09}
{van Leeuwen}, F. 2009, \aap, 497, 209

\bibitem[{{Vogt} {et~al.}(1994){Vogt}, {Allen}, {Bigelow}, {Bresee}, {Brown},
  {Cantrall}, {Conrad}, {Couture}, {Delaney}, {Epps}, {Hilyard}, {Hilyard},
  {Horn}, {Jern}, {Kanto}, {Keane}, {Kibrick}, {Lewis}, {Osborne},
  {Pardeilhan}, {Pfister}, {Ricketts}, {Robinson}, {Stover}, {Tucker}, {Ward},
  \& {Wei}}]{Vogt94}
{Vogt}, S.~S., {Allen}, S.~L., {Bigelow}, B.~C., {et~al.} 1994, in \procspie,
  Vol. 2198, Instrumentation in Astronomy VIII, ed. D.~L. {Crawford} \& E.~R.
  {Craine}, 362

\bibitem[{{Wang} {et~al.}(2014){Wang}, {Chen}, {Lin}, {Pandey}, {Huang},
  {Panwar}, {Lee}, {Tsai}, {Tang}, {Goldman}, {Burgett}, {Chambers}, {Draper},
  {Flewelling}, {Grav}, {Heasley}, {Hodapp}, {Huber}, {Jedicke}, {Kaiser},
  {Kudritzki}, {Luppino}, {Lupton}, {Magnier}, {Metcalfe}, {Monet}, {Morgan},
  {Onaka}, {Price}, {Stubbs}, {Sweeney}, {Tonry}, {Wainscoat}, \&
  {Waters}}]{Wang14}
{Wang}, P.~F., {Chen}, W.~P., {Lin}, C.~C., {et~al.} 2014, \apj, 784, 57

\bibitem[{{West} {et~al.}(2011){West}, {Morgan}, {Bochanski}, {Andersen},
  {Bell}, {Kowalski}, {Davenport}, {Hawley}, {Schmidt}, {Bernat}, {Hilton},
  {Muirhead}, {Covey}, {Rojas-Ayala}, {Schlawin}, {Gooding}, {Schluns},
  {Dhital}, {Pineda}, \& {Jones}}]{West11}
{West}, A.~A., {Morgan}, D.~P., {Bochanski}, J.~J., {et~al.} 2011, \aj, 141, 97

\bibitem[{{Witte} \& {Savonije}(2002)}]{Witte02}
{Witte}, M.~G., \& {Savonije}, G.~J. 2002, \aap, 386, 222

\bibitem[{{Yang} {et~al.}(2015){Yang}, {Chen}, \& {Zhao}}]{Yang15}
{Yang}, X.~L., {Chen}, Y.~Q., \& {Zhao}, G. 2015, \aj, 150, 158

\bibitem[{{Zahn}(1975)}]{Zahn75}
{Zahn}, J.-P. 1975, \aap, 41, 329

\bibitem[{{Zahn}(1977)}]{Zahn77}
---. 1977, \aap, 57, 383

\bibitem[{{Zahn}(1989)}]{Zahn89a}
---. 1989, \aap, 220, 112

\bibitem[{{Zahn} \& {Bouchet}(1989)}]{Zahn89}
{Zahn}, J.-P., \& {Bouchet}, L. 1989, \aap, 223, 112

\end{thebibliography}

%\begin{thebibliography}{}
%
%%\bibitem[Astropy Collaboration et al.(2013)]{2013A&A...558A..33A} Astropy Collaboration, Robitaille, T.~P., Tollerud, E.~J., et al.\ 2013, \aap, 558, A33 
%%\bibitem[Bertin \& Arnouts(1996)]{1996A&AS..117..393B} Bertin, E., \& Arnouts, S.\ 1996, \aaps, 117, 393 
%%\bibitem[Corrales(2015)]{2015ApJ...805...23C} Corrales, L.\ 2015, \apj, 805, 23
%%\bibitem[Ferland et al.(2013)]{2013RMxAA..49..137F} Ferland, G.~J., Porter, R.~L., van Hoof, P.~A.~M., et al.\ 2013, \rmxaa, 49, 137
%%\bibitem[Hanisch \& Biemesderfer(1989)]{1989BAAS...21..780H} Hanisch, R.~J., \& Biemesderfer, C.~D.\ 1989, \baas, 21, 780 
%%\bibitem[Lamport(1994)]{lamport94} Lamport, L. 1994, LaTeX: A Document Preparation System, 2nd Edition (Boston, Addison-Wesley Professional)
%%\bibitem[Schwarz et al.(2011)]{2011ApJS..197...31S} Schwarz, G.~J., Ness, J.-U., Osborne, J.~P., et al.\ 2011, \apjs, 197, 31  
%%\bibitem[Vogt et al.(2014)]{2014ApJ...793..127V} Vogt, F.~P.~A., Dopita, M.~A., Kewley, L.~J., et al.\ 2014, \apj, 793, 127  
%
%\end{thebibliography}



\end{document}